\documentclass[10pt,prl,twocolumn,preprint,nofootinbib]{revtex4}

\usepackage{epsfig}
\usepackage{amsmath,amsfonts,amssymb}
\usepackage{t1enc}
\usepackage{verbatim}
\usepackage{float}
\usepackage{morefloats}

\providecommand{\openone}{\leavevmode\hbox{\small1\kern-3.8pt\normalsize1}}

\usepackage{booktabs}
\usepackage[Q=yes,pverb-linebreak=no]{examplep}

\begin{document}

\title{Probing the CP nature of the Higgs coupling in $t{\bar t}h$ events at the LHC}

\author{
S. Amor dos Santos$^1$,   
M.C.N. Fiolhais$^{1,2}$,
R. Frederix$^3$, 
R. Gon\c{c}alo$^{4}$, 
E. Gouveia$^5$, 
R. Martins$^{5}$, 
A. Onofre$^{5}$,
C.M. Pease$^{5}$,
H. Peixoto$^{6}$,
A. Reigoto$^5$, 
R. Santos$^{5,7,8}$, 
J. Silva$^6$ 
\\[3mm]
{\footnotesize {\it 
$^1$ LIP, Departamento de F\'{\i}sica, Universidade de Coimbra, 3004-516 Coimbra, Portugal\\
$^2$ Department of Science, Borough of Manhattan Community College, City University of New York, \\ 
     199 Chambers St, New York, NY 10007, USA \\
$^3$ Physik Department T31, Technische Universit\"at M\"unchen, James-Franck-Str.~1, D-85748 Garching, Germany \\
$^4$ LIP, Av. Elias Garcia, 14-1, 1000-149 Lisboa, Portugal\\
$^5$ LIP, Departamento de F\'{\i}sica, Universidade do Minho, 4710-057 Braga, Portugal\\
$^6$ Centro de F\'{\i}sica, Universidade do Minho, Campus de Gualtar, 4710-057 Braga, Portugal\\ 
$^7$ Instituto Superior de Engenharia de Lisboa - ISEL, 1959-007 Lisboa, Portugal \\
$^8$ Centro de F\'{\i}sica Te\'{o}rica e Computacional,
    Faculdade de Ci\^{e}ncias, Universidade de Lisboa, Campo Grande, Edif\'{\i}cio C8 1749-016 Lisboa, Portugal \\
}}
}

\begin{abstract}
The determination of the CP nature of the Higgs coupling to top quarks
is addressed in this paper, using $t{\bar t} h$ events produced in
$\sqrt{s} = 13$~TeV proton-proton collisions at the LHC. Dileptonic
final states are employed, with two oppositely charged leptons and
four jets, corresponding to the decays $t\rightarrow bW^+ \rightarrow
b \ell^+\nu_\ell$, $\bar{t}\rightarrow \bar{b}W^- \rightarrow \bar{b}
\ell^-\bar{\nu}_\ell$ and $h\rightarrow b\bar{b}$. Pure scalar ($h=H$), pure pseudo-scalar ($h=A$) and CP-violating Higgs boson signal events, generated with
MadGraph5\Q{_}aMC@NLO, are fully reconstructed through a kinematic
fit. We furthermore generate samples that have both a CP-even and a CP-odd component
in the $t \bar t h$ coupling in order to probe the ratio of the two components.
New angular distributions of the decay products, as well as CP
angular asymmetries, are explored in order to separate the scalar from the
pseudo-scalar components of the Higgs boson and reduce the
contribution from the dominant irreducible background, $t{\bar
  t}b{\bar b}$. Significant differences between the angular
distributions and asymmetries are observed, even after the full
kinematic fit reconstruction of the events, allowing to define the
best observables for a global fit of the Higgs couplings parameters.
\end{abstract}

\maketitle

\section{Introduction}

On July 2012, the discovery of a Higgs boson, predicted by the electroweak symmetry breaking mechanism~\cite{higgsmech} of the Standard Model (SM) of particle physics, with a mass close to 125~GeV, was announced by both ATLAS~\cite{:2012gk} and CMS~\cite{:2012gu} collaborations. Since then, studying the Higgs boson properties has motivated many physics analyses at the LHC. So far, the measured properties of the Higgs boson have shown remarkable consistency with those predicted by the SM~\cite{allexphiggs}. Nevertheless, it is by now clear that the SM cannot explain all the observed physical phenomena. One of the best known examples
is that it fails to explain the matter anti-matter asymmetry of the Universe, for which new sources of CP-violation 
beyond the SM (BSM) are required. One possibility would be to introduce CP violation in the Higgs sector. This is allowed in BSM models, such as supersymmetry and 2-Higgs doublets models (2HDM), where the Higgs boson(s) have no definite CP quantum number resulting in a Yukawa coupling with two components, one CP-even and one CP-odd (see for instance~\cite{Fontes:2015mea}).

Analyses focusing on the Higgs boson decays to photons, $ZZ$ and $WW$, as well as on the $VH$ ($V=W,Z$) associated production have been conducted to measure its spin and parity quantum numbers~\cite{Khachatryan:2014kca,Khachatryan:2016tnr,Aad:2015mxa}. All the results are consistent with a SM-like spin 0, parity even boson, while the pure pseudoscalar scenario has been excluded at a 99.98\% confidence level (CL). However, the possibility of a CP admixture manifestation in the Yukawa couplings remains to be probed
directly. So far only CP-odd components of the Higgs couplings to the weak gauge bosons were shown to be very small. 
Within all fermions, the top quark is expected to have the largest Yukawa coupling. Currently, this coupling can be measured indirectly from loop effects in $gg\rightarrow h$ and $h\rightarrow \gamma\gamma$, which suffer from large systematic uncertainty and require the assumption of no BSM contributions to the loops. This motivates the interest in associated production of the Higgs boson with a top quark pair ($t\bar{t}h$)\cite{ttHtheory}, which allows for a direct measurement of the top Yukawa coupling and provides sensitivity to its CP nature, through the rich kinematics of the events.

The main background contaminating $t\bar{t}h$ searches at the LHC is $pp \to t \bar t+\textrm{jets}$. In particular, if the dominant Higgs decay channel ($h\rightarrow b\bar{b}$) is analysed, $t\bar{t}b\bar{b}$ is a challenging irreducible background. Several $t\bar{t}h$ decay channels have been studied~\cite{Aad:2016zqi,Aad:2014lma,Aad:2015iha,Aad:2015gra,Khachatryan:2015ila,Khachatryan:2014qaa}. The very complex final states, together with the huge backgrounds, make it a particularly difficult Higgs process to study at the LHC. Nevertheless, both the ATLAS and CMS collaborations have reached remarkable sensitivities, with expected upper limits at 95\% CL for the $t\bar{t}H$ signal strength, $\mu$, below 2 in the background-only scenario. The best-fit values obtained for $\mu$ were $1.7\pm 0.8$ by ATLAS~\cite{Aad:2016zqi} and $2.8\pm 1.0$ by CMS~\cite{Khachatryan:2014qaa}. Combined results from both collaborations and from the various Higgs analyses were used to fit the signal strengths of five Higgs production processes, while assuming SM-like Higgs branching ratios~\cite{Khachatryan:2016vau}. The best-fit value obtained for $\mu(t\bar{t}H)$ was $2.3^{+0.7}_{-0.6}$.

In the present work, we address the dileptonic final state of $t\bar{t}$ with the Higgs boson decaying through $h\rightarrow b\bar{b}$. The two leptons in the final state make it a fairly clean channel, with the advantage that they preserve spin information from the top quarks. We investigate possible departures from the SM nature of the Higgs boson by comparing the kinematics of $t\bar{t}h$ signal samples with SM Higgs boson ($h=H$ and $J^{CP}=0^+$) to samples of $t\bar{t}h$ signal with pure pseudo-scalar Higgs boson ($h=A$ and $J^{CP}=0^-$). Furthermore, we use a general Yukawa coupling for the top quark defined as
\begin{equation}
\mathcal{L} = \kappa \, y_t \, \bar t \, (\cos \alpha + i \gamma_5 \sin \alpha) \, t \, h,
\end{equation}
where $y_t$ is the SM Higgs Yukawa coupling and $\alpha$ represents a CP phase. This approach allows us to probe the mixing between the CP-even and
the CP-odd components of the top quark Yukawa coupling to the 125 GeV Higgs. Note that with this Lagrangian $h$ has
no definite CP quantum number. The SM interaction is recovered for $\cos\alpha=\pm1$, while the pure pseudoscalar
is obtained by setting $\cos\alpha=0$.

Several observables in $t\bar{t}h$ events, sensitive to the CP nature of the top Yukawa coupling, have been proposed from which
we will study in detail the ones presented in~\cite{Gunion:1996xu,Boudjema:2015nda,Santos:2015dja} (other proposals including 
observables probing the CP nature of the $\tau^+ \tau^- h$ coupling were also discussed in~\cite{ttHCPtest, Brooijmans:2014eja}). While some rely on leptons in the dileptonic final state, more general obervables are obtained from the particles at production ($t$, $\bar{t}$ and $h$), only accessible experimentally through a reconstruction algorithm.

A full kinematical reconstruction method is applied to recover the four-momenta of the undetected neutrinos from the $W$-bosons decays, and a large set of new angular observables is presented. We will show that the information that is present in the matrix elements partially survives parton showering, detector simulation, event selection and event reconstruction. It has been suggested ~\cite{tth_spin,Artoisenet:2012st} that the different spins of $h$ in signal and $g$ in the $t\bar{t}b\bar{b}$ background ($g$ being a gluon which splits into $b\bar{b}$) can be exploited for background discrimination, through differences in angular distributions. In ~\cite{Santos:2015dja}, we presented a set of interesting observables for that effect, and we will demonstrate similar discriminating power for some of the observables introduced here. Even though we start by considering only the irreducible $t{\bar t}b{\bar b}$ background, without a highly-optimized event reconstruction method, we present results with a complete set of SM backgrounds and argue that our findings are also valid in a more general and realistic case. For other observables in this set, two signal samples, one with a scalar Higgs $H$ and another with a pseudoscalar Higgs $A$ are also differently distributed, suggesting the observables can be used to probe the CP nature of the top Yukawa coupling.

\section{Event generation, simulation and reconstruction}

The $t\bar{t}h$ signal events, as well as the dominant background process ($t \bar t b\bar b$) were generated at next to leading order (NLO) in QCD, using {\scshape MadGraph5\Q{_}aMC@NLO}~\cite{Alwall:2014hca} with the NNPDF2.3 PDF sets~\cite{Ball:2012cx}. The SM signal was generated using the default \texttt{sm} model in {\scshape MadGraph\Q{_}aMC@NLO}. The samples in which the Higgs has a non-zero CP-odd component were generated using the \texttt{HC\Q{_}NLO\Q{_}X0} model, described in \cite{Artoisenet:2013puc}. Signal samples were generated for values of $\cos\alpha$ ranging from -1 to 1 (in steps of 0.1). The model also allows the adjustment of effective couplings between the Higgs boson and vector bosons. Since $t\bar{t}h$ associated production with subsequent $h\rightarrow b\bar{b}$ decay is considered, those were all set to 0 (with the exceptions of $H_{\gamma\gamma}$, $A_{\gamma\gamma}$, $H_{Z\gamma}$ and $A_{Z\gamma}$). For this analysis not only contributions from the dominant background ($t \bar t b\bar b$), but also from other SM processes, were taken into account.
Samples of $t\bar{t}+jets$ (where {\it jets} stands for up to 3 additional $c$-jets or light-flavoured jets), 
$t\bar{t}V+jets$ (where $V$ can either be $Z$ or $W^{\pm}$ and {\it jets} can go up to 1 additional jet), 
single top quark production ($t$-channel, $s$-channel and $Wt$ with up to 1 additional jet), 
diboson ($WW,WZ,ZZ+jets$ with up to 3 additional jets), 
$W+jets$ and $Z+jets$ (with up to 4 additional jets), 
and $Wb\bar{b}+jets$ and $Zb\bar{b}+jets$ (with up to 2 additional jets), were generated at LO with {\scshape MadGraph5\Q{_}aMC@NLO}~\cite{Alwall:2014hca}. 
While the $t\bar{t}+jets$ sample was normalised to the QCD next-to-next-to leading order (NNLO) cross section with next-to-next-to leading logarithmic (NNLL) resummation of soft gluons~\cite{Czakon:2011xx, Botje:2011sn, Martin:2009bu,Gao:2013xoa,Ball:2012cx},  the single top quark production cross section was scaled to the approximate NNLO theoretical predictions ~\cite{Kidonakis:2010tc, Kidonakis:2011wy}, assuming the NNPDF2.3 PDF sets and scaled according to the generated top quark mass, following the prescription defined in~\cite{Czakon:2013goa}.

The full spin correlations information of the $t\rightarrow bW^+ \rightarrow b \ell^+\nu_\ell$, $\bar{t}\rightarrow \bar{b}W^- \rightarrow \bar{b}
\ell^-\bar{\nu}_\ell$ and $h\rightarrow b\bar{b}$ decays, with $\ell^\pm\in\{e^\pm,\mu^\pm\}$, is preserved
by using {\scshape MadSpin}~\cite{Artoisenet:2012st} to perform the decay chain of top quarks and Higgs bosons.
All events were generated for LHC $pp$ collisions, with a centre-of-mass energy of 13~TeV, with non-fixed renormalization and factorisation scales set to the sum of the transverse masses of all final state particles and partons.
The masses of the top quark ($m_t$), the $W$ boson ($m_W$) and Higgs bosons (for both scalar, $m_H$, and pseudo-scalar, $m_A$) were set to $173$~GeV, $80.4$~GeV and $125$~GeV, respectively. 

The events were then passed through {\scshape Pythia6}~\cite{Sjostrand:2006za} for parton shower and hadronization. Matching between the generator and the parton shower was performed using the MLM~\cite{Alwall:2007fs} scheme for LO events and the MC@NLO~\cite{Frixione:2002ik} matching for NLO events.
The {\scshape Delphes}~\cite{deFavereau:2013fsa} package was then used for a fast simulation of a general-purpose collider experiment, using the default ATLAS parameter card. During detector simulation, jets and charged leptons are reconstructed, as well as the transverse missing energy. The efficiencies and resolutions of the detector subsystems are parametrised in segments of $p_T$ (or $E$) and $\eta$. Particle tracking only occurs in the $\vert\eta\vert\leq2.5$ region, and its efficiency for a particle with $p_T=1$~GeV is, at least, 85\% for charged hadrons and 83\% (98\%) for electrons (muons). The momentum resolution of a track is at most 5\%. Calorimeters are segmented in $(\eta,\phi)$ rectangular cells. In the region with $\vert\eta\vert\leq2.5$, the cells have dimensions $(\eta,\phi)=(0.1, 10^{\circ})$, and for $2.5<\vert\eta\vert\leq4.9$, their size is $(\eta,\phi)=(0.2, 20^{\circ})$. Electron and muon identification efficiencies are 95\% in the central region $\vert\eta\vert\leq1.5$, 85\% in the intermediate region $1.5<\vert\eta\vert\leq2.5$ ($2.7$ for muons), and zero for $\vert\eta\vert>2.5$ ($2.7$ for muons) or $p_T<10$~GeV. Energy resolution for an electron with $E=25$~GeV and with $\vert\eta\vert\leq3.0$ is 1.5\%, and it drops asymptotically to 0.5\% for higher energies. The muon momentum resolution is worse for higher $p_T$ and higher $\vert\eta\vert$, with its maximum at 10\%, for $p_T>100$~GeV and $1.5<\vert\eta\vert\leq2.5$. Jet reconstruction uses the anti-$k_t$ algorithm~\cite{Cacciari:2008gp} with $R$ parameter set to 0.6. The efficiency for $b$-tagging is given separately for $b$-jets and $c$-jets, as an asymptotically increasing function of $p_T$. For $b$-jets ($c$-jets), the $b$-tagging efficiency is limited to 50\% (20\%) in the $\vert\eta\vert\leq1.2$ region and to 40\% (10\%) in the $1.2<\vert\eta\vert\leq2.5$ region. It is zero for jets with $p_T\leq10$~GeV or $\vert\eta\vert>2.5$. For any other jet, a constant $b$-tagging misidentification rate was set to 0.1\%.

The analysis of the generated and simulated events was performed with \mbox{\scshape MadAnalysis 5}~\cite{Conte:2012fm} in the expert mode~\cite{Conte:2014zja}. Events are selected if at least four reconstructed jets and exactly two oppositely-charged leptons with transverse momentum $p_T\ge20$~GeV and pseudo-rapidity $\vert \eta \vert \le2.5$ are present. After selection, 16\% (17\%) of $t\bar{t}H$ ($t\bar{t}A$) signal events are accepted. No cuts are applied to the events' transverse missing energy ($\slash\kern-.6emE_{T}$). The full kinematic reconstruction of the four-momenta of the undetected 
neutrinos is performed by imposing energy-momentum conservation and mass constraints to signal and background events~\cite{Santos:2015dja}. Mass values are randomly generated for the intermediate particles $W^+$, $W^-$, $t$ and $\bar{t}$, using probability density functions (p.d.f.s) obtained from the corresponding generator-level mass distributions. Firstly, a two-dimensional p.d.f. for $m_t$ and $m_{\bar{t}}$ is used to generate random mass values for the top quarks. 
Secondly, $m_{W^+}$ and $m_{W^-}$ are generated from the two-dimensional p.d.f.s of ($m_t,m_{W^+}$) and ($m_{\bar{t}},m_{W^-}$), respectively, such that possible correlations are preserved in the reconstruction. 
The following mass constraints are then applied to the $t\bar{t}$ system,
\begin{eqnarray}
\label{equ:a0}
\ensuremath{( p_{\ell +} + p_{\nu} )^{2} &=& m_{W^+}^{2}}, \\
\label{equ:a1}
\ensuremath{( p_{\ell -} + p_{\bar{\nu}} )^{2} &=& m_{W^-}^{2}}, \\
\label{equ:a2}
\ensuremath{( p_{W^+} + p_{b} )^{2} &=& m_{t}^{2}}, \\
\label{equ:a3}
\ensuremath{( p_{W^-} + p_{\bar{b}} )^{2} &=& m_{\bar{t}}^{2}}.
\end{eqnarray}
The $p_b$ and $p_{\bar{b}}$ correspond to the four-momenta of the two $b$-jets, respectively from the $t$ and $\bar{t}$ decays. The $p_{\ell+}$ and $p_{\ell-}$ ($p_{\nu}$ and $p_{\bar{\nu}}$) correspond to the four-momenta of the positive and negative charged leptons (neutrino and anti-neutrino), respectively from the decaying $W^+$ and $W^-$, which in turn have momenta $p_{W^+}$ and $p_{W^-}$.
In order to reconstruct the neutrino and anti-neutrino four-momenta (six unknowns, since we set $m_{\nu}=m_{\bar{\nu}}=0$), we assume they fully account for the missing transverse energy, i.e.,
\begin{eqnarray}
\label{equ:a4}
\ensuremath{p_{x}^{\nu} + p_{x}^{\bar{\nu}} &=& \slash\kern-.6emE_{x}}, \\
\label{equ:a5}
\ensuremath{p_{y}^{\nu} + p_{y}^{\bar{\nu}} &=& \slash\kern-.6emE_{y}}.
\end{eqnarray}
The $\ensuremath{\slash\kern-.6emE_{x}$ and $\slash\kern-.6emE_{y}}$
represent the $x$ and $y$ components of the transverse missing energy. If a solution is not found for the particular choice of top quark and $W$-boson masses, the generation of mass values is repeated, up to a maximum of 500, until at least one solution is found. If still no solution is found, the event is discarded as not compatible with the topology under study.


The kinematic reconstruction based on equations \eqref{equ:a0}-\eqref{equ:a5} may result in more than one possible solution for a particular event and choice of masses. We calculate, for each solution, the likelihood ($L_{t\bar{t}h}$) of it being consistent with a $t\bar{t}h$ dileptonic event. This likelihood is computed as the product of one-dimensional probability density functions (p.d.f.) built from $p_T$ distributions of the neutrino, anti-neutrino, top quark, anti-top quark, and $t\bar{t}$ system, respectively $P({p_T}_{\nu})$, $P({p_T}_{\bar{\nu}})$, $P({p_T}_{t})$, $P({p_T}_{\bar{t}})$ and $P({p_T}_{t\bar{t}})$, all obtained from fits to the corresponding parton level distributions. The two-dimensional p.d.f. of the top quark masses, $P(m_t,m_{\bar{t}})$, and the one-dimensional p.d.f. of the Higgs candidate mass, $P(m_h)$, are also included. The latter is obtained at reconstruction level, using a $\Delta R$ criterion\footnote{$\Delta R = \sqrt{\Delta\Phi^2+\Delta\eta^2}$, where $\Delta\Phi$ ($\Delta\eta$) correspond to the difference in $\Phi$ ($\eta$) between two objects.} to match jets to the truth-level $b$ and $\bar{b}$ partons from the $h$ decay.
\begin{multline}
\label{equ:a7}
\ensuremath{L_{t\bar{t}h} ~~\sim~~ \frac{1}{{p_T}_{\nu} {p_T}_{\bar{\nu}}} P({p_T}_{\nu}) P({p_T}_{\bar{\nu}}) \times \\ \times P({p_T}_{t}) P({p_T}_{\bar{t}}) P({p_T}_{t\bar{t}}) P(m_t,m_{\bar{t}}) P(m_h)}.
\end{multline}

The momenta of the neutrino and anti-neutrino must accomodate any energy losses in the event (QCD radiation, as well as detector effects) in order to reconstruct the top quarks and $W$ bosons masses. This may result in larger estimated neutrino and anti-neutrino $p_T$ after reconstruction, relatively to their $p_T$ at parton level. In order to compensate for this effect, the factor $1/({p_T}_{\nu} \times {p_T}_{\bar{\nu}})$ is introduced in the likelihood, thus favouring solutions with lower neutrino and anti-neutrino $p_T$ that better match parton level. The solution with the largest value of $L_{t\bar{t}h}$ is chosen as the correct one. 
A solution is found for 70\% of truth-matched $t\bar{t}H$ and $t\bar{t}A$ signal events.

At reconstruction level (without truth-match), the number of combinations of jets available to reconstruct the top and anti-top quarks, together with the Higgs boson, can be overwhelming. Choosing one of the wrong combinations of jets for reconstructing signal events gives rise to combinatorial background, one of the main challenges of this analysis. To reduce the number of possible combinations only the 6 highest $p_T$ jets are used (it was confirmed that in more than 95\% of all signal events, for both $t\bar{t}H$ and $t\bar{t}A$, jets produced from the top quarks and Higgs boson decays are within the 6 highest $p_T$ jets). Furthermore, the jet combinations were required to verify $m_{\ell^+b_t}<150$~GeV, $m_{\ell^-\bar{b}_{\bar{t}}}<150$~GeV and 50~GeV$\leq m_{b_H\bar{b}_H} \leq200$~GeV, where $b_t$ and $\bar{b}_{\bar{t}}$ refer to the jets assigned in reconstruction to the hadronization of the $b$ and $\bar{b}$ quarks from the $t$ and $\bar{t}$ decays, respectively.

\begin{figure*}
\begin{center}
\begin{tabular}{ccc}
\hspace*{-5mm}\epsfig{file=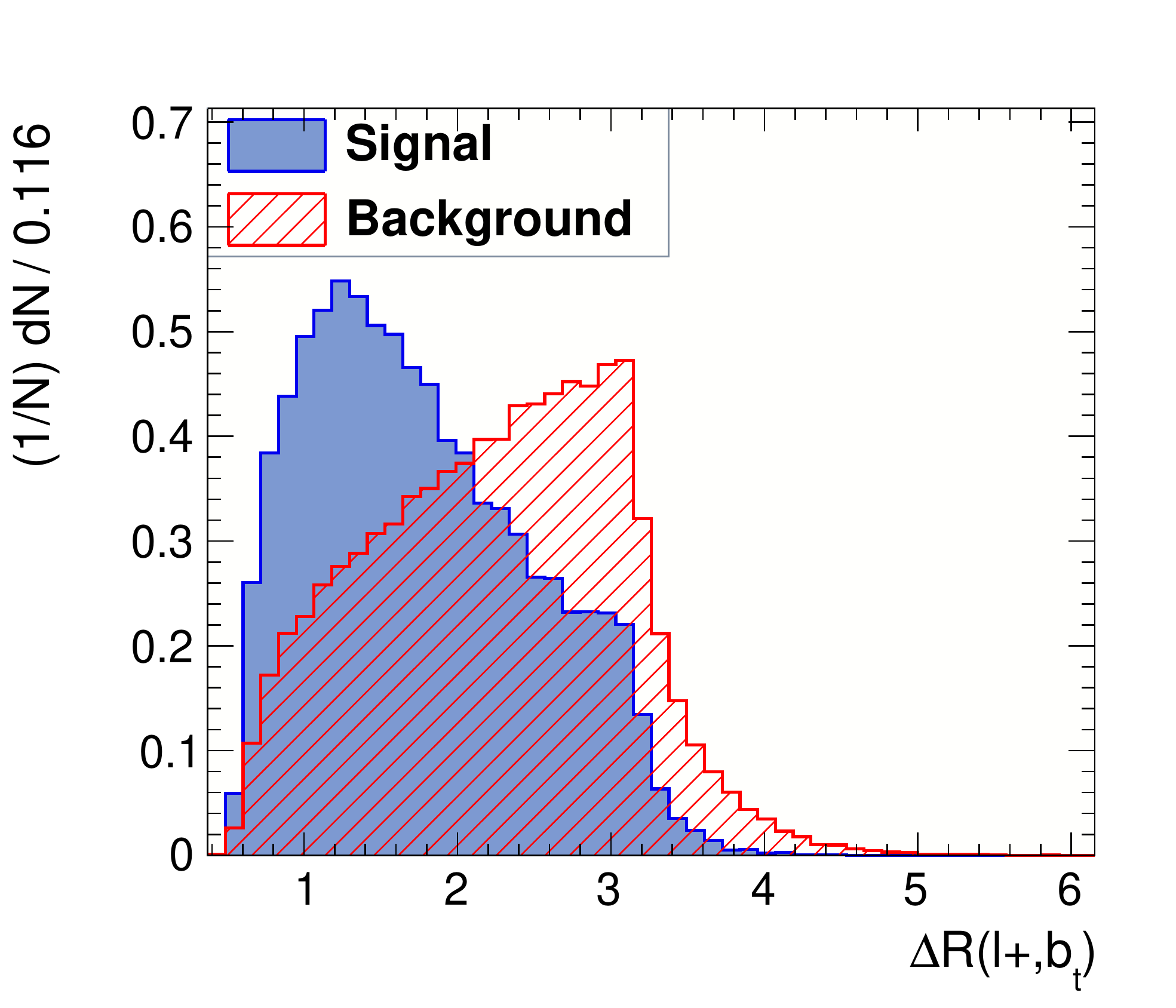,height=4.8cm,clip=} & 
			\epsfig{file=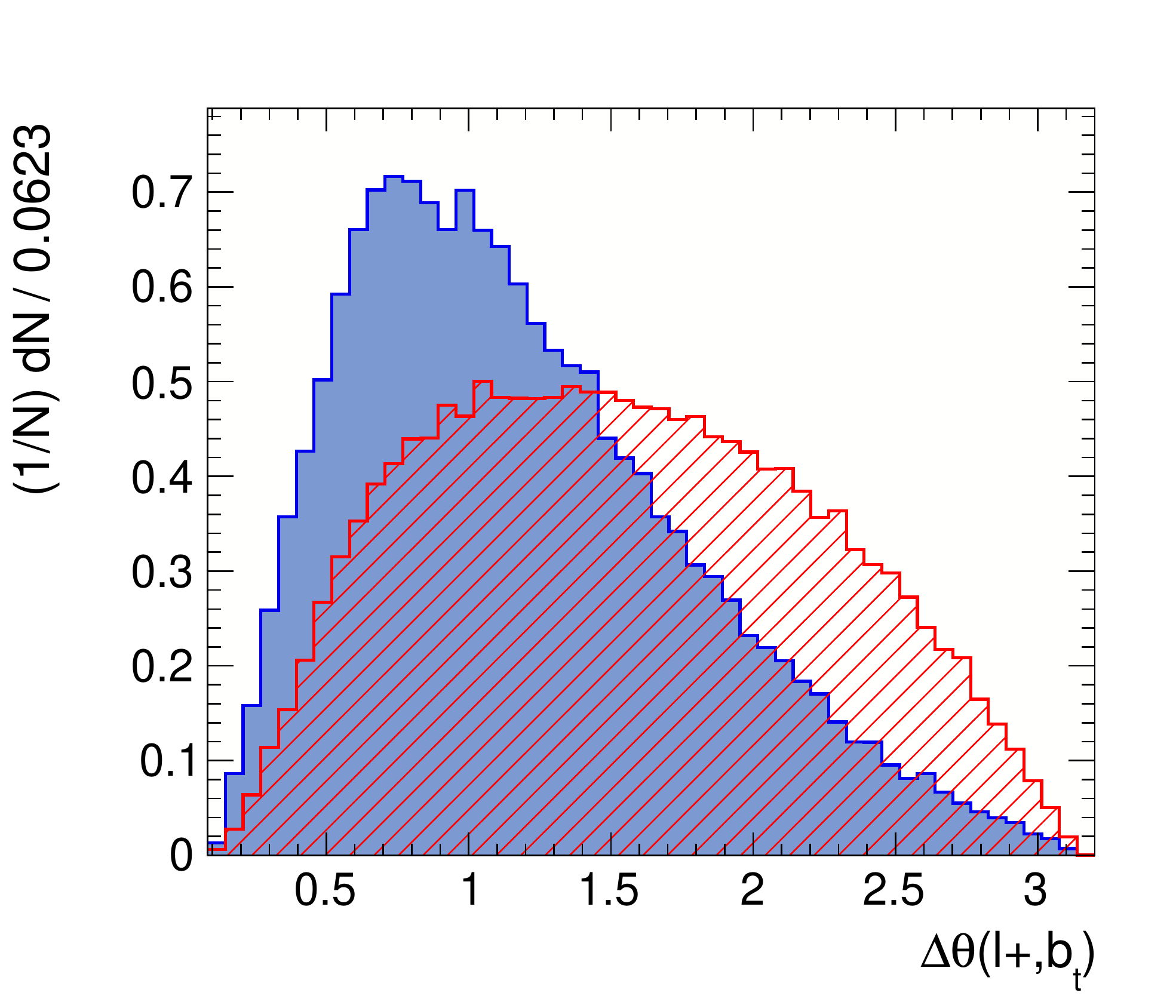,height=4.8cm,clip=} &
			\epsfig{file=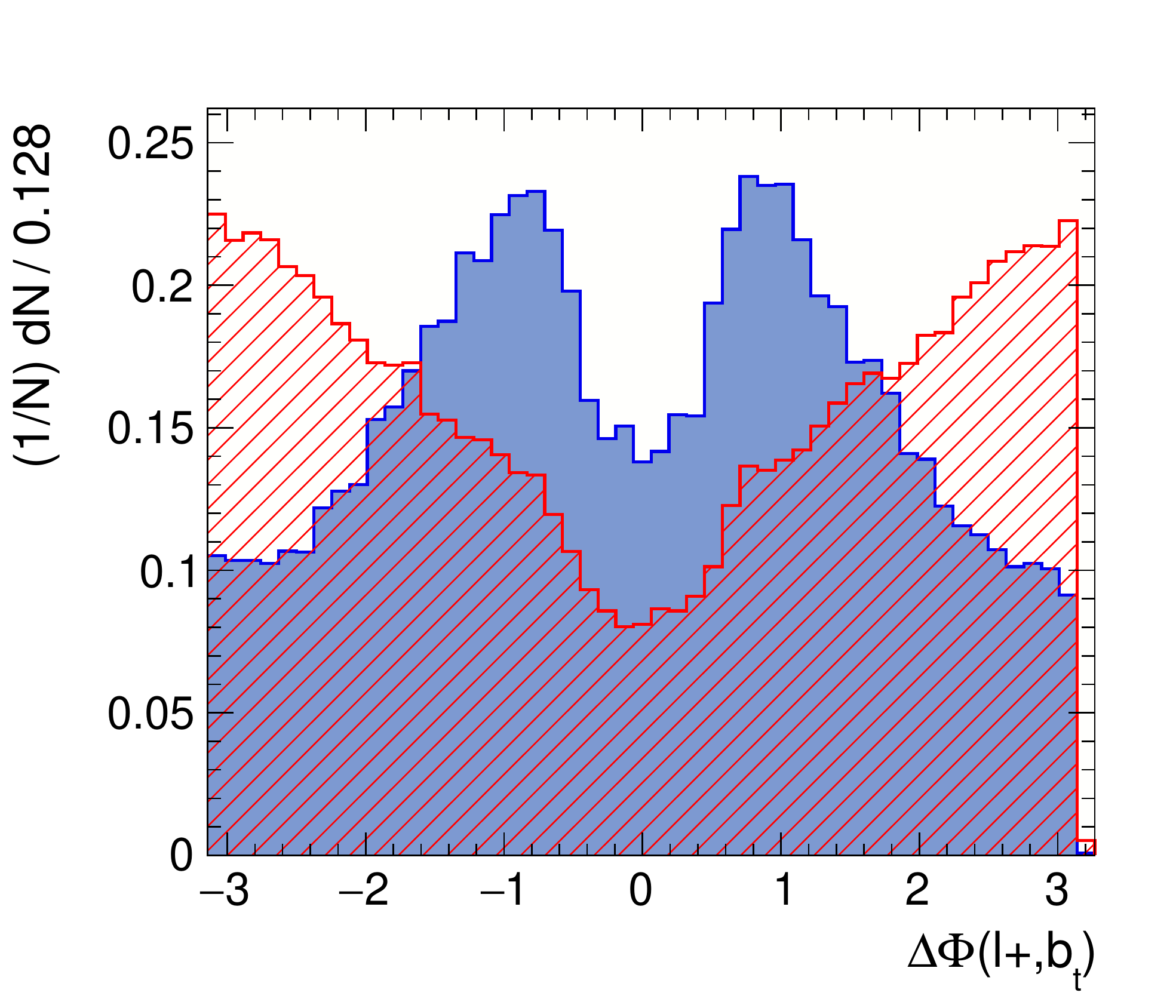,height=4.8cm,clip=}  \\[-2mm]
\hspace*{-5mm}\epsfig{file=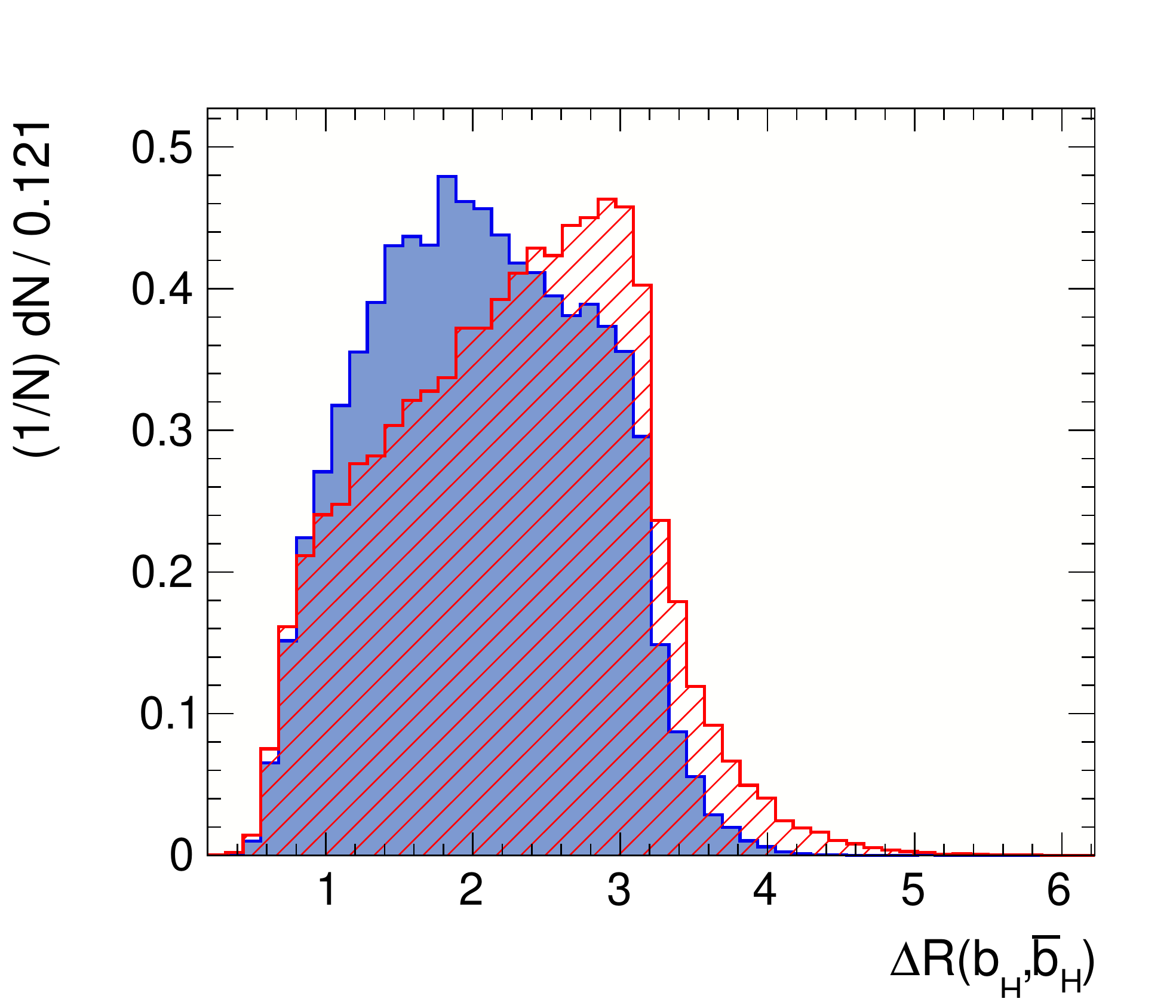,height=4.8cm,clip=}  & 
			\epsfig{file=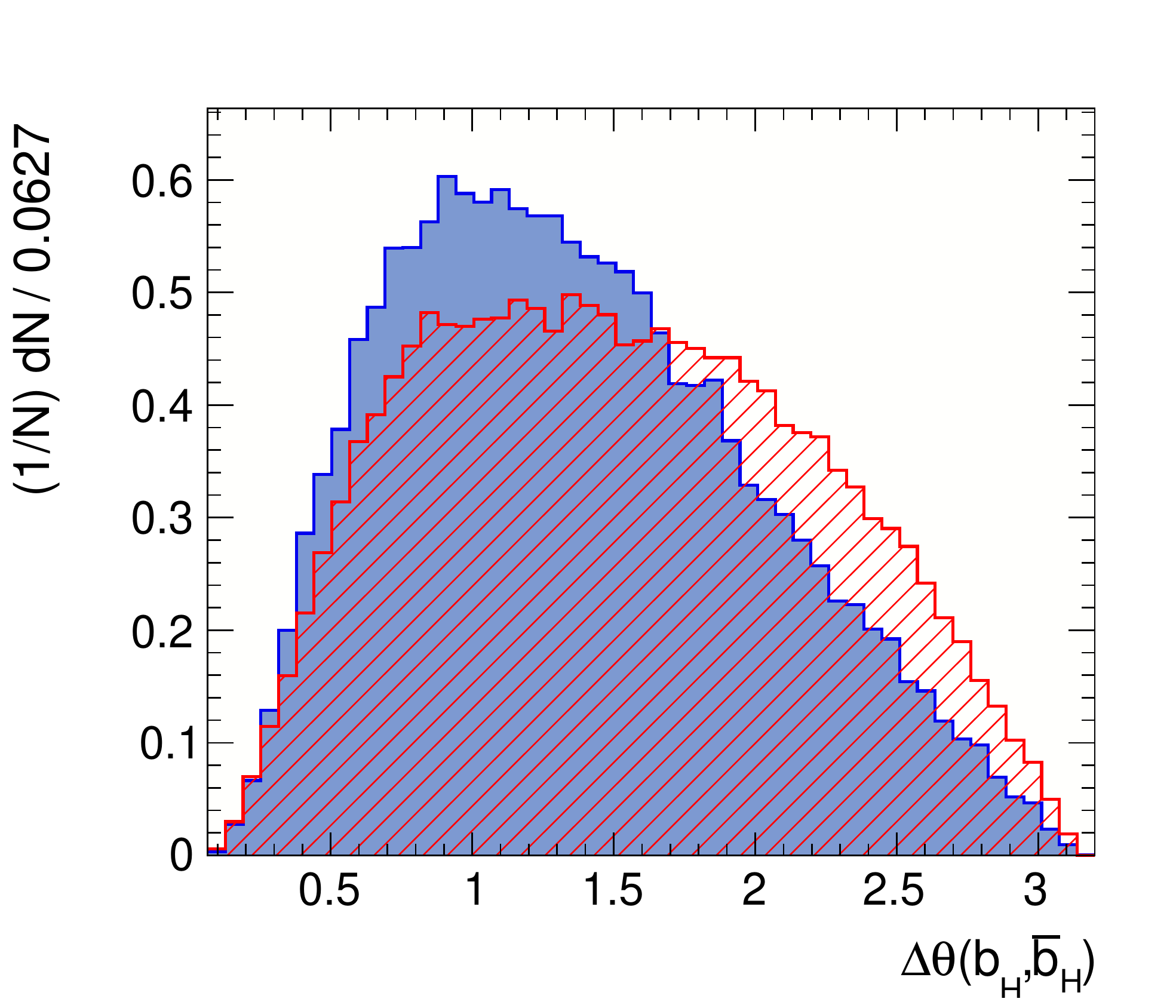,height=4.8cm,clip=} &
			\epsfig{file=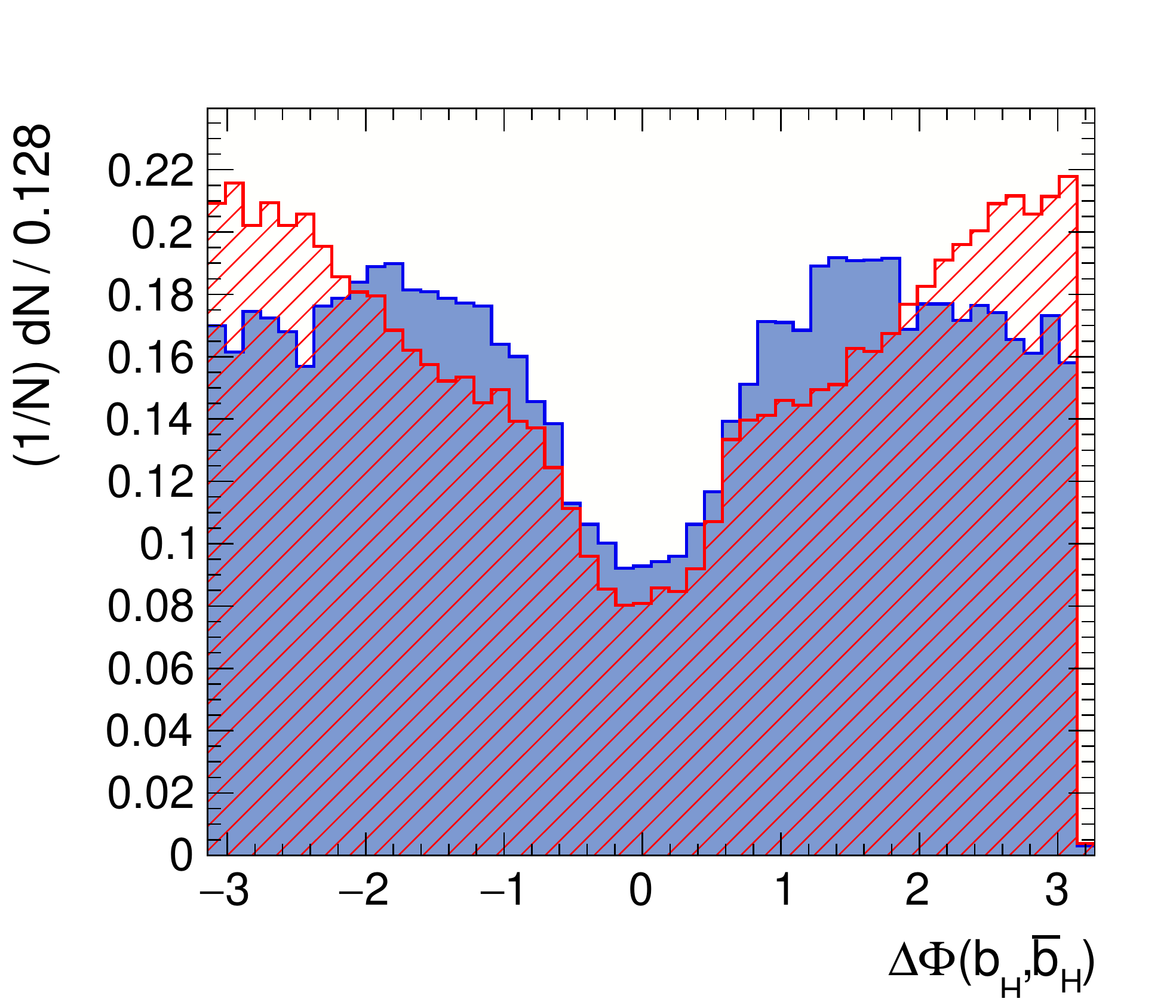,height=4.8cm,clip=}  \\[-2mm]
\end{tabular}
\vspace*{-0.2cm}
\caption{Distributions of TMVA input variables for right (filled blue, labelled "Signal") and wrong combinations (red shaded, labelled "Background") of jets and leptons from the same parent decaying particle: 
$\Delta R (\ell^+,b_t)$ (top left) and $\Delta R (b_H,\bar{b}_H)$ (bottom left); 
$\Delta \theta(\ell^+,b_t)$ (top middle) and $\Delta \theta(b_H,\bar{b}_H)$ (bottom middle); 
$\Delta \Phi(\ell^+,b_t)$ (top right) and $\Delta \Phi(b_H,\bar{b}_H)$ (bottom right), see text for details. }
\label{fig:TMVAinput}
\end{center}
\end{figure*}

\begin{figure*}
\begin{center}
\begin{tabular}{ccc}
\hspace*{-5mm} \epsfig{file=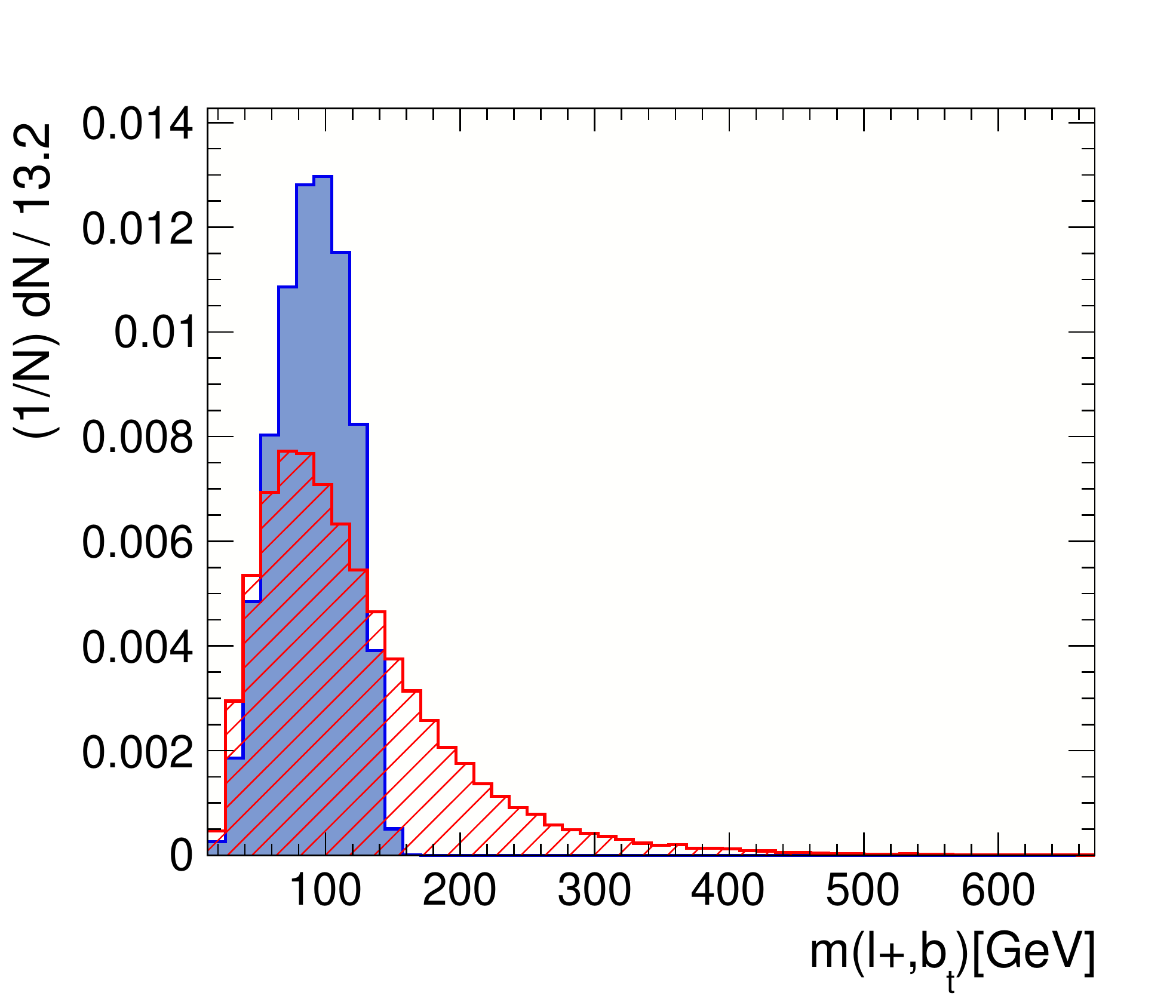,height=4.8cm,clip=} & 
                           \epsfig{file=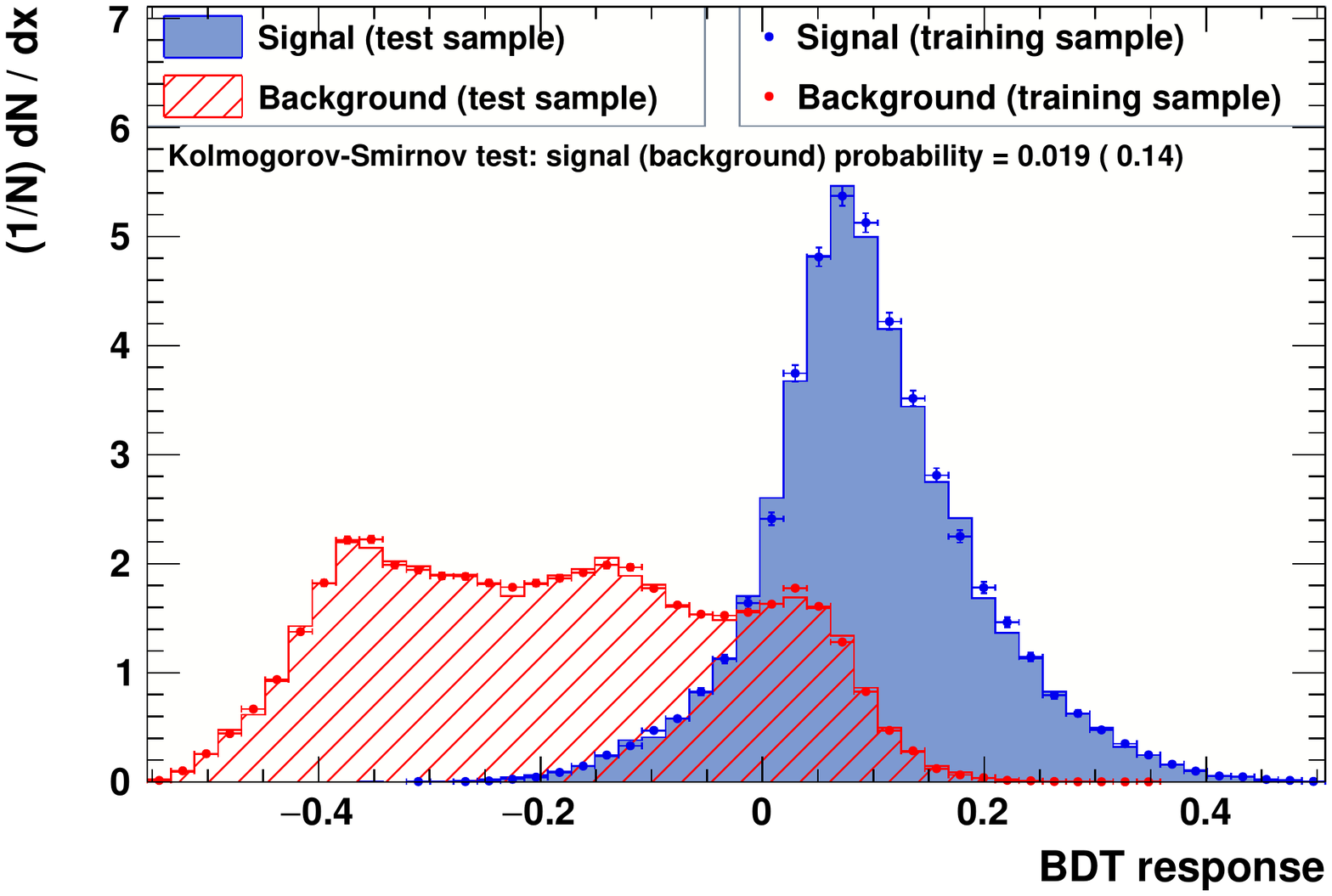,height=4.7cm,clip=} &
                           \epsfig{file=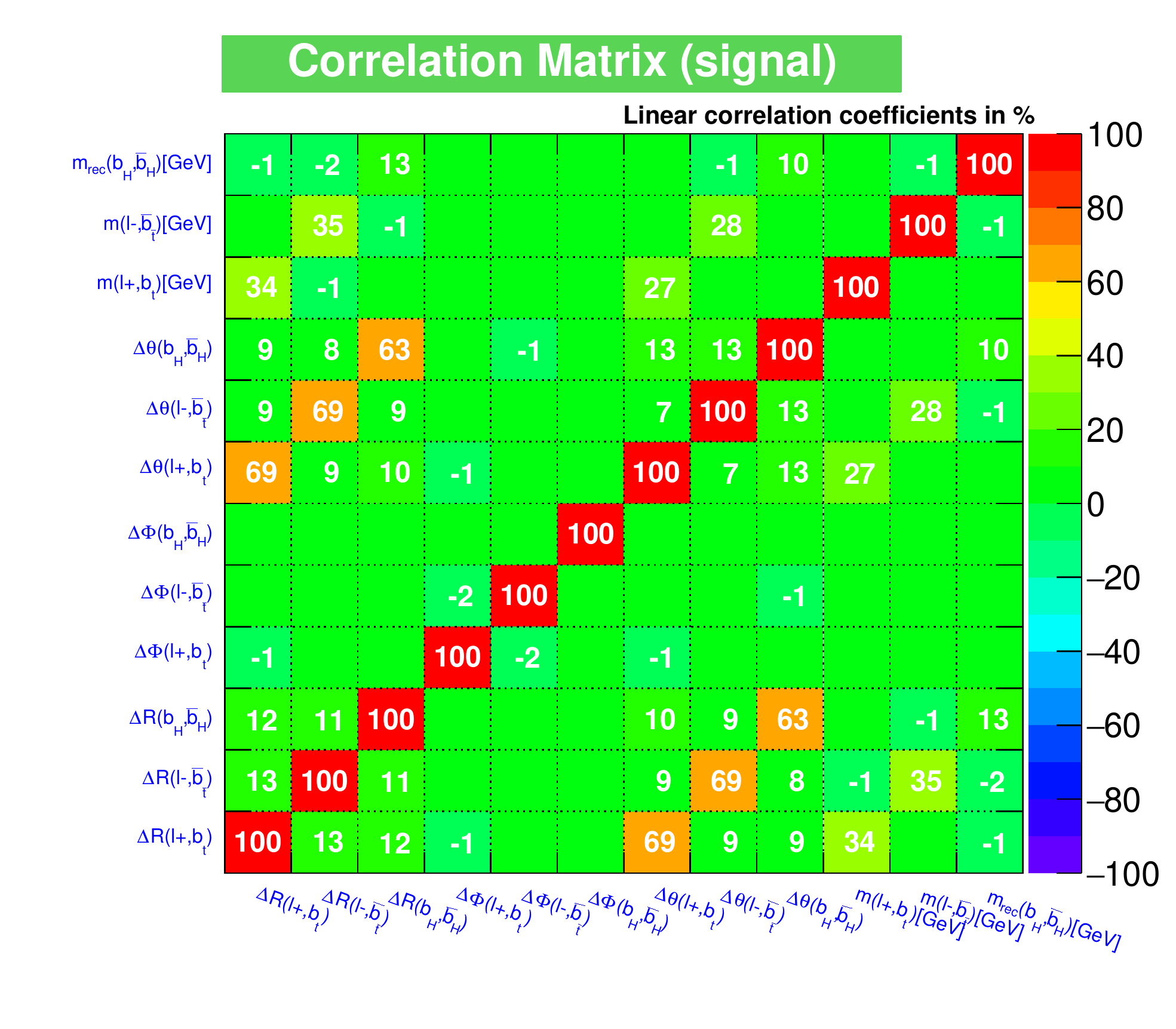,height=4.8cm,clip=} \\[-2mm] 
\hspace*{-5mm} \epsfig{file=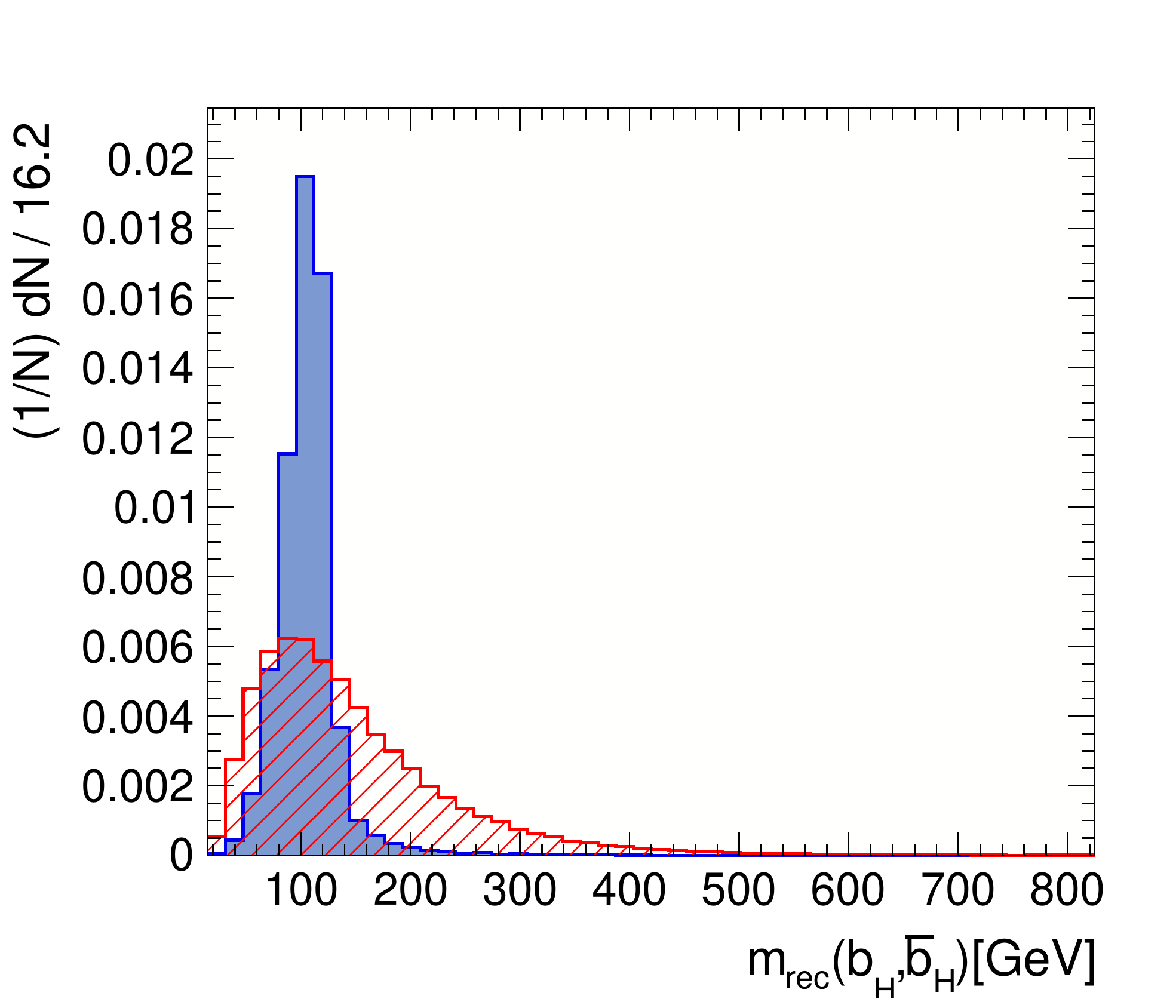,height=4.8cm,clip=} &
                           \epsfig{file=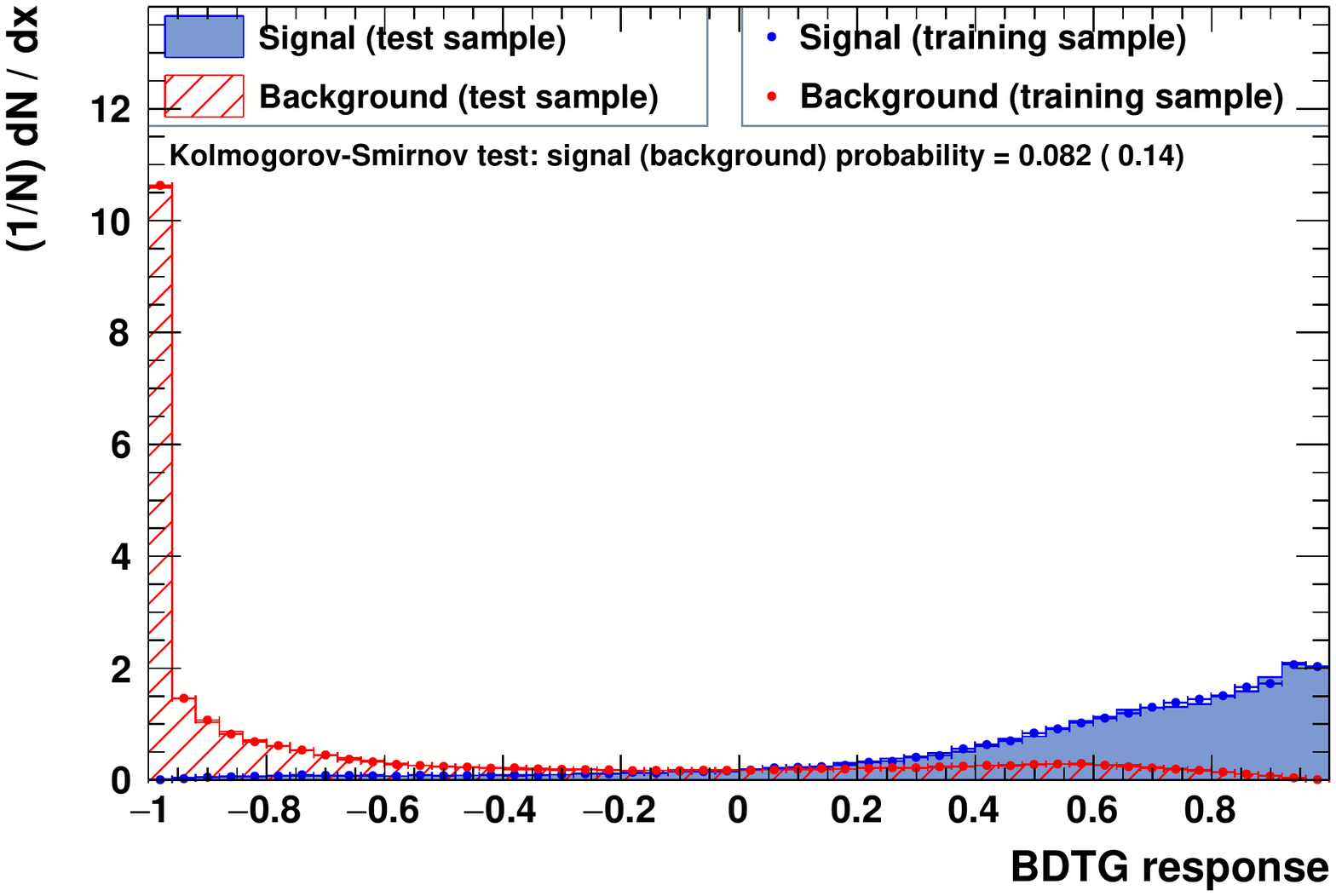,height=4.7cm,clip=} &
                           \epsfig{file=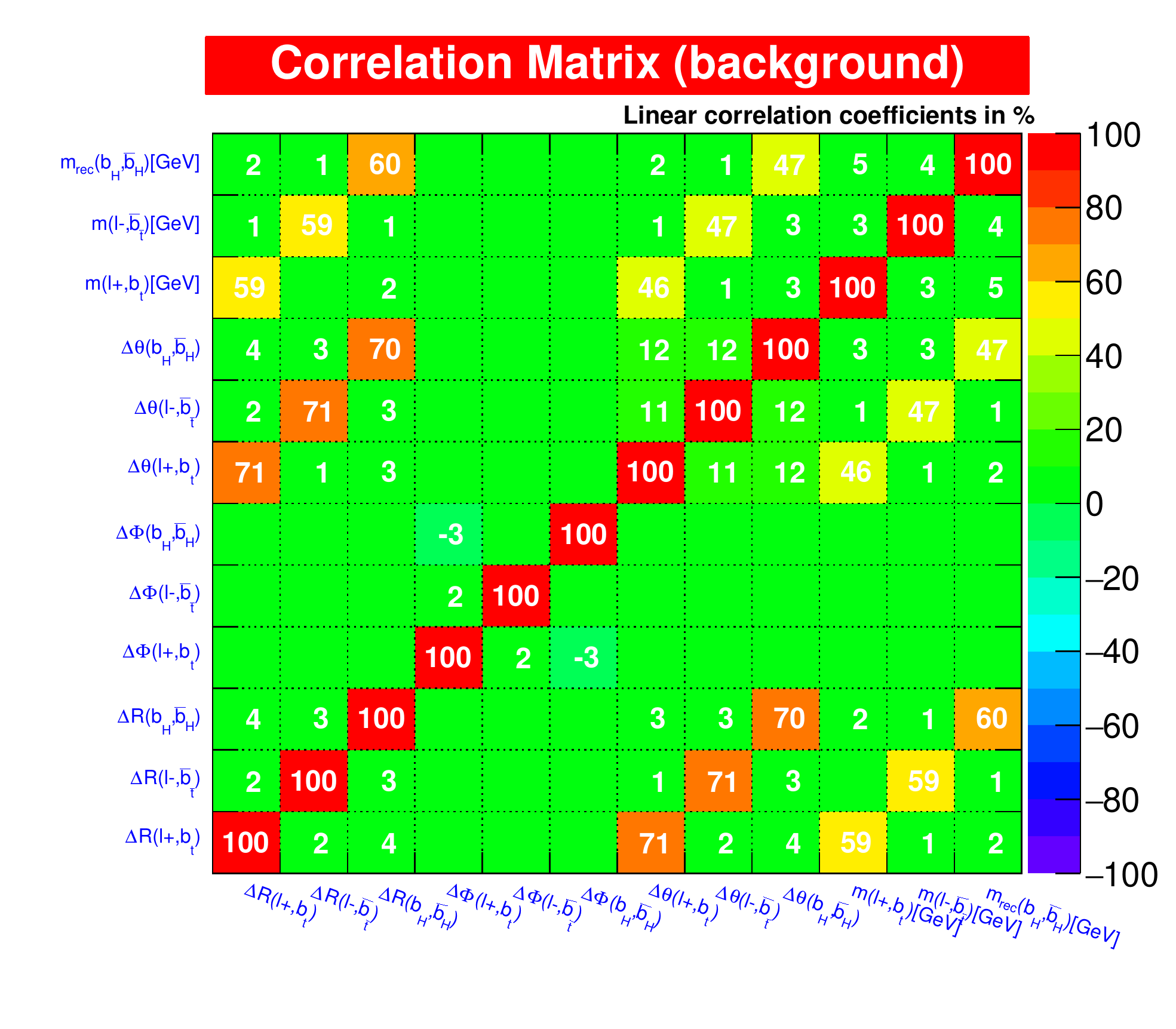,height=4.8cm,clip=} \\[-2mm]
\end{tabular}
\caption{Mass distributions (left) for right (filled blue, signal) and wrong (red shaded, background) combinations of jets and leptons from the same parent decaying particle:
(upper-left) the $m(\ell^+,b_t)$ and (lower-left) $m(b_H, \bar{b}_H)$;
(middle-top) BDT and (middle-bottom) BDTG TMVA methods response for signal and background;
(right-top) TMVA input variables correlations for signal and (right-bottom) background.}
\label{fig:TMVAinput2}
\end{center}
\end{figure*}

\begin{figure*}
\begin{center}
\begin{tabular}{ccc}
\epsfig{file=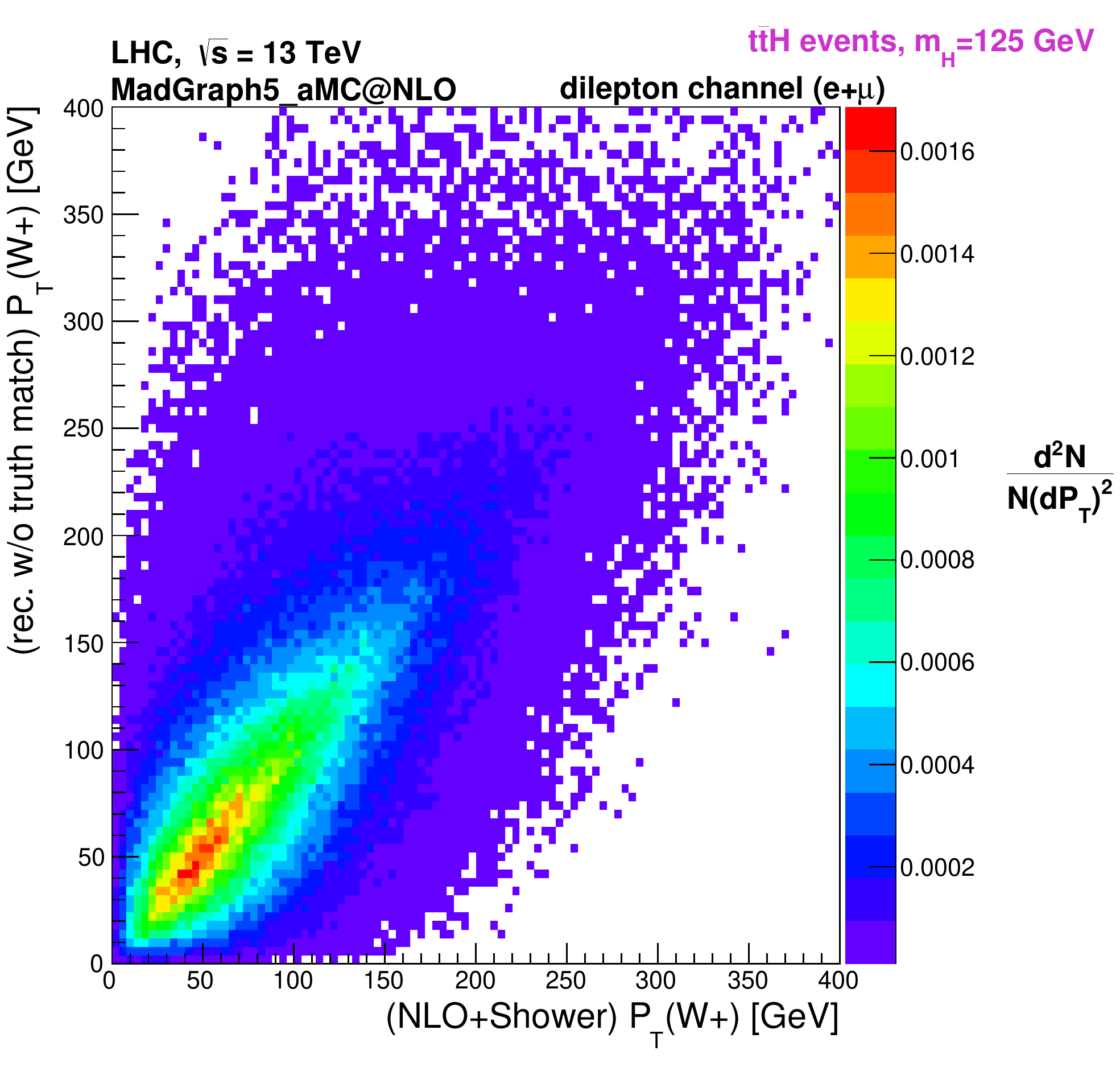,height=6.3cm,width=8cm,clip=} & \quad & 
\epsfig{file=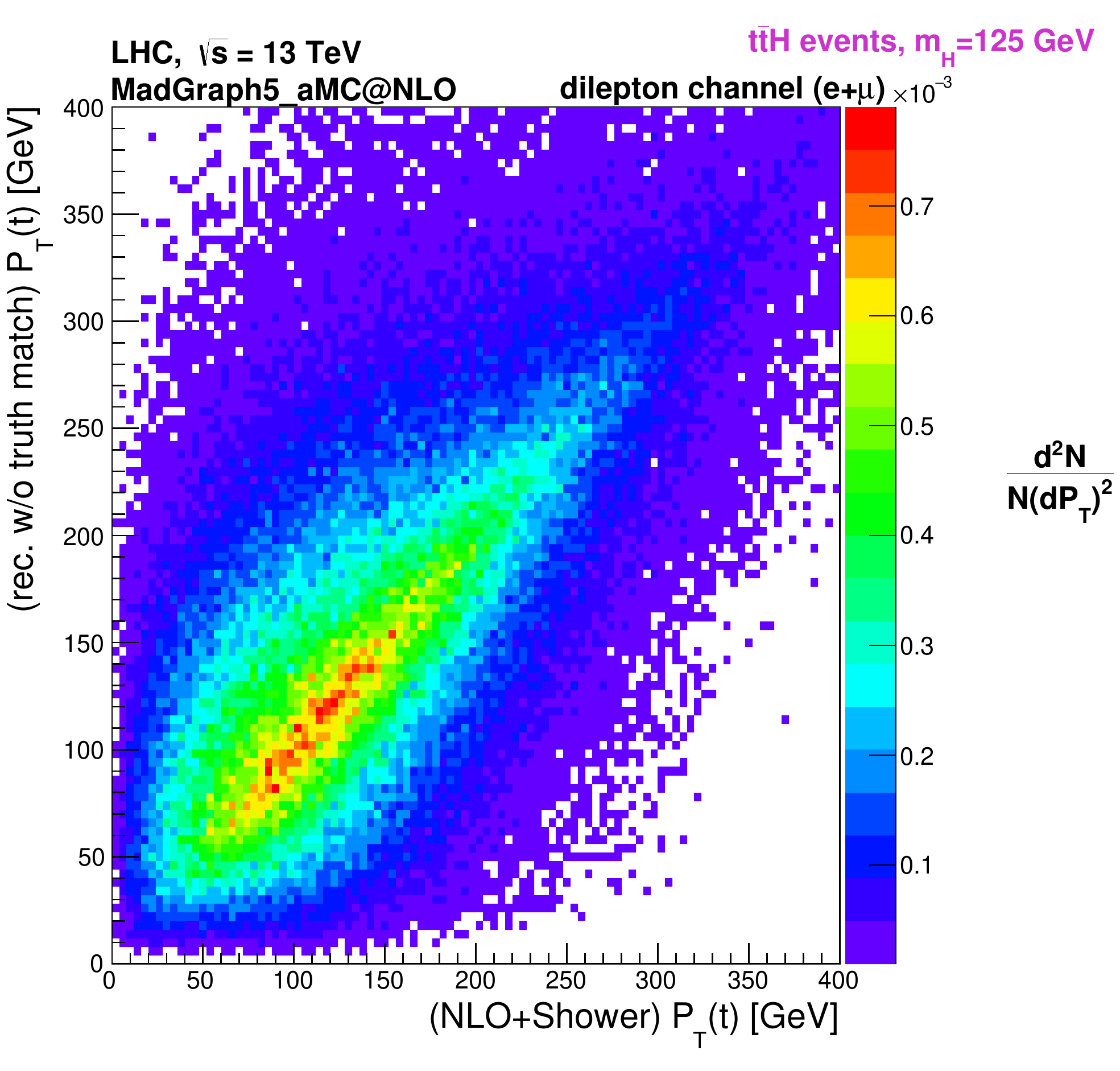,height=6.3cm,width=8cm,clip=} \\[-2mm]
\epsfig{file=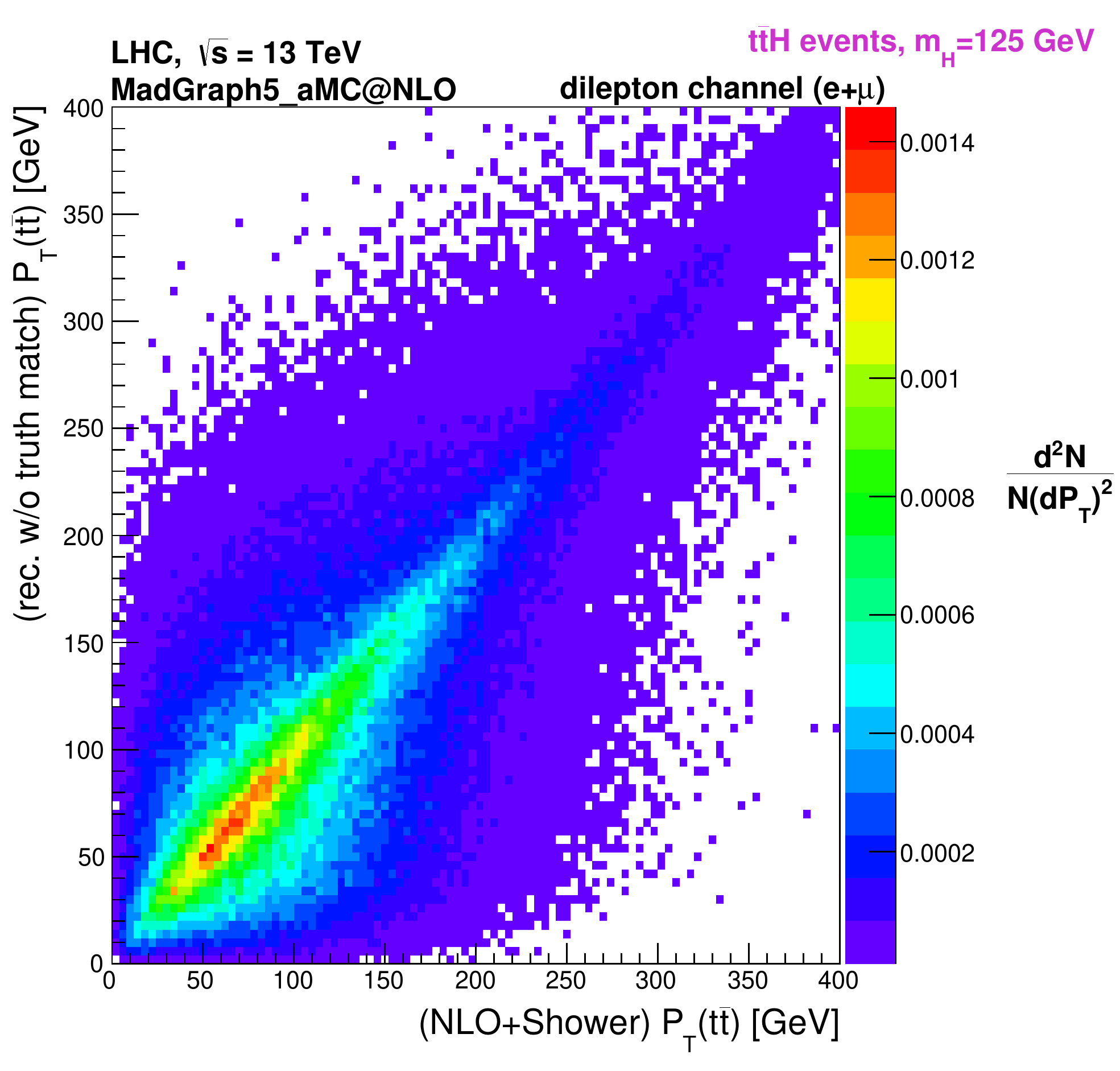,height=6.3cm,width=8cm,clip=} & \quad & 
\epsfig{file=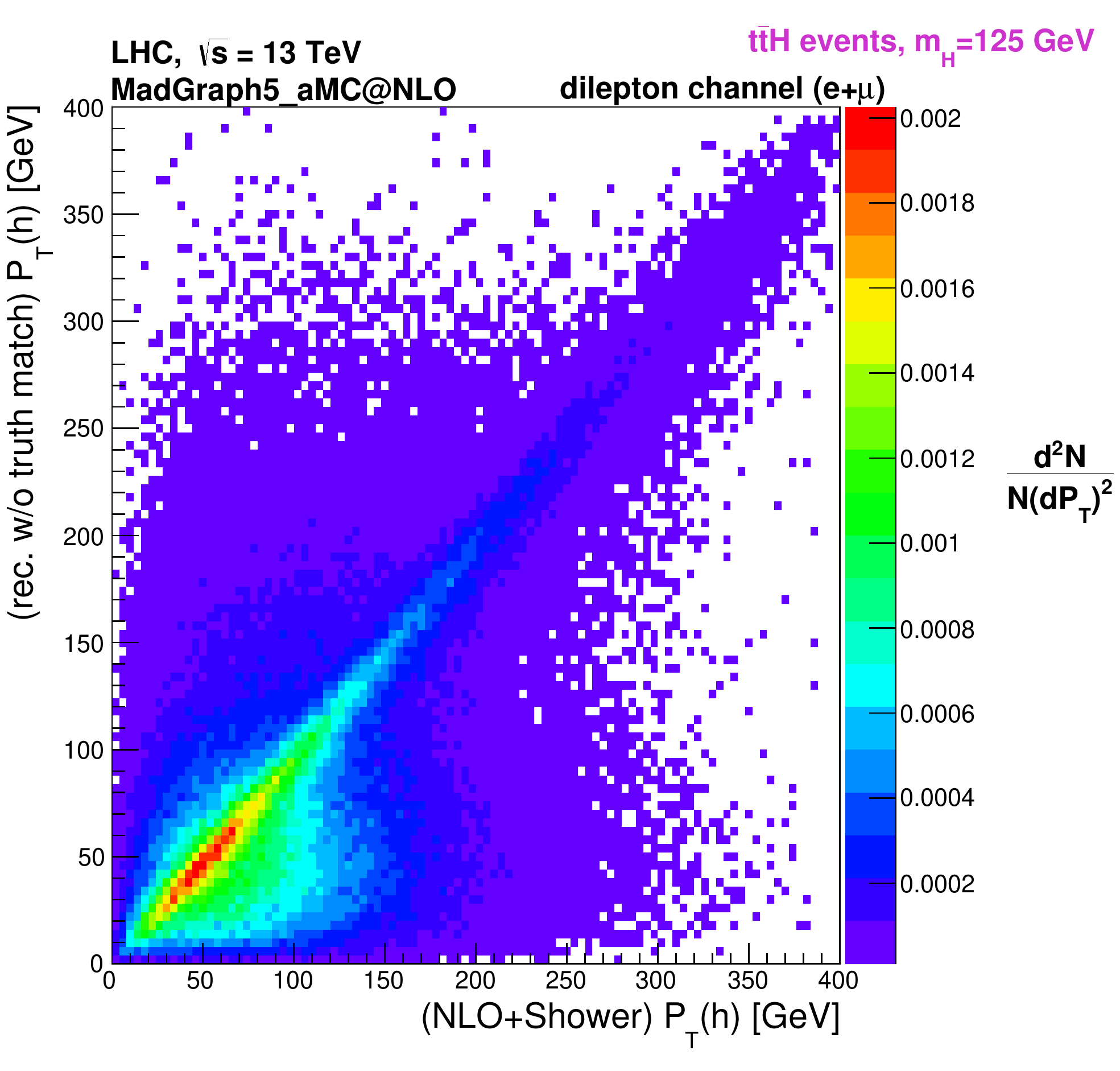,height=6.3cm,width=8cm,clip=} \\[-2mm]
\end{tabular}
\caption{Two-dimensional distributions of $p_T$ in $t\bar{t}H$ events. The horizontal axes represent variables recorded at NLO+Shower, and the vertical axes represent the corresponding variables recorded at reconstruction level without truth-match. Upper-left: distribution for $W^+$. A similar distribution is obtained for $W^-$, but is not shown here. Upper-right: distribution for $t$. A similar distribution is obtained for $\bar{t}$, but is not shown here. Lower-left: distribution for $t\bar{t}$. Lower-right: distribution for $H$.}
\label{fig:genexp2D}
\end{center}
\end{figure*}

\begin{figure*}
\begin{center}
\begin{tabular}{ccc}
\hspace*{-15mm}\epsfig{file=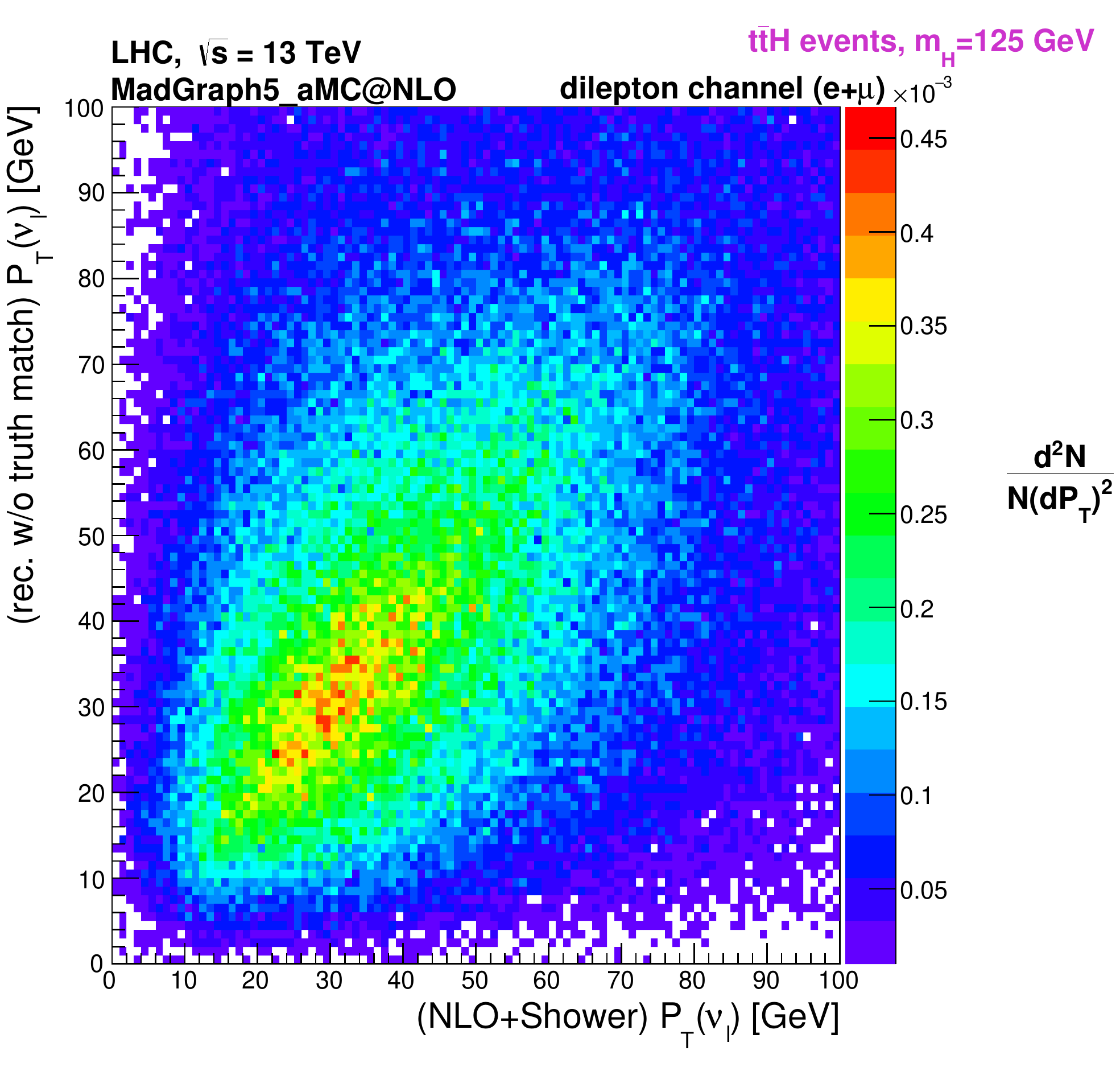,height=6.3cm,width=8cm,clip=} & \quad \hspace*{-6mm}& 
\epsfig{file=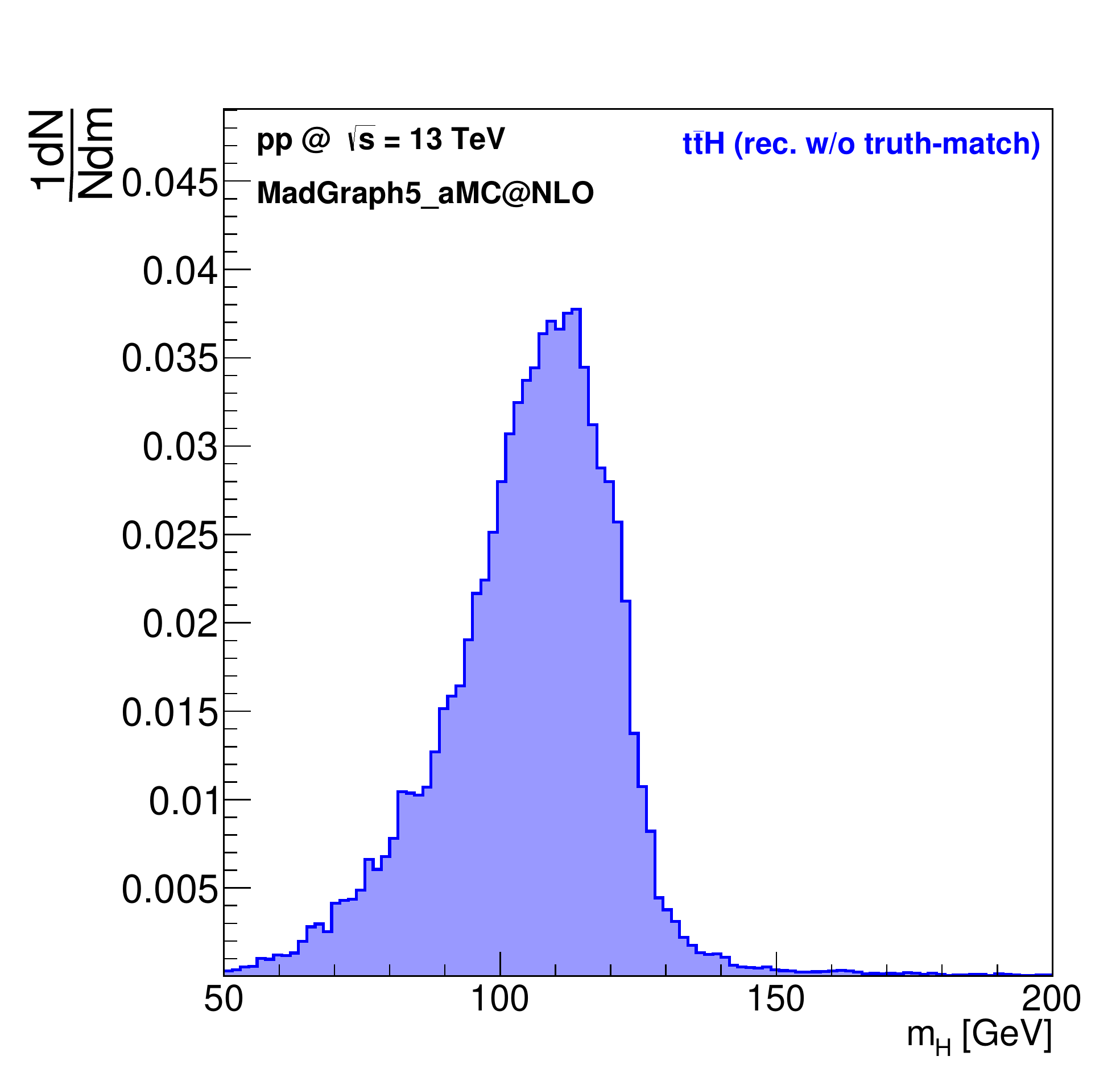,height=6.3cm,width=7cm,clip=} \\[-2mm]
\end{tabular}
\vspace*{-0.2cm}
\caption{Two-dimensional distribution of the neutrino $p_T$ in $t\bar{t}H$ events (left): the NLO+Shower $p_T$ (x-axis) against the reconstructed $p_T$ without truth-match (y-axis) is shown. Distribution of the reconstructed Higgs boson mass without truth-matched jets in $t\bar{t}H$ events (right).}
\label{fig:genexp}
\end{center}
\end{figure*}

At reconstruction level (without truth match), in order to preferentially pick the correct combination among the ones surviving the previous requirements, several multivariate methods were trained, using TMVA~\cite{TMVA2007}. The correct and wrong jet combinations were labeled respectively signal and combinatorial background in the following procedure. Nine parton level variables were used as input for the methods: $\Delta R$, lab-frame angles $\Delta\theta$ and $\Delta\Phi$ between the particle pairs ($b_t,\ell^+$), ($\bar{b}_{\bar{t}},\ell^-$) and ($b_H,\bar{b}_H$). The invariant masses of the systems composed of these pairs were also included, but were computed at reconstruction level with truth-match, to take into account detector resolution effects. A sample of $t\bar{t}h$ events (with $h=H$) was used to create both the signal and combinatorial background samples for this training and testing. For the signal sample, the variables were computed once per event, using the correct jet combination. For the combinatorial background sample, three different variable entries took place per event, each one corresponding to a wrong permutation of the 4 $b$ and $\bar{b}$ partons. These three permutations are chosen such that all the variables computed in each permutation are different from the ones in any other, including the correct one. In Figure \ref{fig:TMVAinput} and Figure \ref{fig:TMVAinput2} (left), distributions of the input variables are shown for the signal and combinatorial background training samples. The correlations between variables are shown in Figure \ref{fig:TMVAinput2} (right), for the signal (top) and combinatorial background (bottom) samples. Two boosted decision trees were the most performant, one with an adaptive boost (BDT) and the other with a gradient boost (BDTG). The latter being slightly better, it was used in the full kinematic reconstruction of events in order to increase the correct jet assignment. Figure \ref{fig:TMVAinput2} (middle column) shows the distributions of the BDT (top) and BDTG (bottom) discriminants for the signal and for the combinatorial background, for both the training and test samples. The jet combination chosen is the one returning the highest value of the BDTG discriminant, maximizing signal purity. 
After event selection, 62\% (61\%) of $t\bar{t}H$ ($t\bar{t}A$) signal events are successfully reconstructed. In 31\% (34\%) of the $t\bar{t}H$ ($t\bar{t}A$) signal events, the reconstruction without truth-match results in the same jet combination as the truth-matched one. 
Figure \ref{fig:genexp2D} shows, after $t\bar{t}H$ reconstruction without truth match, two-dimensional $p_T$ distributions of the $W^+$ (top-left), the top quark (top-right), the $t\bar{t}$ system (bottom-left) and Higgs boson (bottom-right). The correlation between the parton level $p_T$ distributions ($x$-axis) and reconstructed ones without truth-match ($y$-axis), is clearly visible. 
The neutrino reconstructed $p_T$ is compared with the parton level at NLO+Shower in Figure~\ref{fig:genexp} (left) and the distribution of the reconstructed Higgs boson mass is shown in Figure~\ref{fig:genexp} (right).  In spite of the wider spread of values in the neutrino $p_T$ distribution, a direct consequence of the reconstruction of two neutrinos in each one of the events, good correlation between the NLO+Shower distribution and the reconstructed neutrino $p_T$ is observed. The distribution of the Higgs mass has an R.M.S. of order 20~GeV. Although reconstruction could be improved by using more elaborate methods, this stays outside the scope of the paper.

\section{$t\bar{t}H$, $t\bar{t}A$ and $t\bar{t}b\bar{b}$ Angular Distributions}

As was done in \cite{Santos:2015dja}, we define $\theta^X_Y$ as the angle between the direction of the $Y$ system in the rest frame of $X$ and the direction of the $X$ system, in the rest frame of its parent system. 
For the reconstruction of the signal angular distributions, we consider the decay chain that starts with the $t\bar{t}h$ system, labeled $(123)$, and goes through successive 
two-body decays i.e., $(123)\rightarrow1+(23)$, $(23)\rightarrow2+(3)$ and $(3)\rightarrow4+5$. Three families of observables are constructed: $f(\theta^{123}_1)g(\theta^{3}_4)$, 
$f(\theta^{123}_1)g(\theta^{23}_3)$ and $f(\theta^{23}_3)g(\theta^{3}_4)$, 
with $f,g=\{\sin,\cos\}$. The $(123)$ system momentum direction is measured with respect to the laboratory frame. Particles 1 to 3 can either be the $t$ or the $\bar{t}$ quarks, or even the Higgs boson, without repetition. 
Particle 4 can be any of the products of the decay of the top quarks and the Higgs boson, including the intermediate $W$ bosons. The boost of particle 4 to the centre-of-mass of particle 3 can be performed in two different ways:
either (i) using the laboratory four-momentum of both particles 3 and 4 ({\it direct} boost), or (ii) boosting particles 3 and 4 sequentially through all intermediate centre-of-mass systems until particle 4 is evaluated in the centre-of-mass frame of particle 3 ({\it sequential} boost or {\it seq.} boost). Due to Wigner rotations, the directions of particle 4 resulting from each of these boosting procedures are different. The observables addressed in this work were studied using both the sequential and direct prescriptions.

\subsection{NLO versus LO Comparison}

\begin{figure*} 
\begin{center}
\begin{tabular}{ccc}
\hspace*{-5mm} \epsfig{file=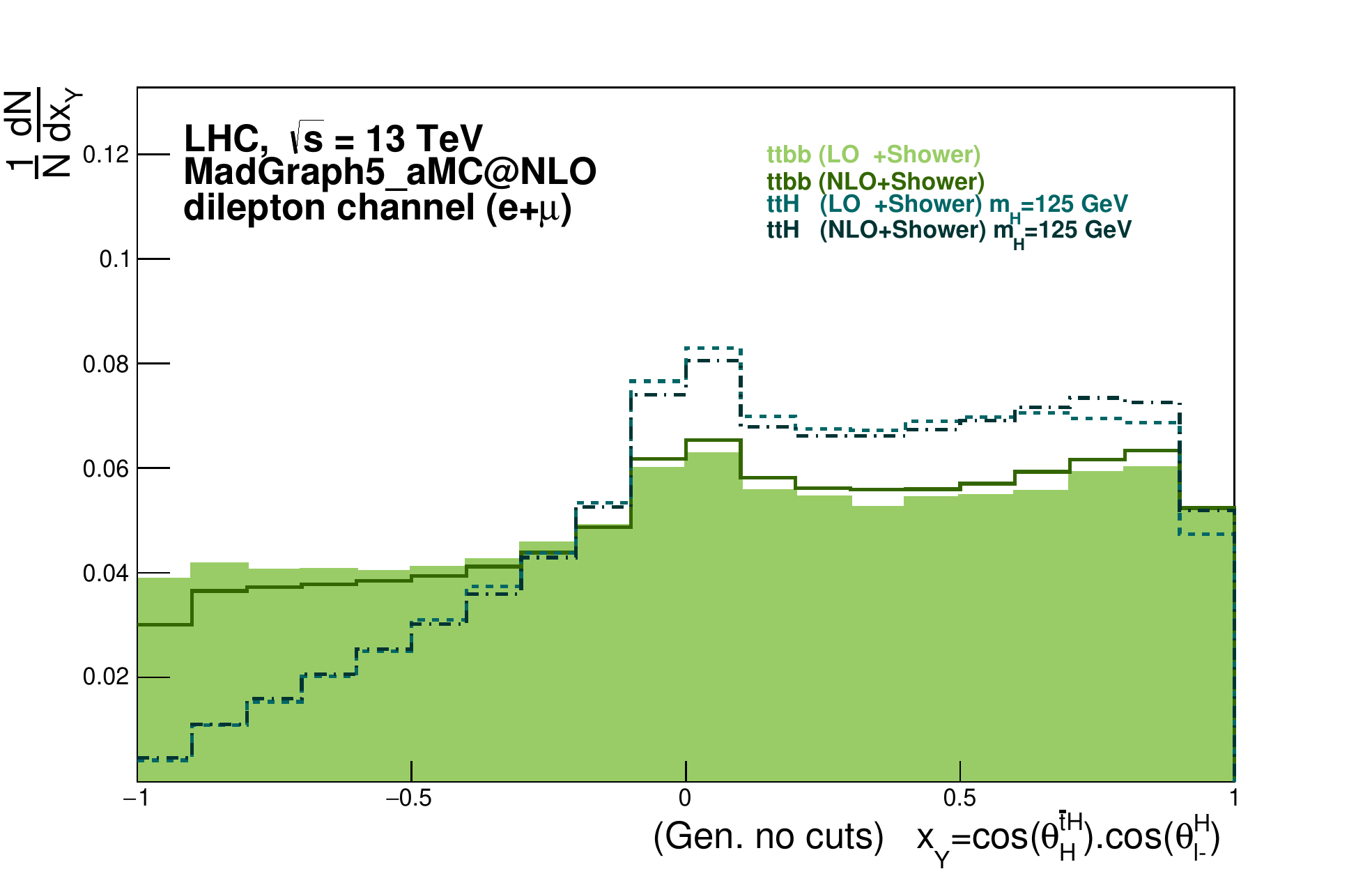,height=3.7cm,clip=}  &
 			  \epsfig{file=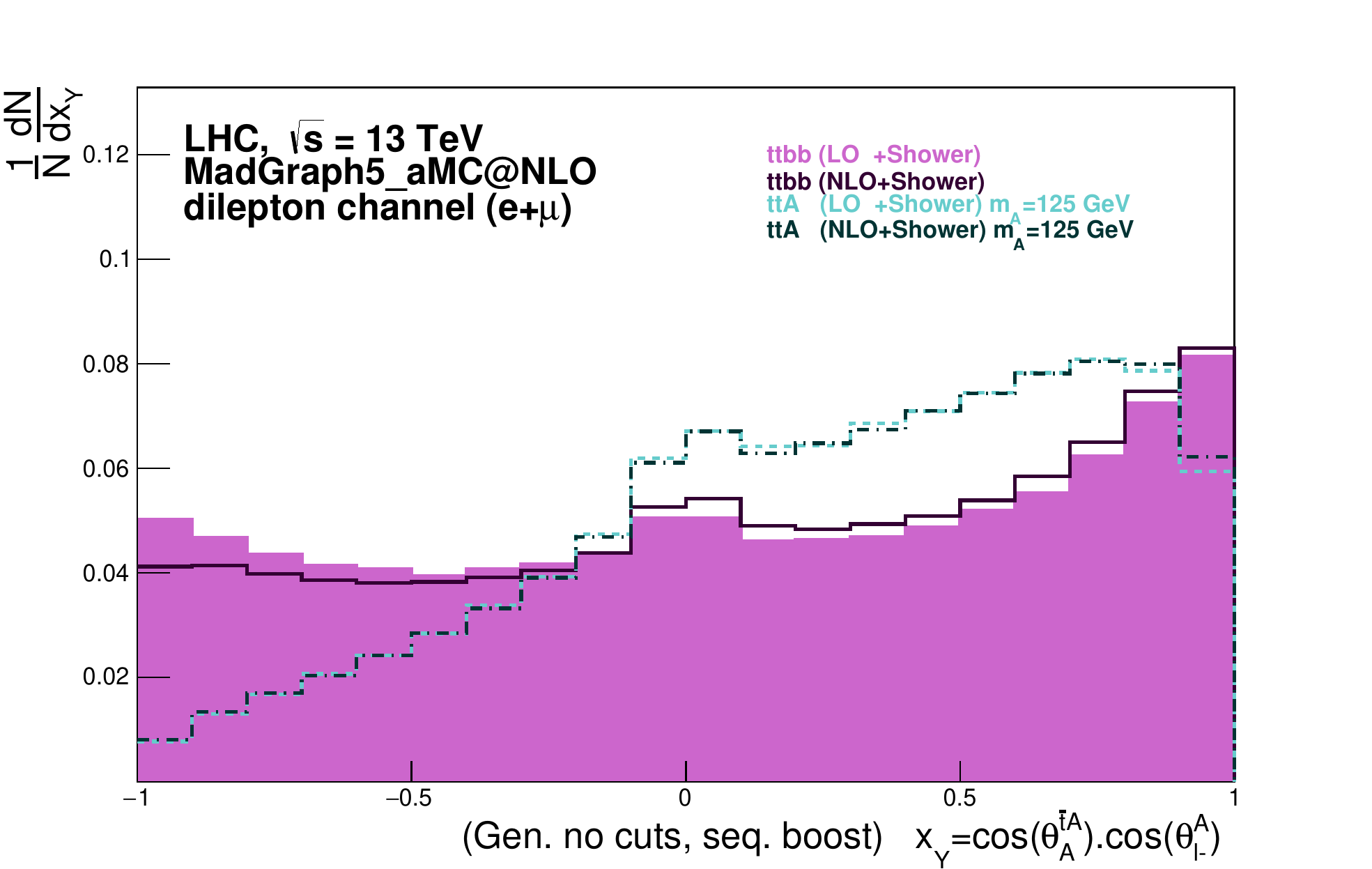,height=3.7cm,clip=} &  
			  \epsfig{file=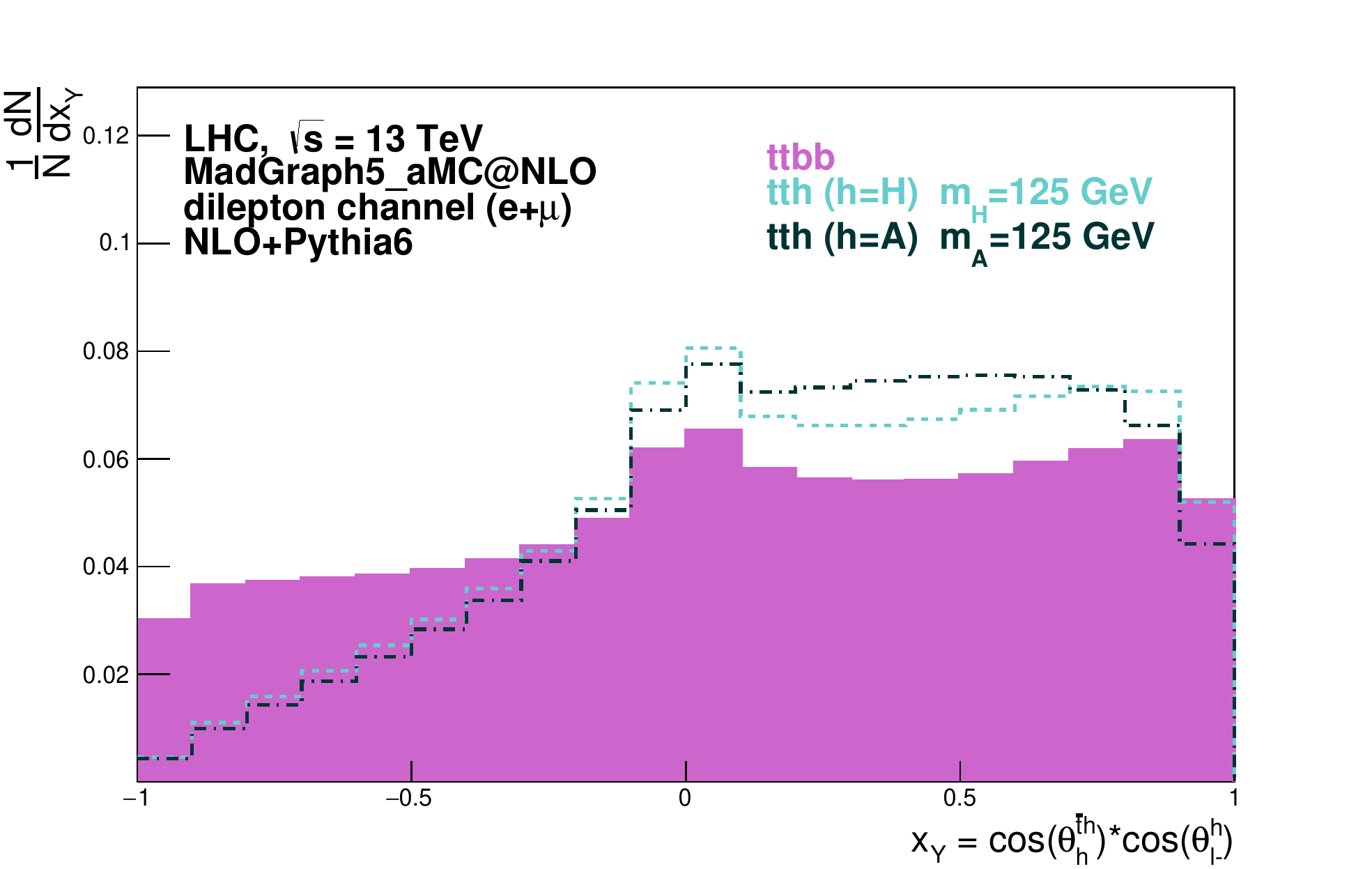,height=3.7cm,clip=} \\[0mm]
\hspace*{-5mm} \epsfig{file=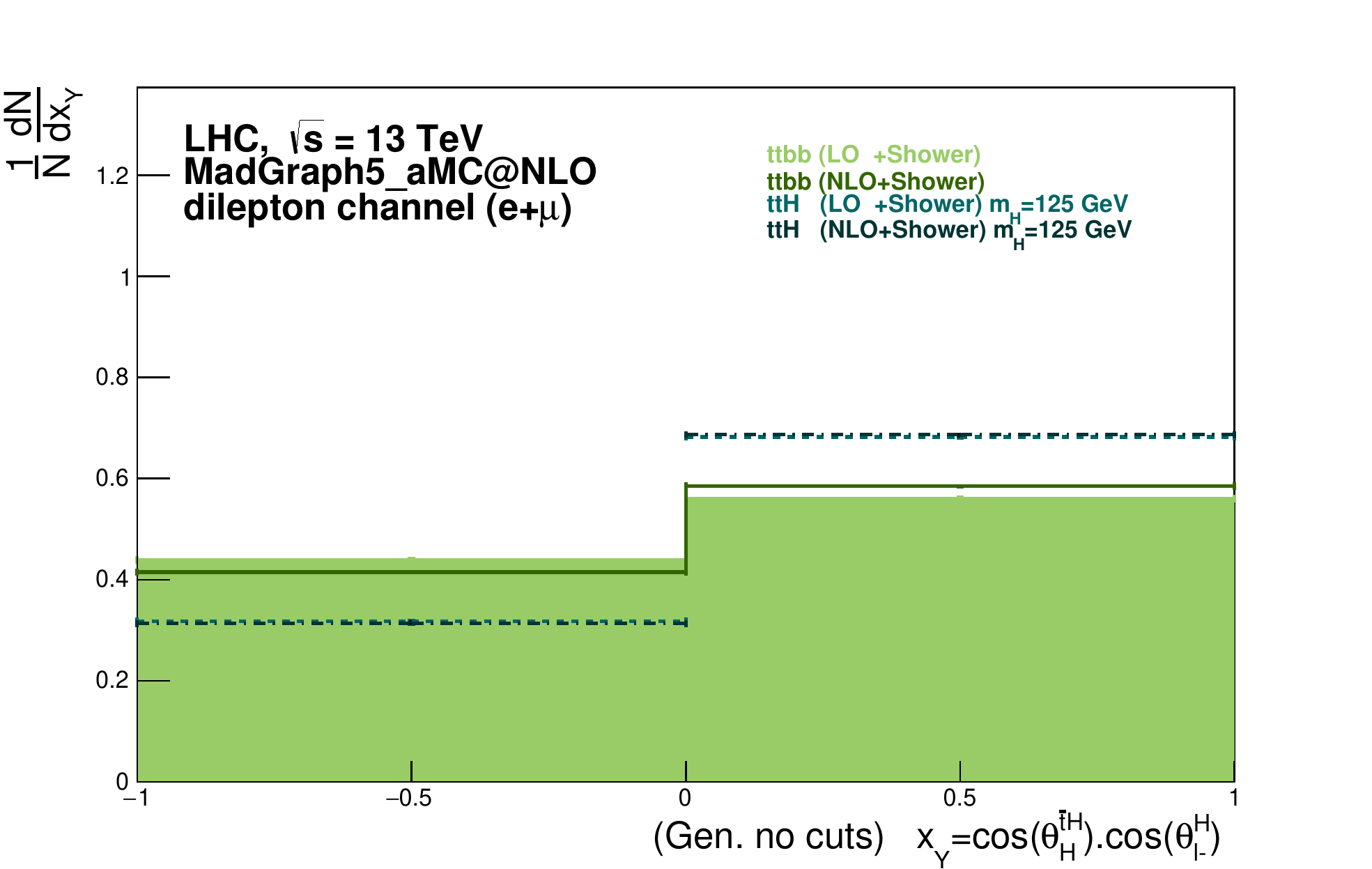,height=3.7cm,clip=} & 
			  \epsfig{file=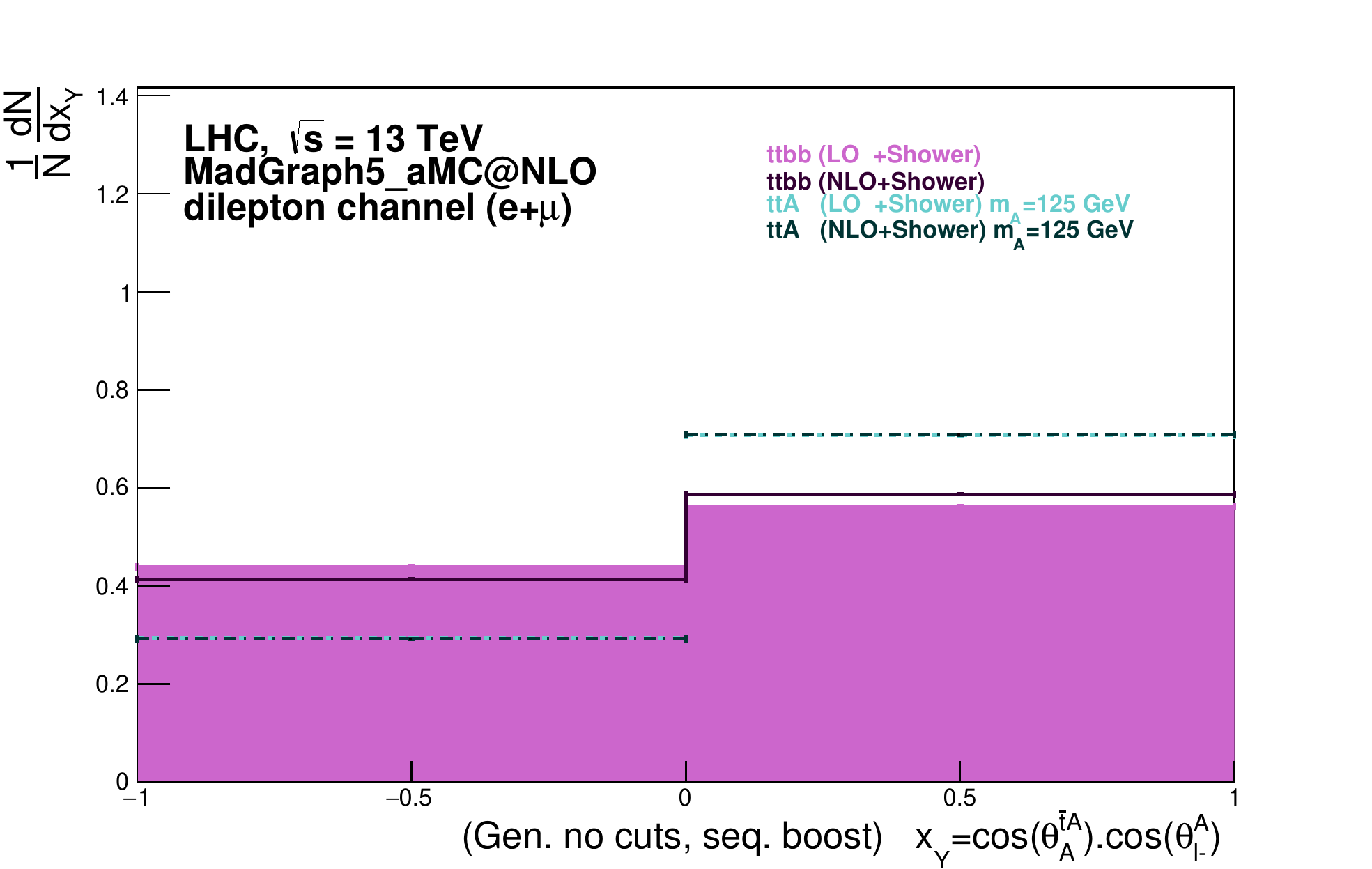,height=3.7cm,clip=} &
			  \epsfig{file=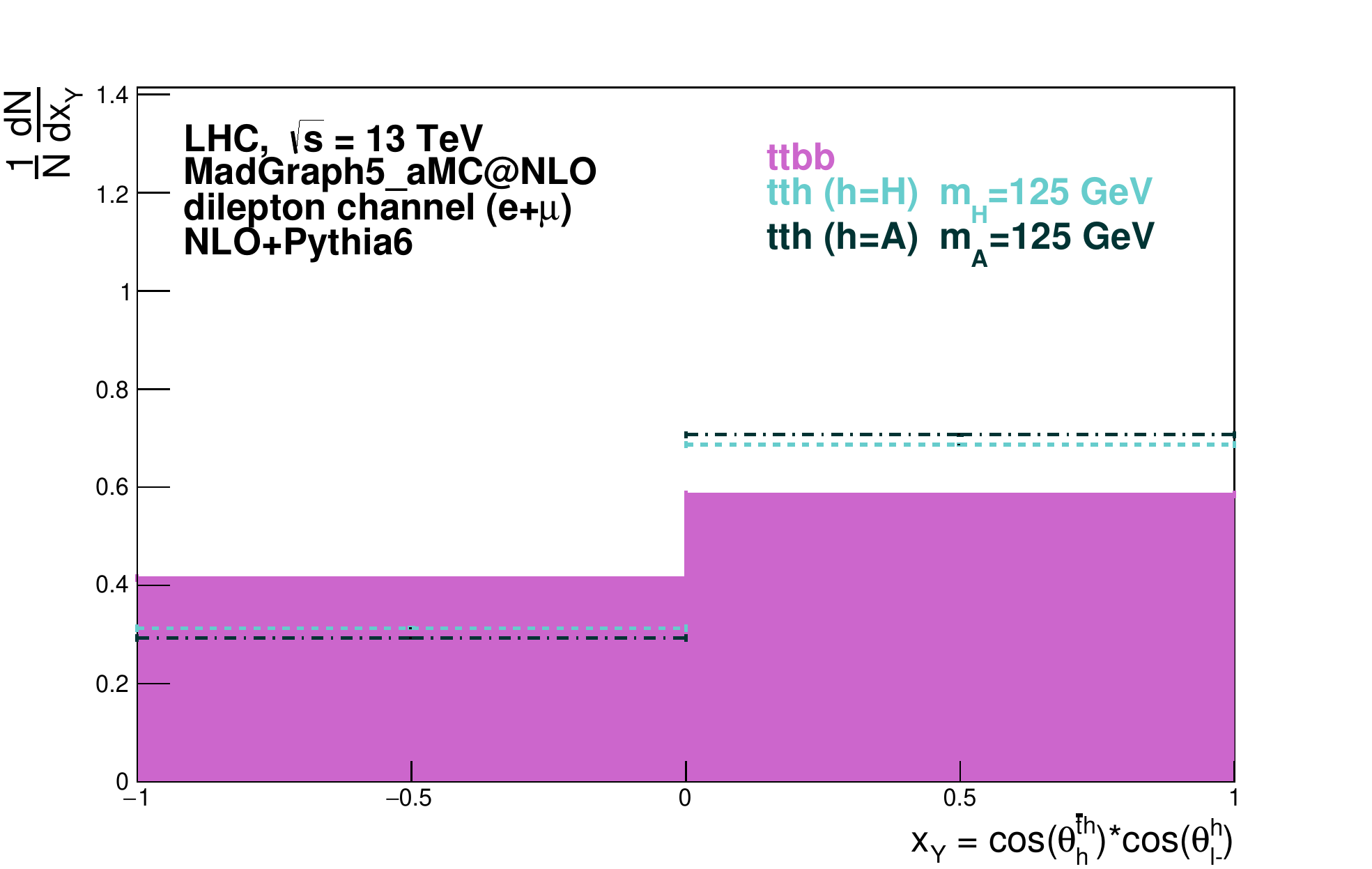,height=3.7cm,clip=} \\[-4mm]
\end{tabular}
\caption{NLO+Shower versus LO+Shower behaviour of the distribution of $x_Y$=$\cos{(\theta^{\bar{t}H}_{H})}\cos{(\theta^{H}_{\ell^-})}$ at parton Level with shower effects, without any selection cuts nor reconstruction, for the SM signal $t\bar{t}H$ (left) and for the $t\bar{t}A$ signal (middle), each one against the main background $t\bar{t}b\bar{b}$. Notice that, for the middle plots, the sequential boost prescription was employed for the lepton. The differences between the LO+Shower and NLO+Shower angular distributions are shown on top. Asymmetries around $x_Y=0$ are visible in the 2 binned distributions (bottom). The $t\bar{t}H$, $t\bar{t}A$ and $t\bar{t}b\bar{b}$ angular distributions at NLO are compared (top right), and the corresponding 2 binned distributions show the asymmetries (bottom right).}
\label{fig:ExpAng1}
\end{center}
\end{figure*}

The impact of NLO corrections on the angular distributions are shown in Figure~\ref{fig:ExpAng1} (left), by comparing with the LO distributions of $x_Y$=$\cos{(\theta^{\bar{t}H}_{H})}\cos{(\theta^{H}_{\ell^-})}$, at parton level (including shower effects) without any cuts, both for the SM $t\bar{t}H$ signal and $t\bar{t}b\bar{b}$ background events. NLO (LO) corrections with the impact of shower effects  are labelled NLO+Shower (LO+Shower) through out the text. The same distributions are shown for the $t\bar{t}A$ signal in Figure~\ref{fig:ExpAng1} (middle), with the exception that the sequential prescription was used for the $\ell^{-}$. Clear differences are visible between the direct and sequential prescriptions in particular for the background. Figure~\ref{fig:ExpAng1} (right) shows a comparison between $t\bar{t}H$, $t\bar{t}A$ and $t\bar{t}b\bar{b}$ at NLO+Shower, where the different possible natures of the signal ($t\bar{t}H$ or $t\bar{t}A$) do not seem to affect significantly the shape of the distribution. In the bottom plots, the corresponding distributions with 2 bins are shown, displaying the differences in forward-backward asymmetries.

\subsection{$t\bar{t}H$ and $t\bar{t}A$ Signals at NLO+Shower}

\begin{figure*} 
\begin{center}
\begin{tabular}{ccc}
\hspace*{-5mm} \epsfig{file=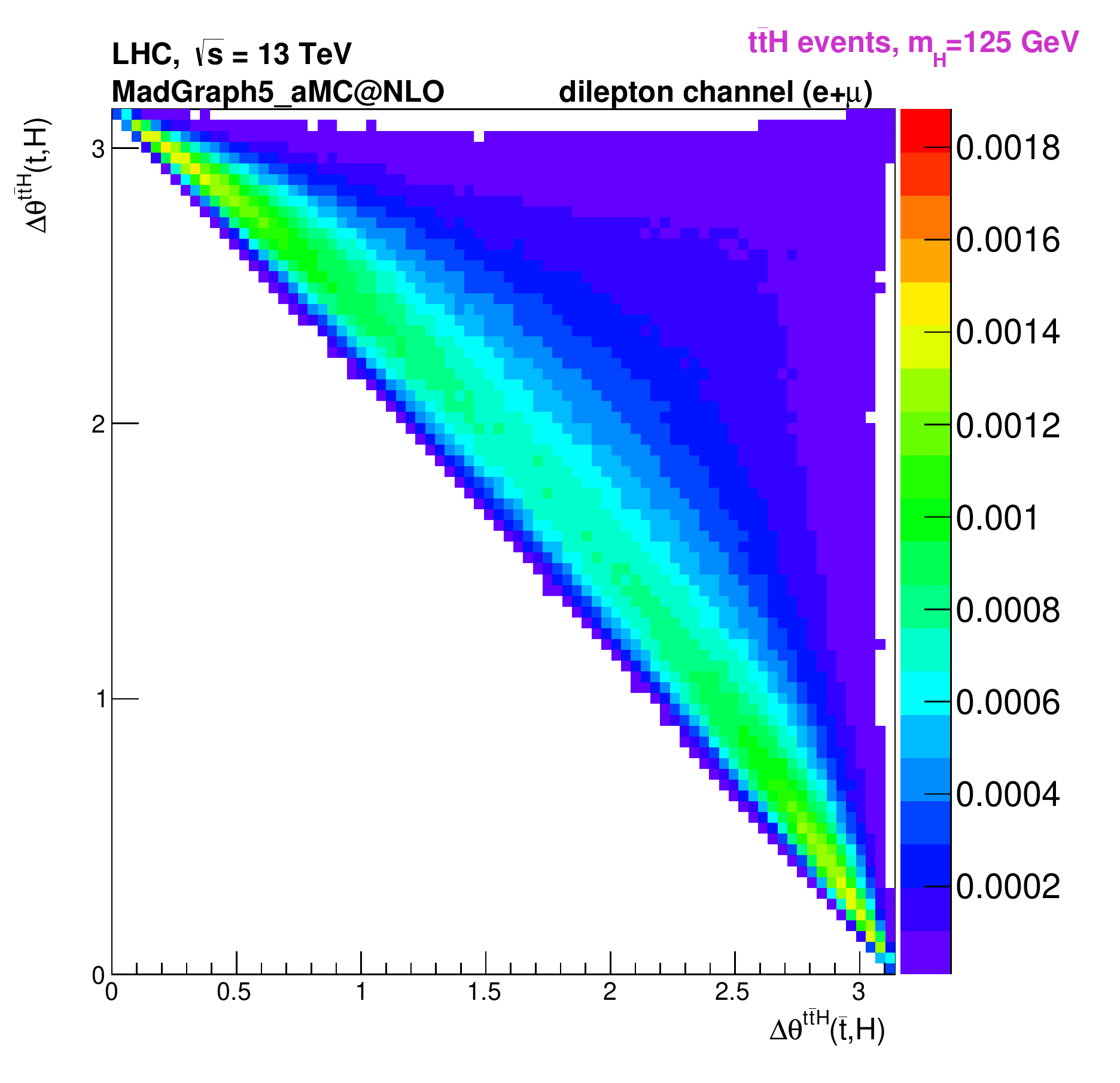,height=8.5cm,clip=} & \quad & 
                          \epsfig{file=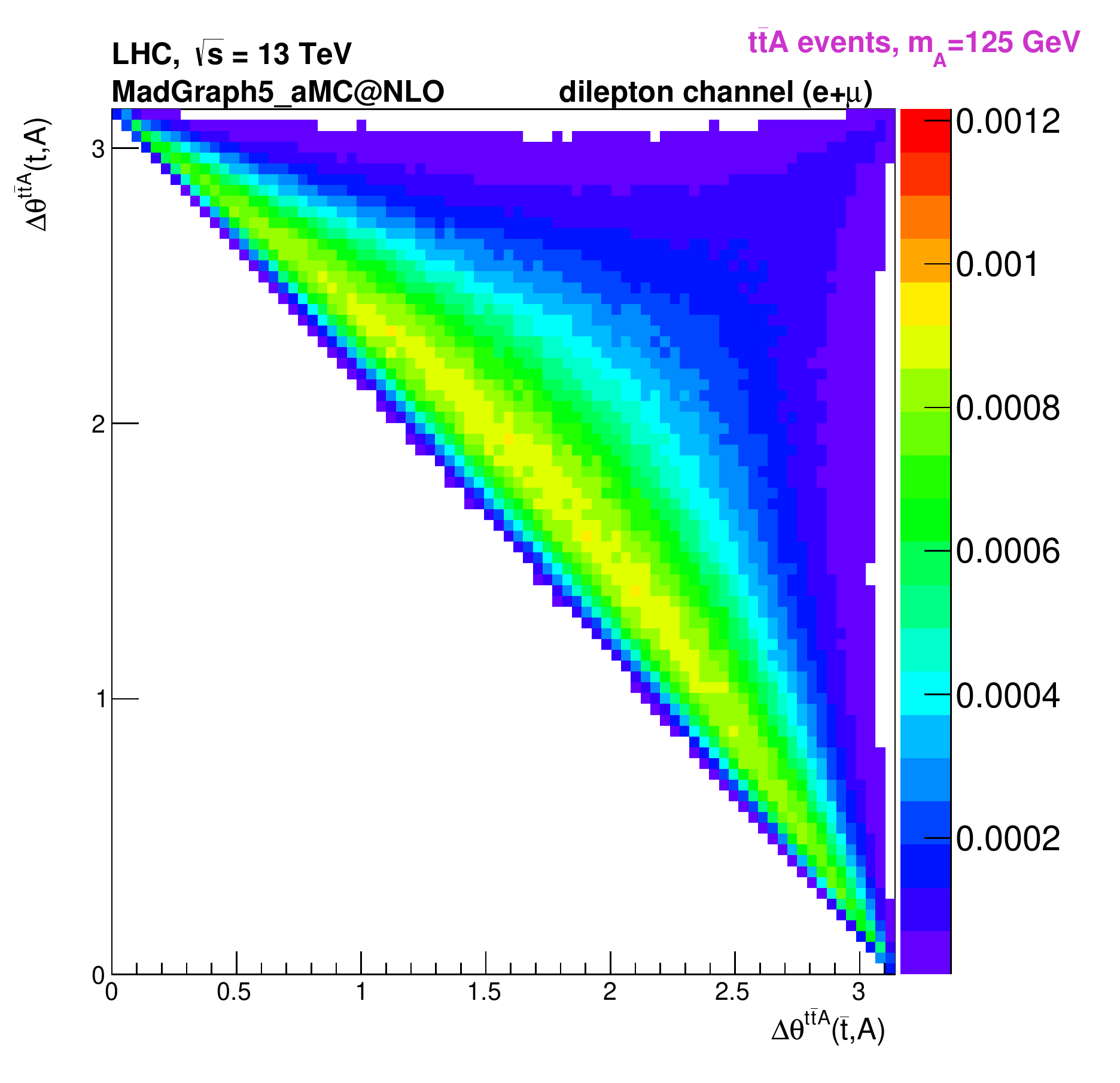,height=8.5cm,clip=} \\[-4mm]
\end{tabular}
\caption{Angle between the $t$ quark and Higgs boson ($x$-axis) at NLO+Shower effects plotted against the angle between the $\bar{t}$ quark and Higgs boson ($y$-axis), in the $t\bar{t}H$ centre-of-mass system. The SM Higss boson ($H$) distribution (left) and the pure pseudo-scalar Higgs boson ($A$) distribution (right) are shown.}
\label{fig:ttHangles00}
\end{center}
\end{figure*}

Exploring kinematic differences between $t\bar{t}H$ and $t\bar{t}A$ is of utmost importance in order to find a set of good discriminating variables that may be sensitive to the nature of top quark Yukawa coupling. In fact, differences between the scalar and pseudo-scalar are visible through angles between particle directions  ($t$, $\bar{t}$ and $h$), already at production. Figure~\ref{fig:ttHangles00} (left) shows, at NLO+Shower, the angle between the top quark and Higgs boson directions ($x$-axis) versus the angle between the anti-top quark and Higgs boson directions ($y$-axis), all evaluated in the $t\bar{t}H$ centre-of-mass system. The same distribution is shown for the pseudo-scalar signal $t\bar{t}A$ in Figure~\ref{fig:ttHangles00} (right). In Figure~\ref{fig:ttHangles01}, the angle between the top quark direction in the $t\bar{t}h$ centre-of-mass frame and the $t\bar{t}h$ direction in the lab frame ($y$-axis), is plotted against the angle between the Higgs direction, in the $\bar t h$ rest frame, and the direction of three decay products, all boosted to the $h$ rest frame ($x$-axis): (left) $b$ quark from Higgs boson, (middle) $\ell^+$ from top quark and (right) $\ell^-$ from $\bar{t}$. In the top (bottom) row, the $t\bar{t}H$ ($t\bar{t}A$) signal is shown, without any cuts. Differences between the scalar and pseudo-scalar signals are clearly visible.

\subsection{Angular Distributions after Reconstruction}

\begin{figure*} 
\begin{center}
\begin{tabular}{ccc}
\hspace*{-5mm} \epsfig{file=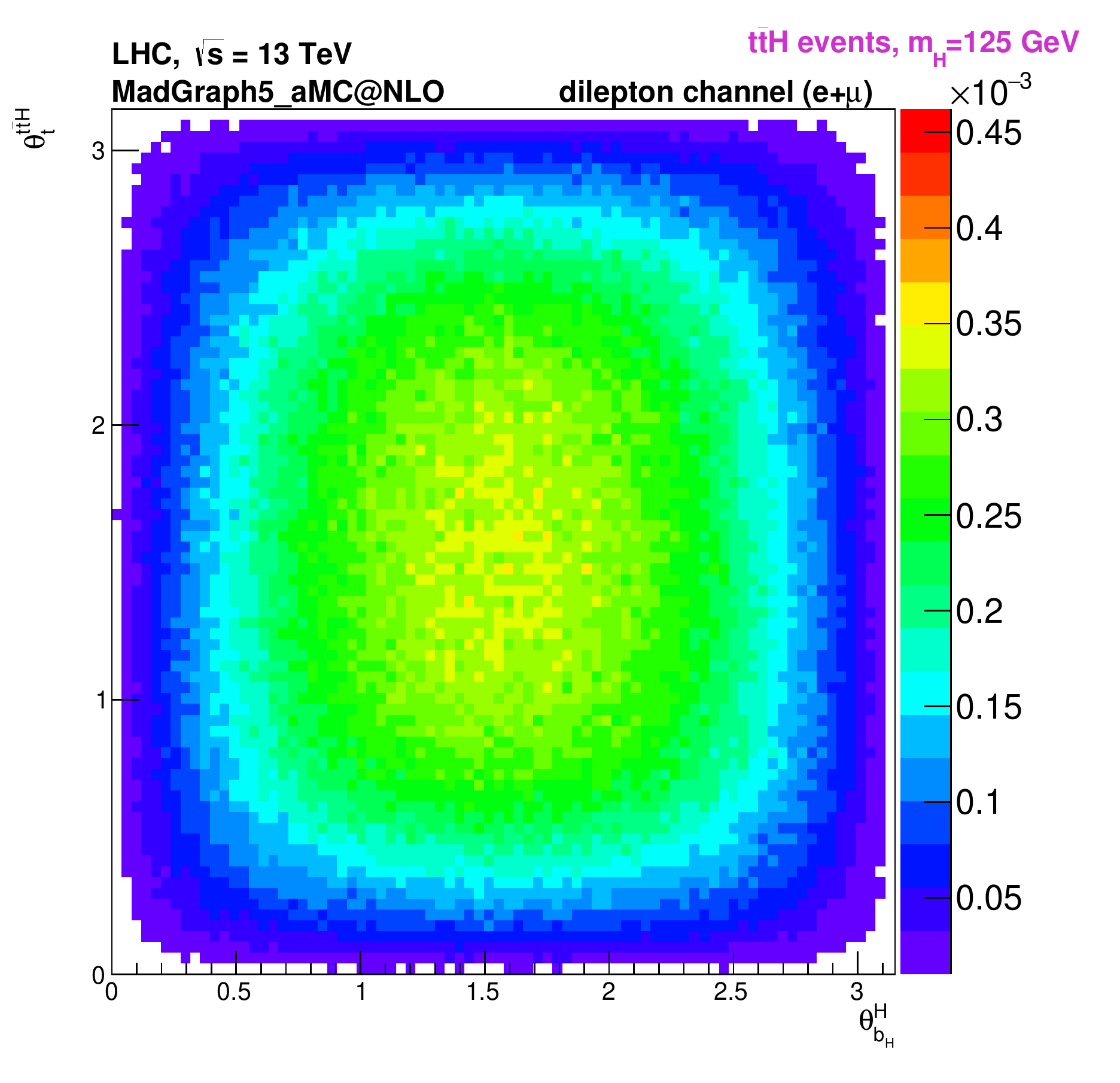,height=5.0cm,clip=}  &
 			 \epsfig{file=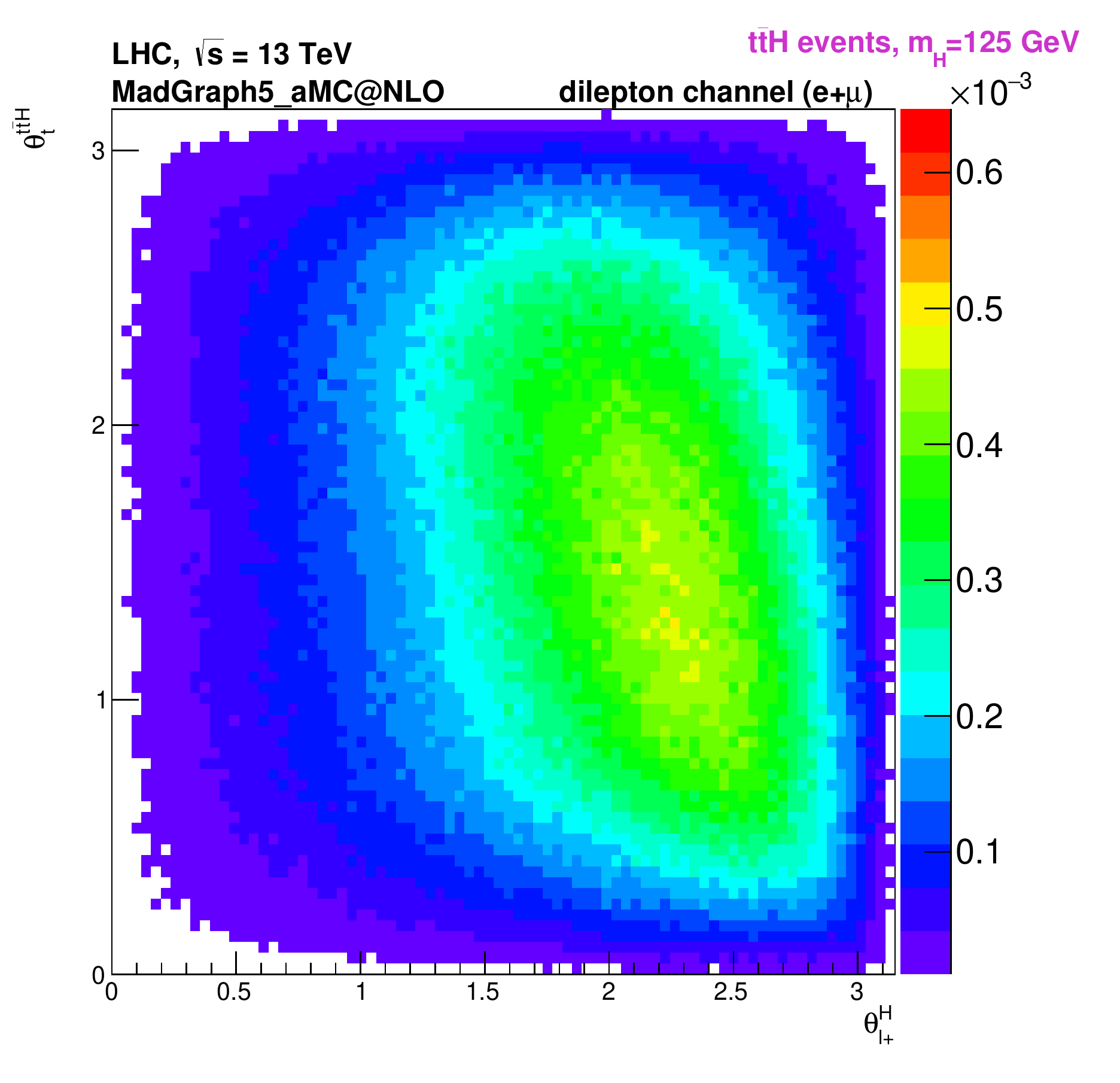,height=5.0cm,clip=} &  
			 \epsfig{file=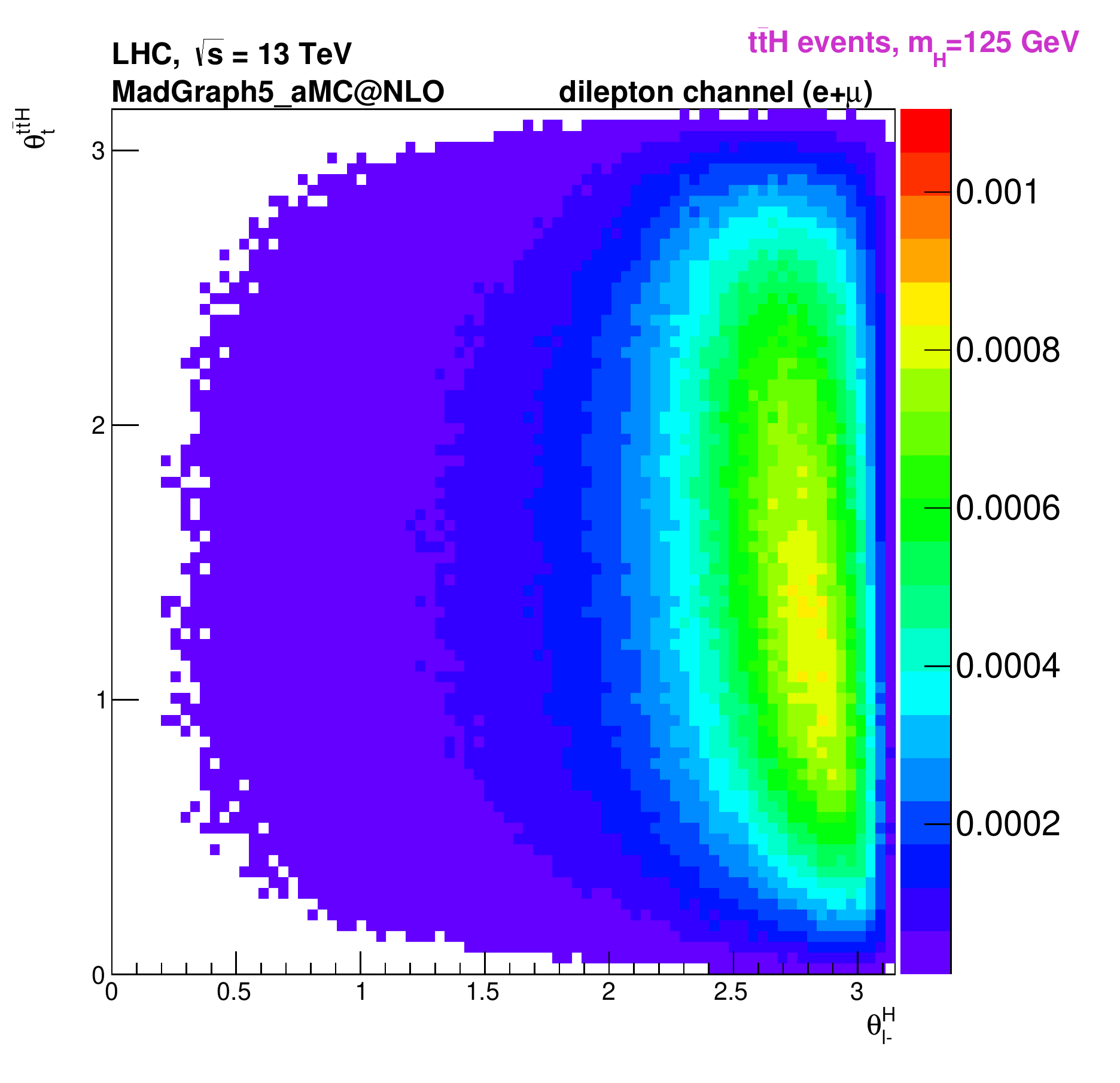,height=5.0cm,clip=} \\[-2mm]
\hspace*{-5mm} \epsfig{file=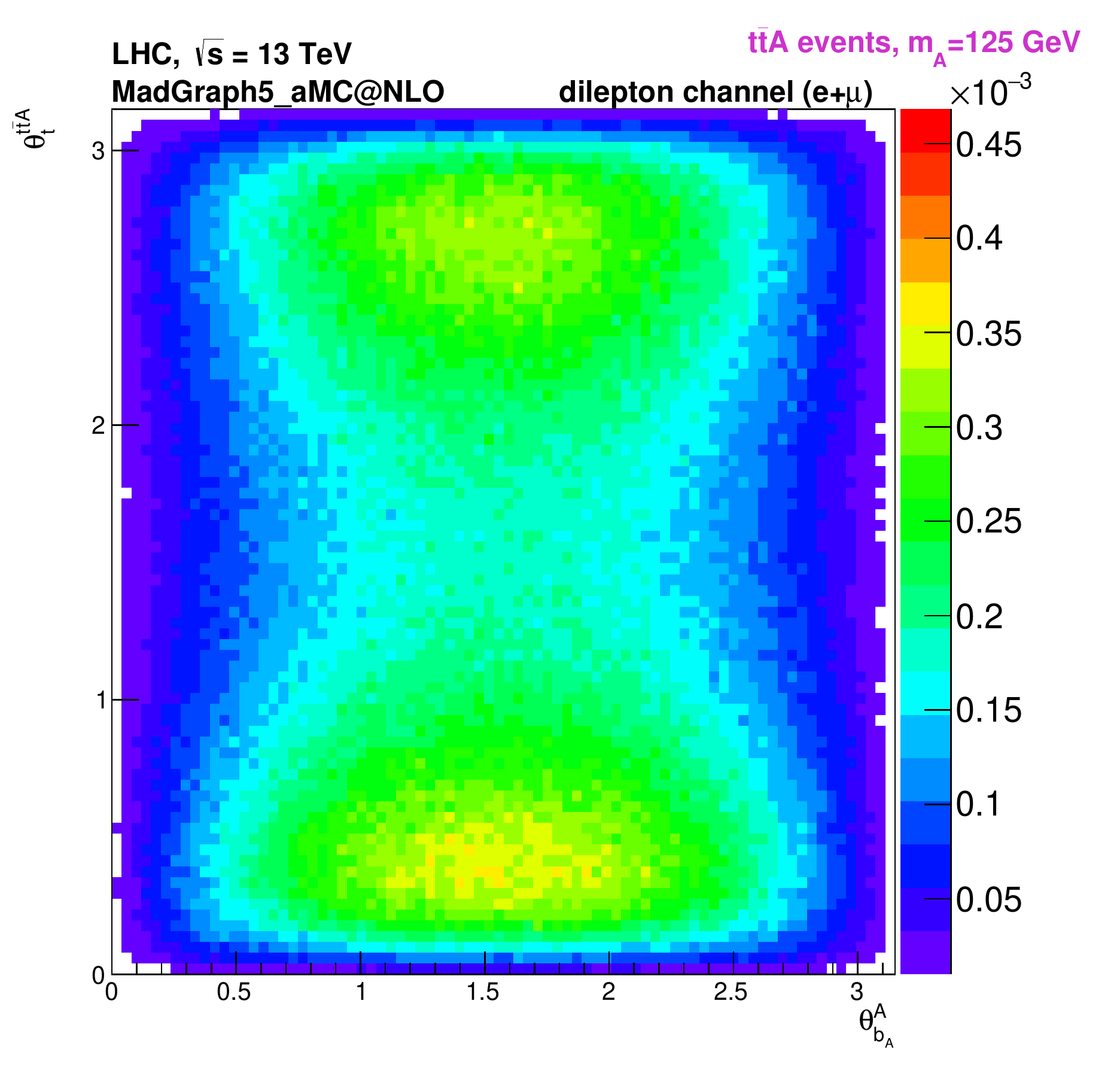,height=5.0cm,clip=}  & 
			 \epsfig{file=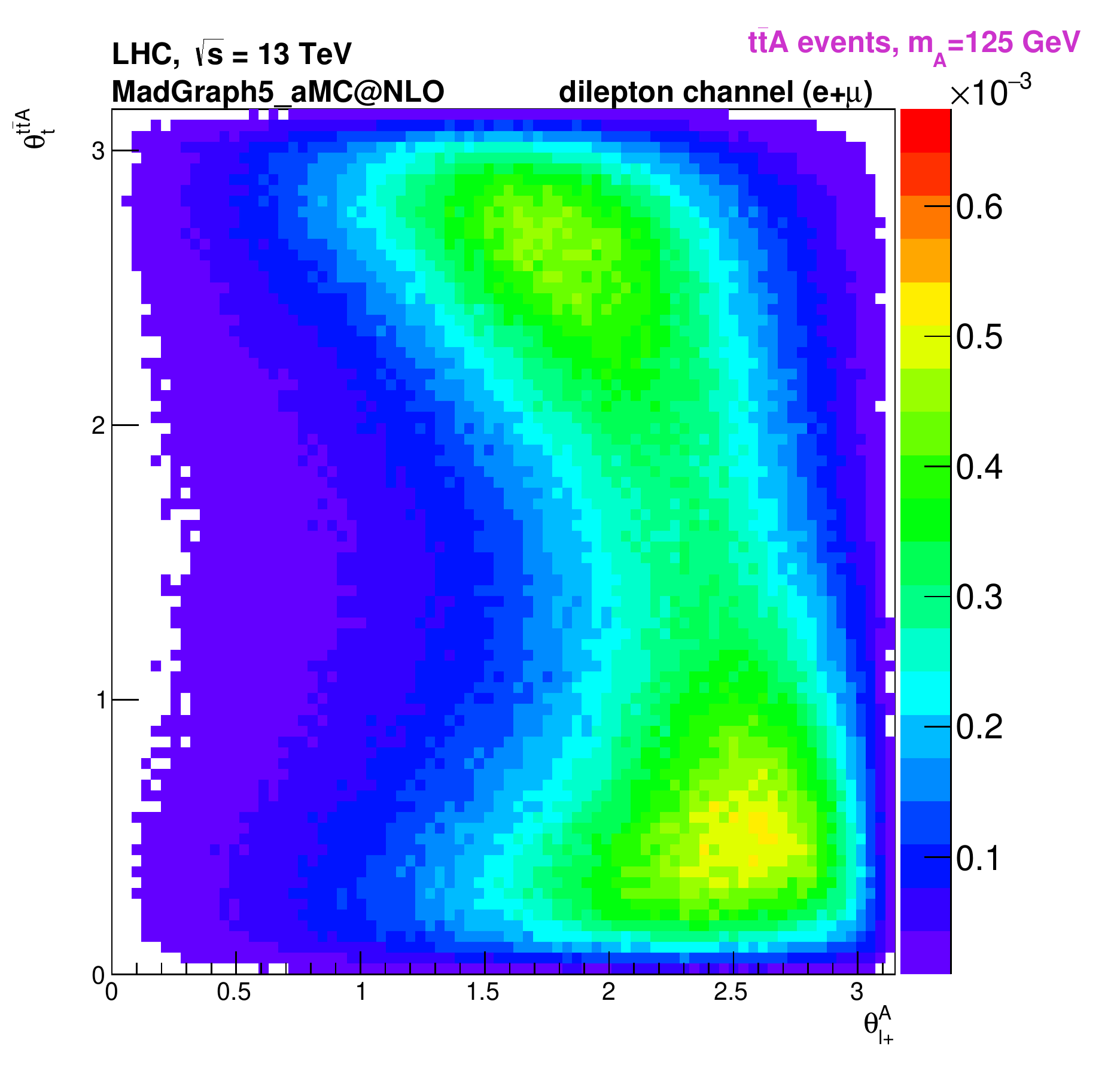,height=5.0cm,clip=} &
			 \epsfig{file=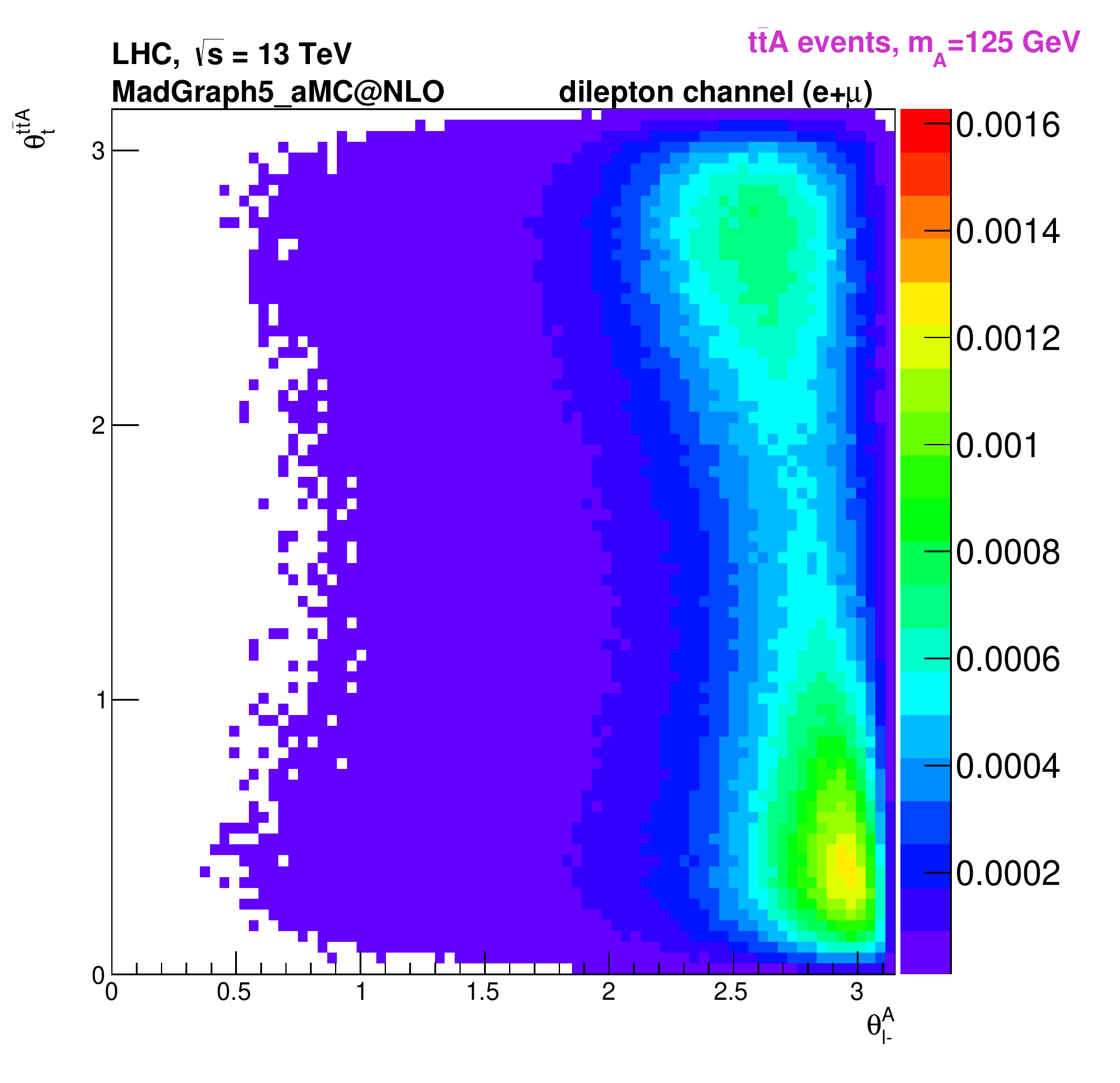,height=5.0cm,clip=}\\[-4mm]
\end{tabular}
\caption{Two dimensional distribution at NLO+Shower of the angle between the top quark, in the $t\bar{t}h$ centre-of-mass frame, and the $t\bar{t}h$ direction in the lab frame ($y$-axis) plotted against the angle between the Higgs direction, in the $\bar t h$ rest frame, and the direction of several decay products (all boosted to the Higgs centre-of-mass): (left) $b$ quark from $h$, (middle) $\ell^+$ from top quark and (right) $\ell^-$ from $\bar{t}$. The top (bottom) distributions correspond to $t\bar{t}H$ ($t\bar{t}A$), without any cuts.}
\label{fig:ttHangles01}
\end{center}
\end{figure*}

\begin{figure*} 
\vspace*{-1cm}
\begin{center}
\begin{tabular}{ccc}
\hspace*{-5mm} \epsfig{file=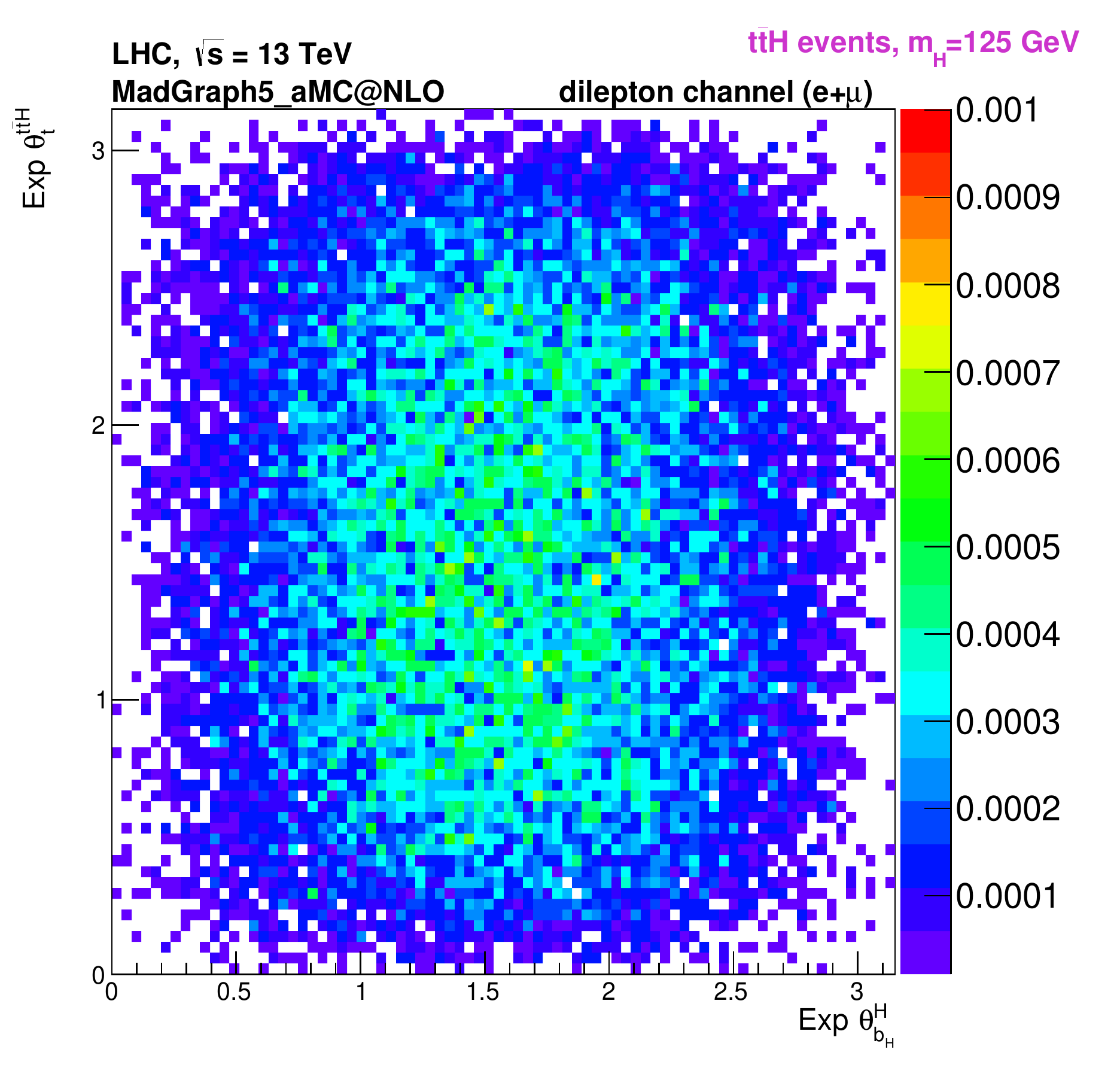,height=5.0cm,clip=}  &
 			 \epsfig{file=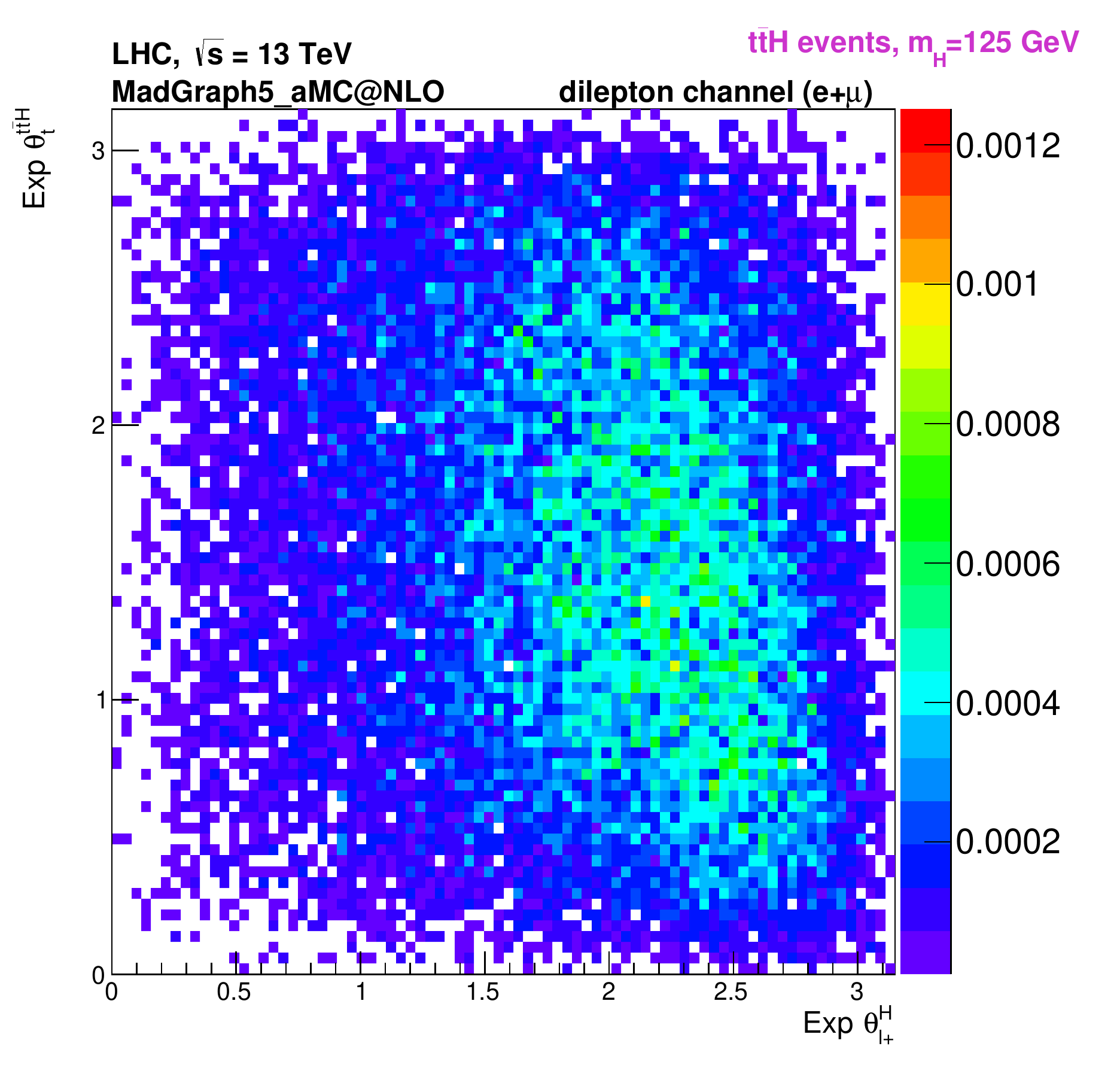,height=5.0cm,clip=} &  
			 \epsfig{file=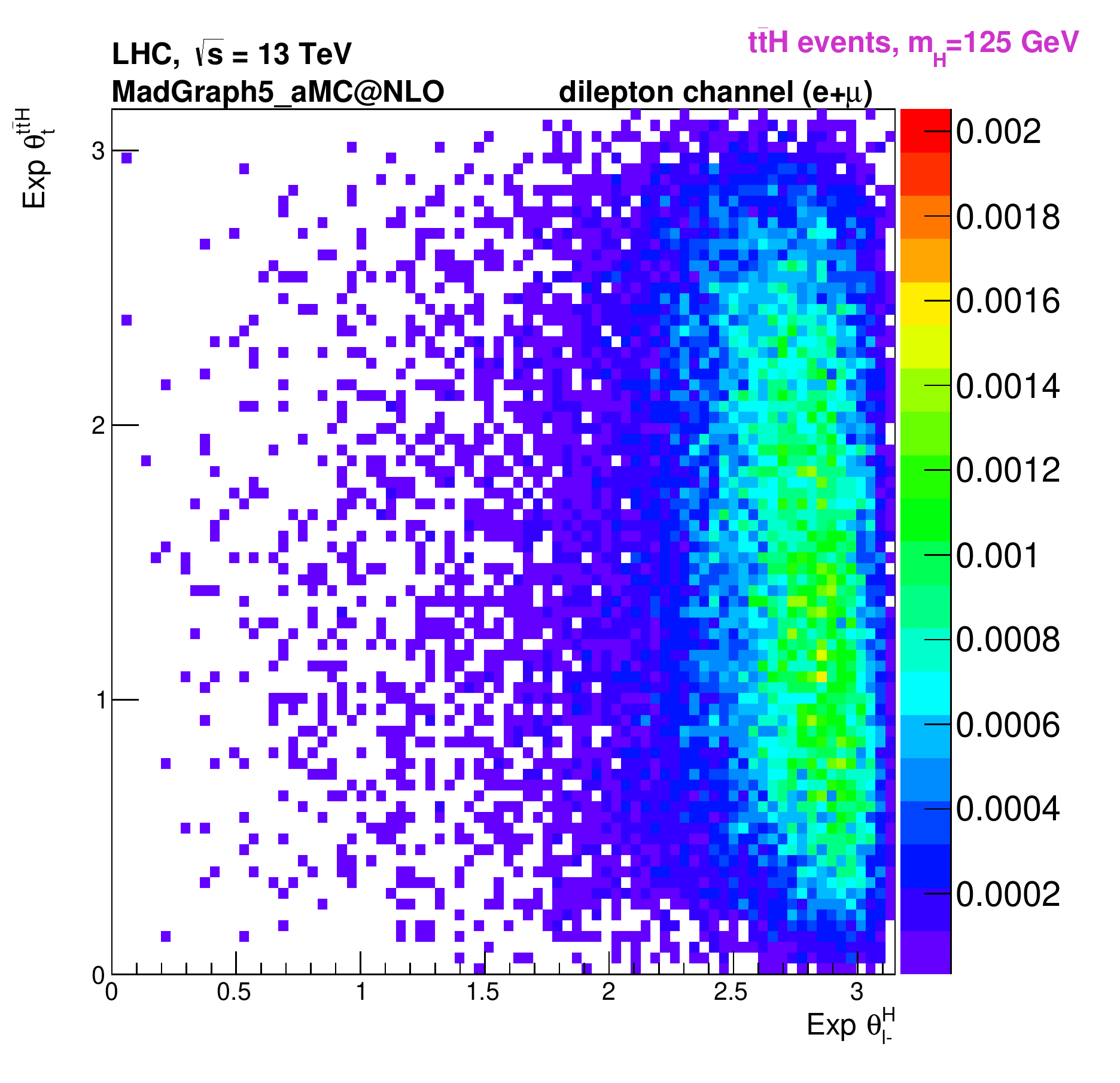,height=5.0cm,clip=} \\[-2mm]
\hspace*{-5mm} \epsfig{file=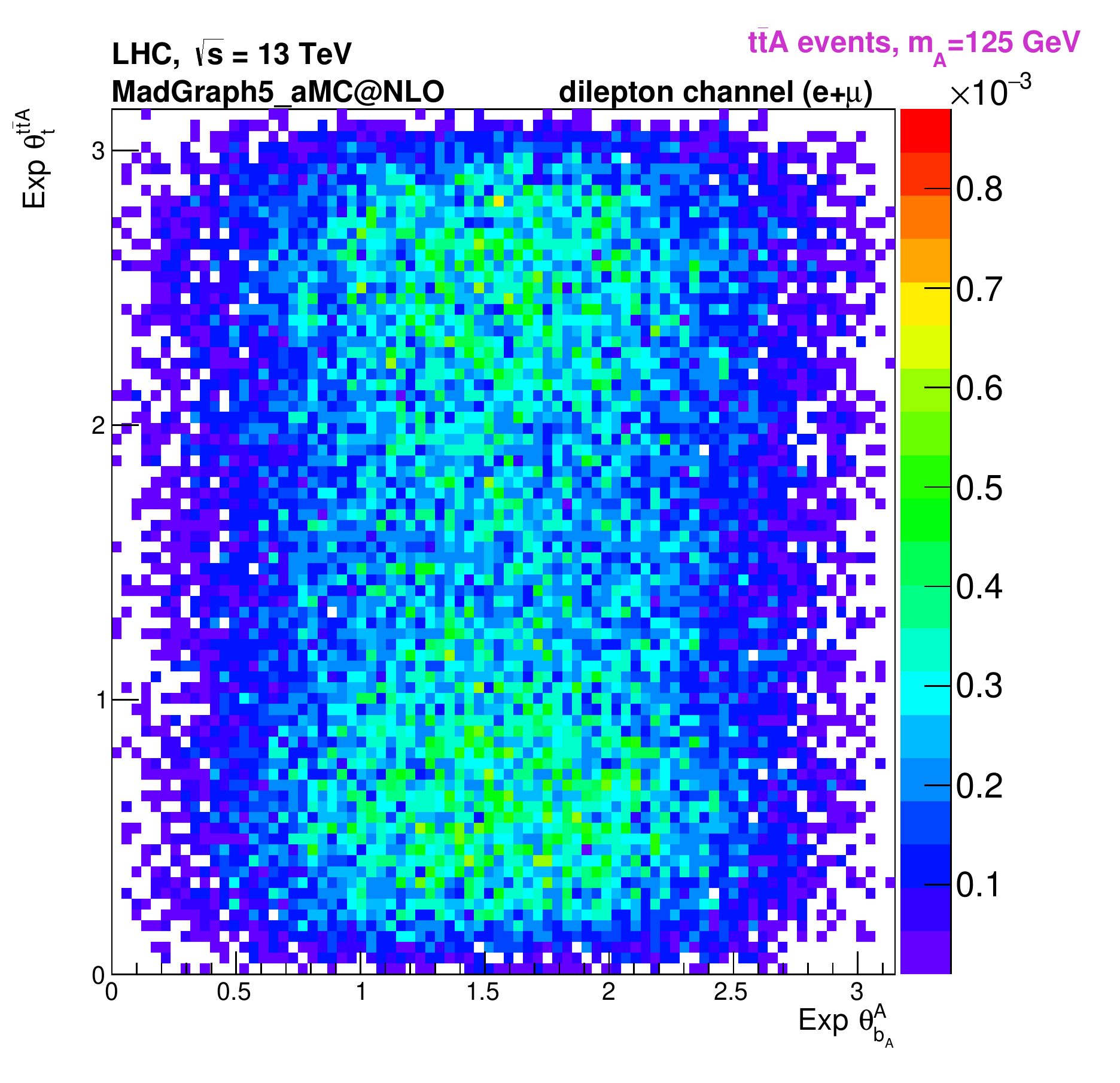,height=5.0cm,clip=}  & 
			 \epsfig{file=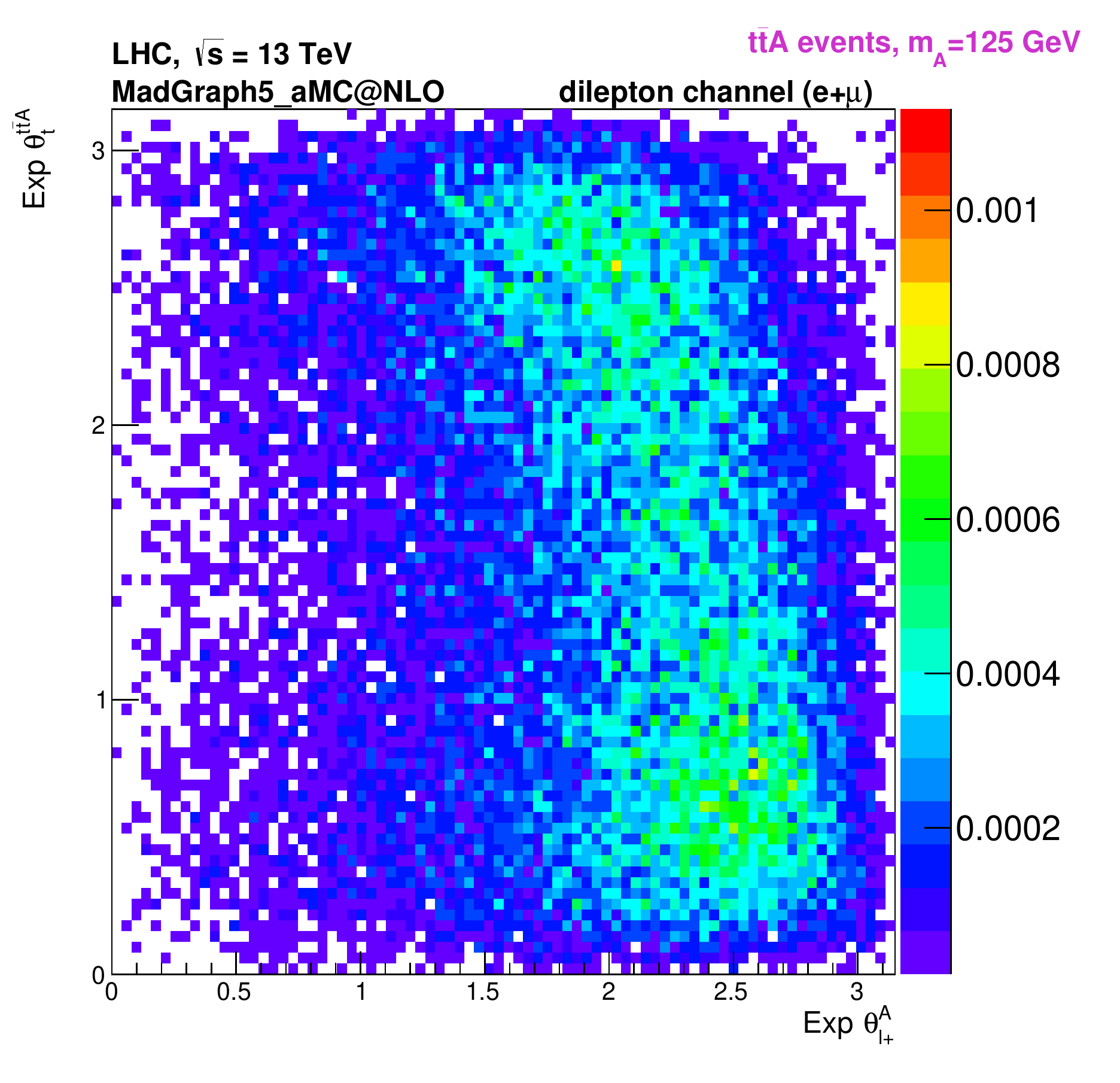,height=5.0cm,clip=} &
			 \epsfig{file=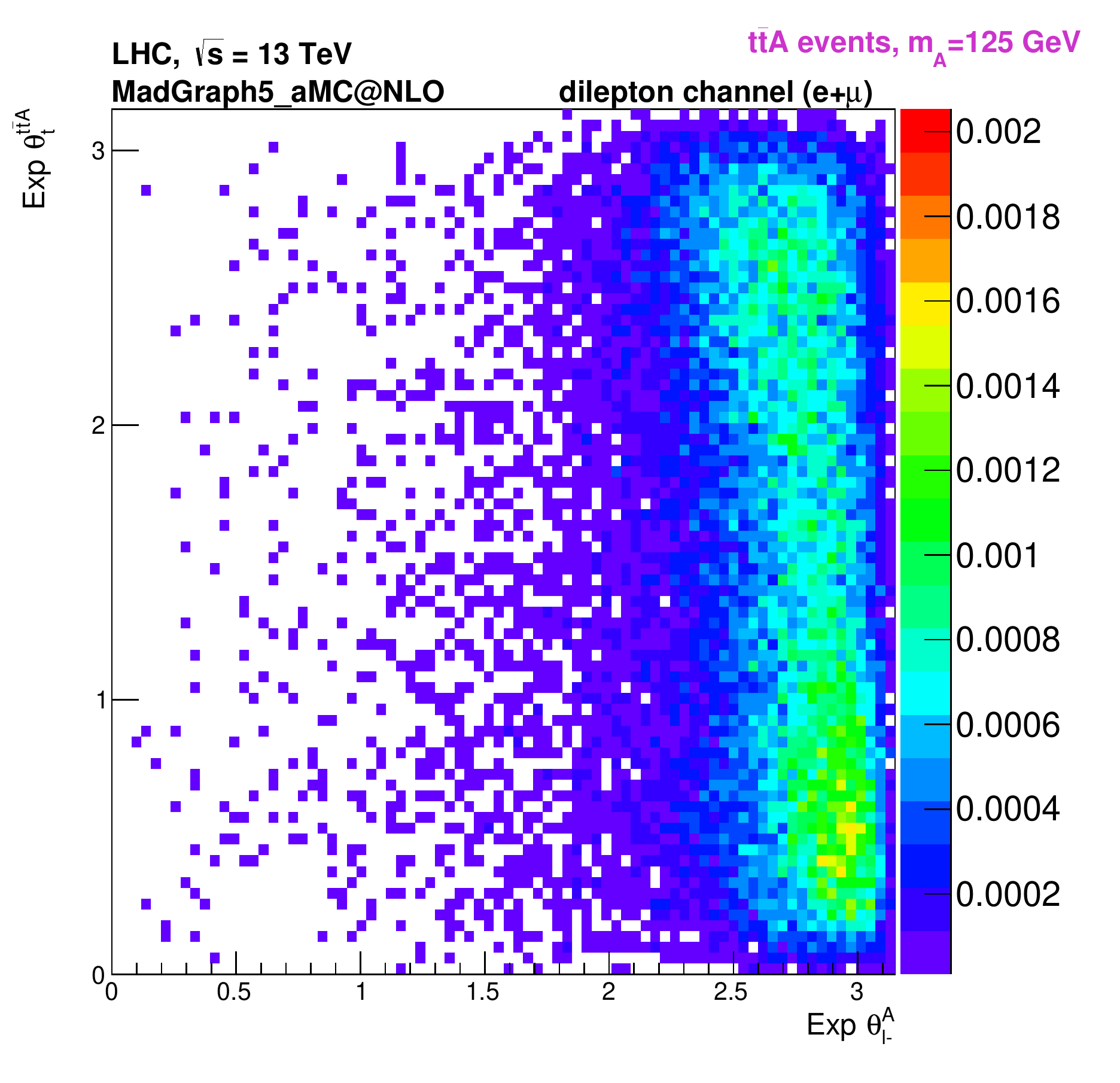,height=5.0cm,clip=}\\[-4mm]
\end{tabular}
\caption{The same as Figure~\ref{fig:ttHangles01}, after all selection cuts and full kinematic reconstruction.}
\label{fig:ttHangles02}
\end{center}
\end{figure*}

Signal distributions are distorted due to cuts from the necessary selection cuts applied to events and the kinematic fit. The shape of the distributions, although affected by the significant reduction on the total number of events, is nevertheless largely preserved. In Figure~\ref{fig:ttHangles02} the same angular distributions as those shown in Figure~\ref{fig:ttHangles01} are represented, after selection cuts and full kinematic reconstruction. The density of points shows a similar pattern of that from Figure~\ref{fig:ttHangles01}. Even after kinematic reconstruction, clear differences between the different signal natures are visible.

Forward-backward asymmetries associated to each of the observables under study, were defined according to~\cite{Santos:2015dja}
\begin{eqnarray}
\ensuremath{A^Y_{FB}= \frac  {    \sigma(x_Y>0)-\sigma(x_Y<0)   }{ \sigma( x_Y>0)+\sigma(x_Y<0) } },
\label{equ:a9}
\end{eqnarray}
where $\sigma(x_Y>0)$ and $\sigma(x_Y<0)$ correspond to the total cross section with $x_Y$ above and below zero, respectively. The asymmetries are evaluated at NLO+Shower and after the kinematic fit, for different choices of the variable $x_Y$ (found to provide a significant difference between the signals and dominant backgound): 

\begin{tabular}{ccl}
 & \quad & $\cos{(\theta^{\bar{t}h}_{h})}\cos{(\theta^{h}_{\ell^-})}$ for $A^{\ell-(h)}_{FB}$, \\[1mm]  
\end{tabular}  

\begin{tabular}{ccl} 
  & \quad & $b_4 = (p^z_t . p^z_{\bar{t}}) / (|\vec{p}_{t}| . |\vec{p}_{\bar{t}}| )$, as defined in \cite{Gunion:1996xu}, for $A^{b_4}_{FB}$, \\[1mm] 
  & \quad & $\sin{(\theta^{t\bar{t}h}_{h})}\sin{(\theta^{\bar{t}}_{\bar{b}_{\bar{t}}})}$ for $A^{\bar{b}_{\bar{t}}({\bar{t}})}_{FB}$({\it seq.}~boost), \\[1mm]
  & \quad & $\sin{(\theta^{t\bar{t}h}_{h})}\cos{(\theta^{\bar{t}}_{b_h})}$ for $A^{b_h(\bar{t})}_{FB}$({\it seq.}~boost), \\[1mm]
  & \quad & $\sin{(\theta^{t\bar{t}h}_{t})}\sin{(\theta^{h}_{W+})}$ for $A^{W+({h})}_{FB}$({\it seq.}~boost), \\[1mm]
  & \quad & $\sin{(\theta^{t\bar{t}h}_{\bar{t}})}\sin{(\theta^{h}_{b_h})}$ for $A^{b_h(h)}_{FB}$({\it seq.}~boost) and \\[1mm]     
  & \quad & $\sin{(\theta^{t\bar{t}h}_{h})}\sin{(\theta^{t\bar{t}}_{\bar{t}})}$ for $A^{\bar{t}(t\bar{t})}_{FB}$.  \\[3mm]   
\end{tabular} 

\begin{figure*} 
\begin{center}
\begin{tabular}{ccc}
\hspace*{-5mm} \epsfig{file=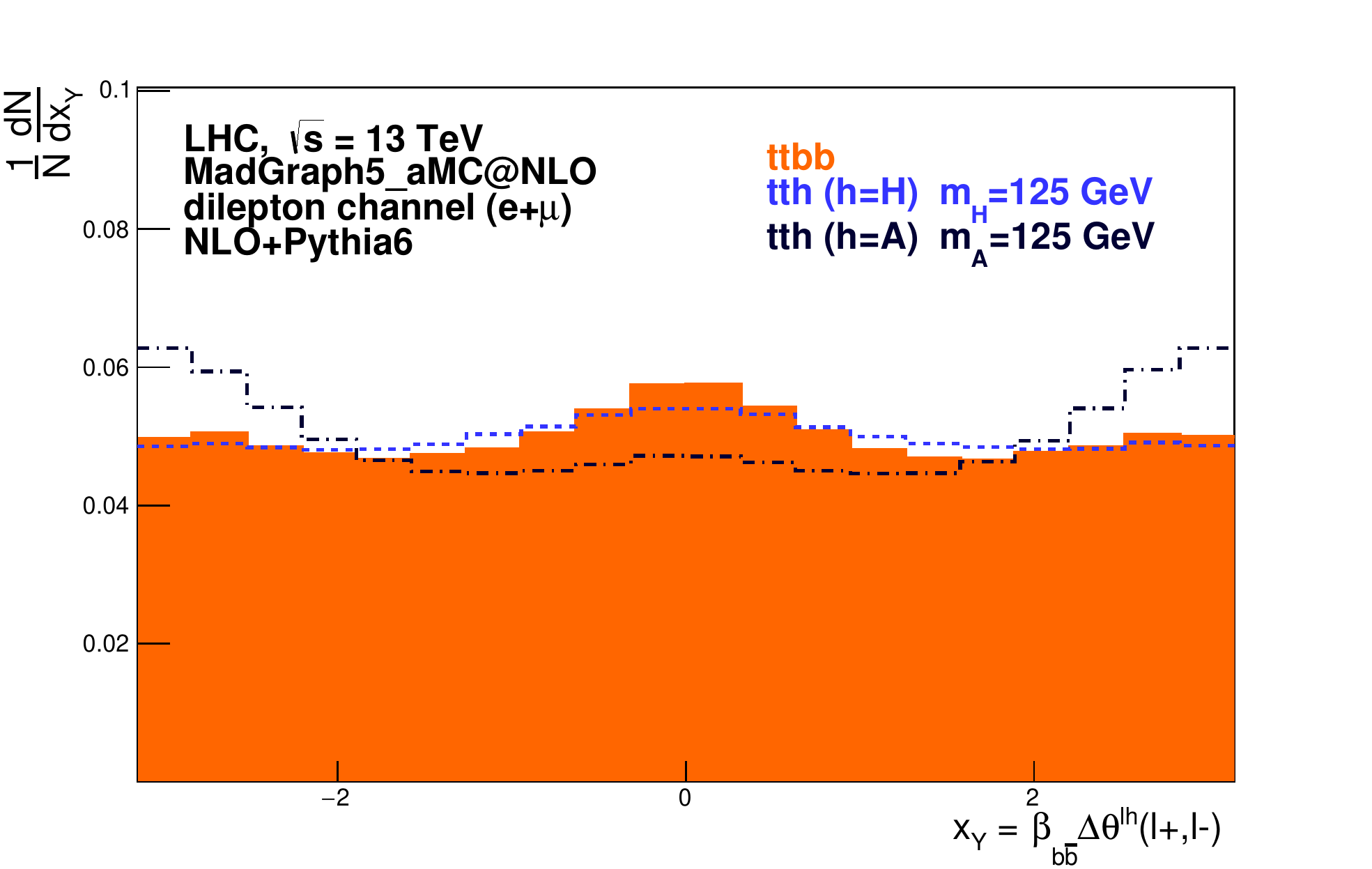,height=3.8cm,clip=}  &
 			 \epsfig{file=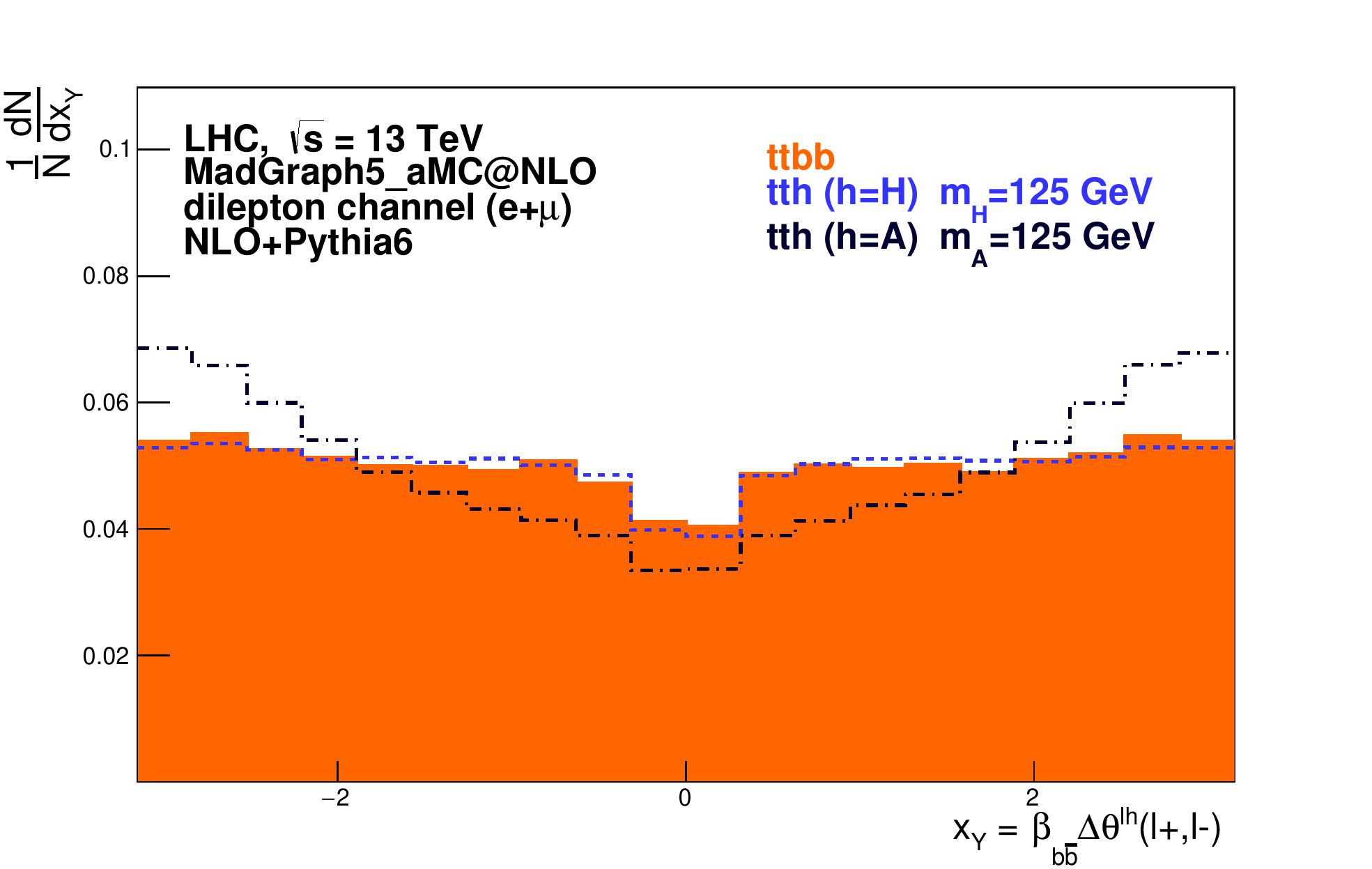,height=3.8cm,clip=} &  
			 \epsfig{file=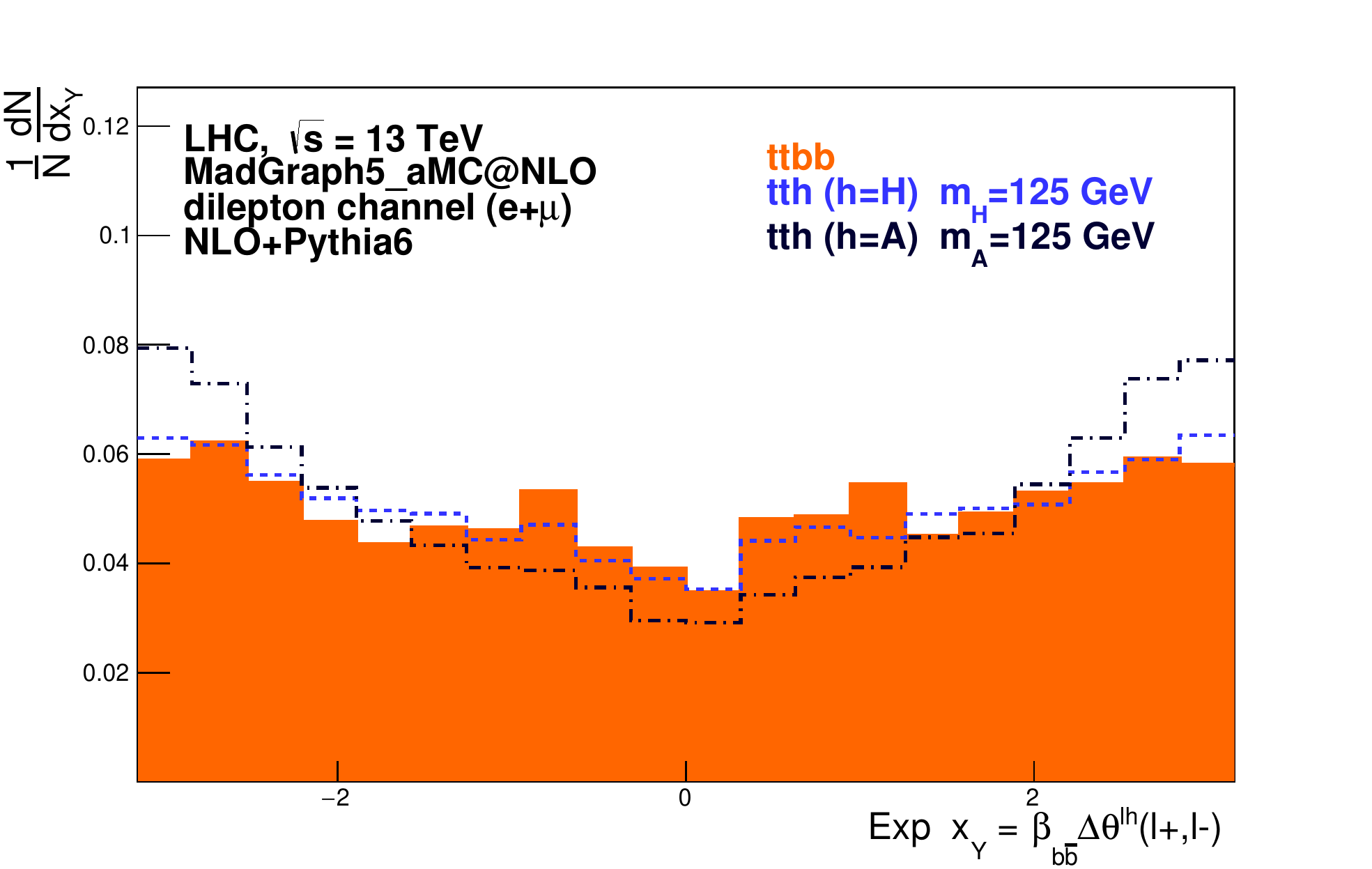,height=3.8cm,clip=} \\[0mm]
\hspace*{-5mm} \epsfig{file=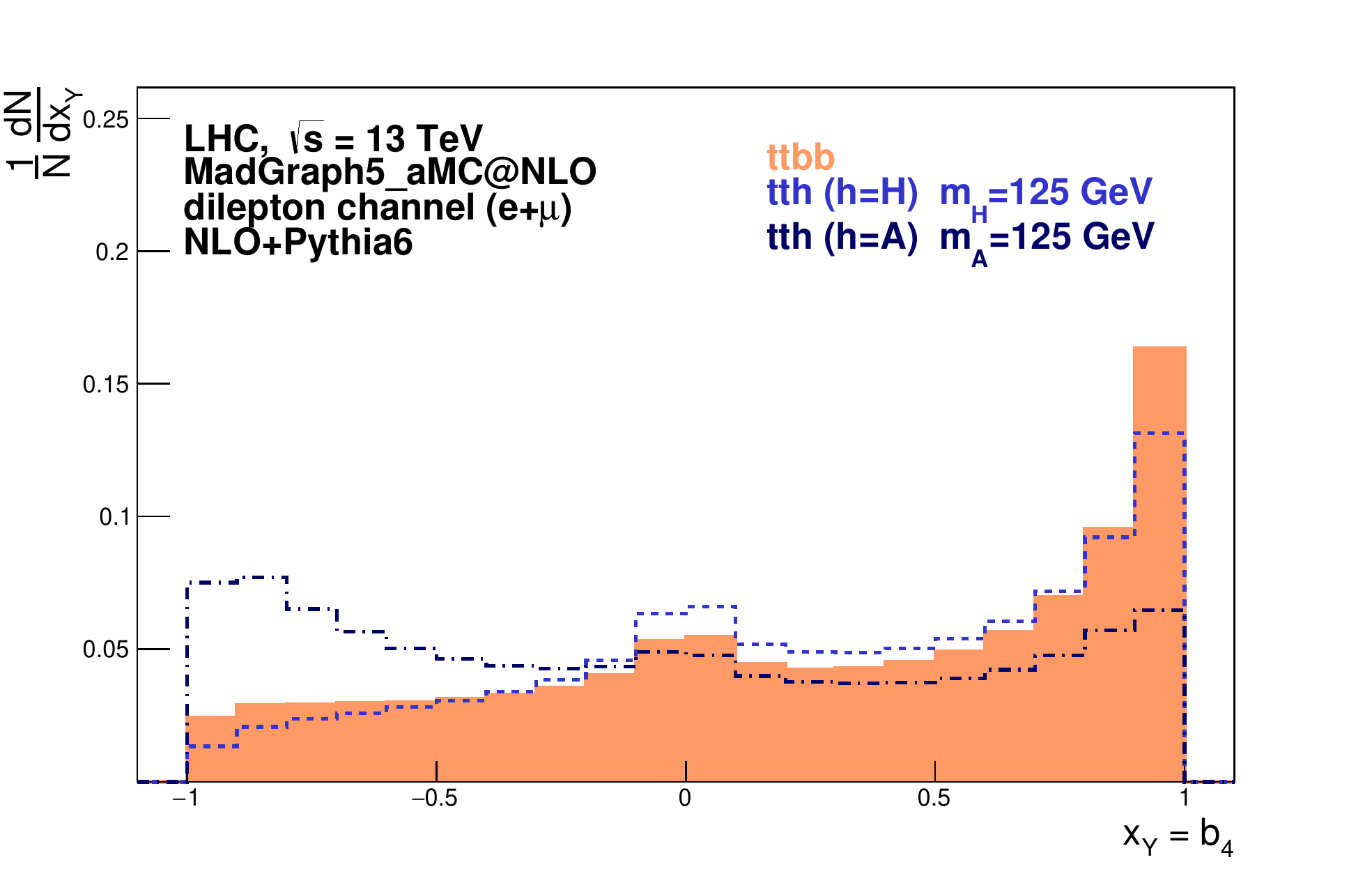,height=3.8cm,clip=}  & 
			 \epsfig{file=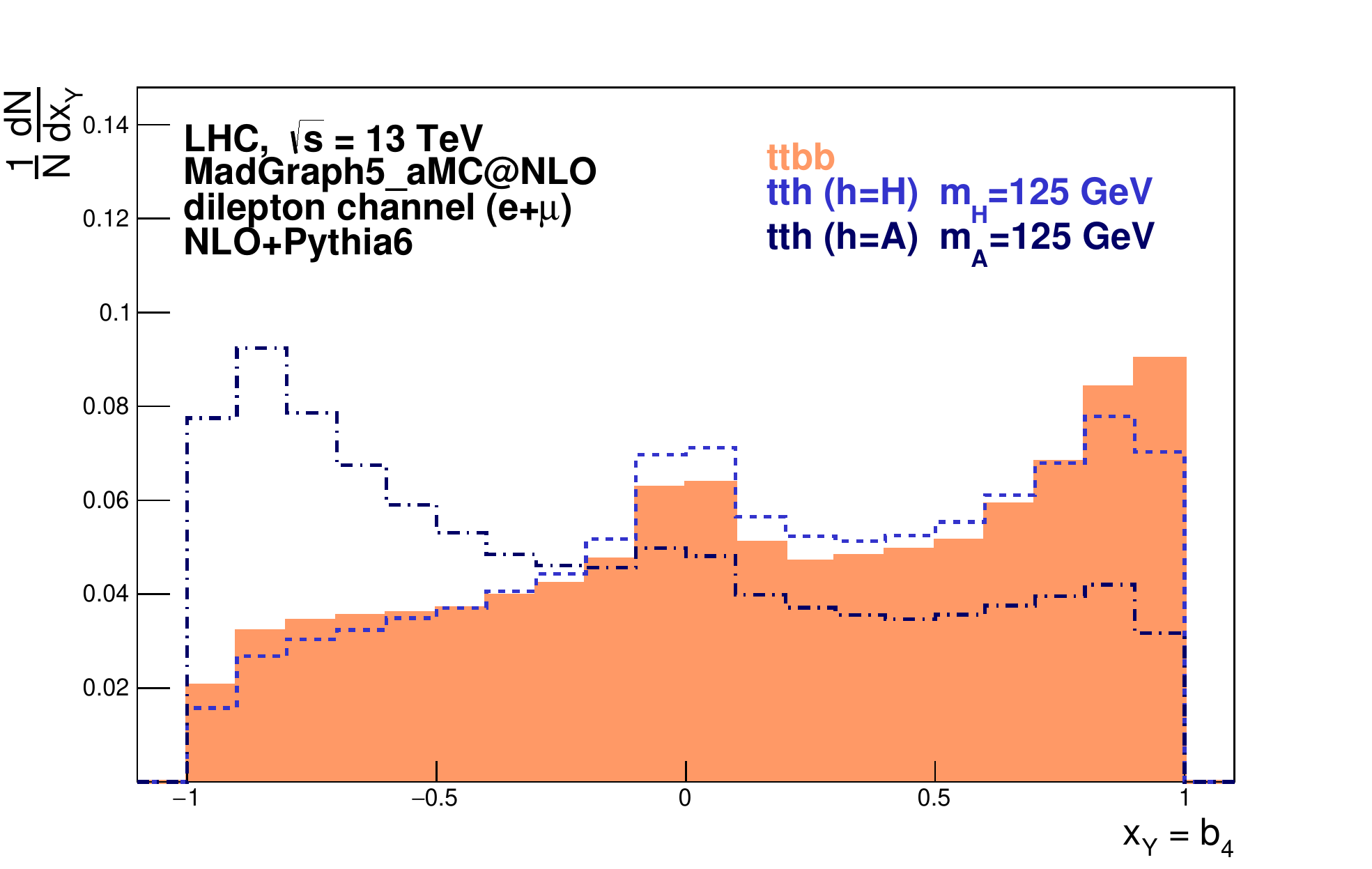,height=3.8cm,clip=} &
			 \epsfig{file=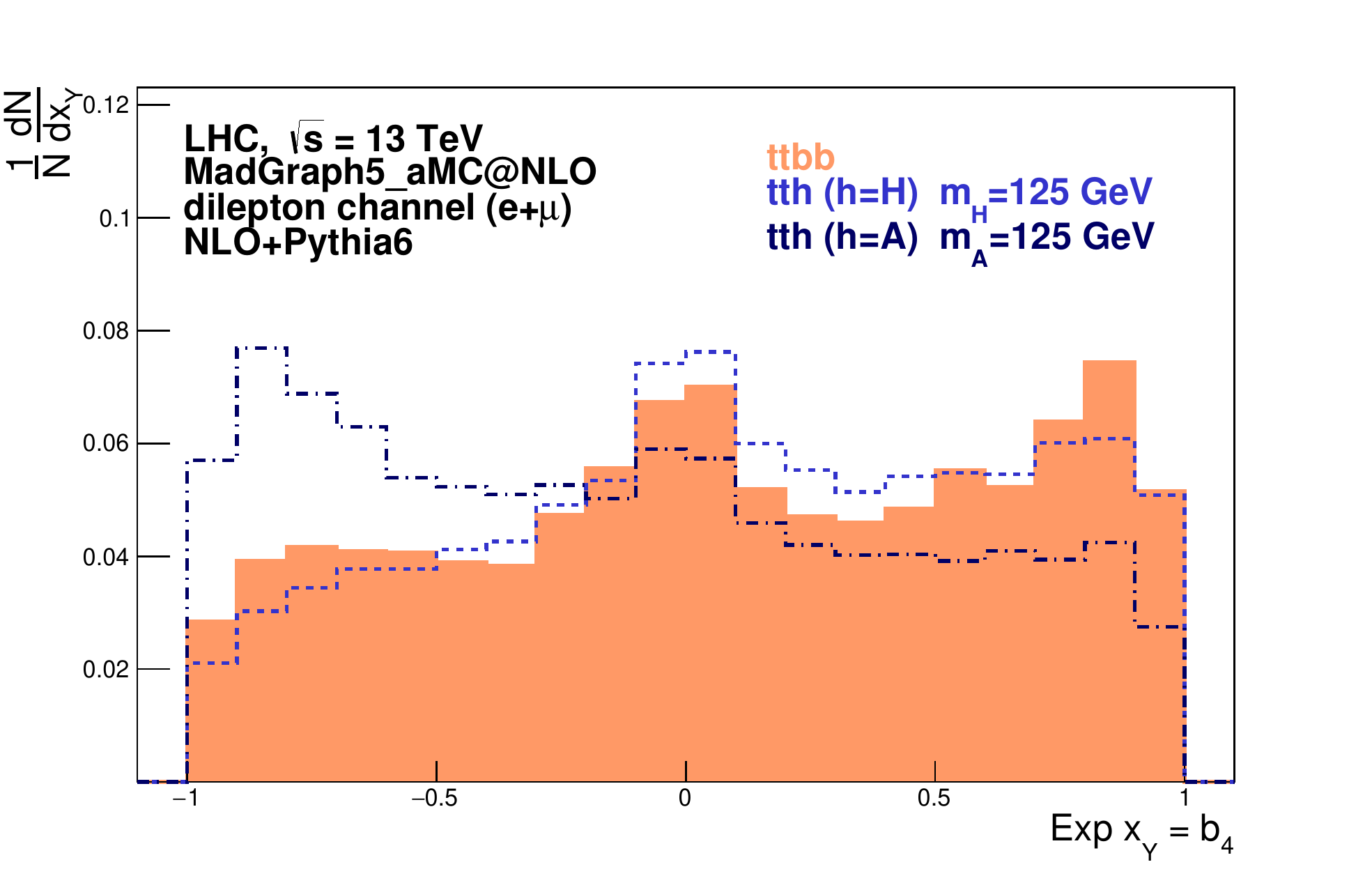,height=3.8cm,clip=}\\[-4mm]
\end{tabular}
\caption{Normalised $\beta\Delta\theta^{\ell h}(\ell+,\ell-)$ distributions at NLO+Shower without cuts (top left), with cuts (top middle) and after cuts and full kinematic reconstruction (top right). 
The NLO+Shower $b_4$ distribution is also shown at parton level without cuts (bottom left), with cuts (bottom middle) and after cuts and full kinematic reconstruction (bottom right). The dashed line represents the $t\bar{t}h$ SM model signal ($h=H$ and $CP=+1$) and the dashed-dotted line corresponds to the pure pseudo-scalar distribution $t\bar{t}h$ ($h=A$ and $CP=-1$). The shadowed region corresponds to the NLO+Shower $t\bar{t}b\bar{b}$ dominant background.}
\label{fig:NewCPAsym}
\end{center}
\end{figure*}

\begin{figure*} 
\vspace*{-1cm}
\begin{center}
\begin{tabular}{ccc}
\hspace*{-5mm} \epsfig{file=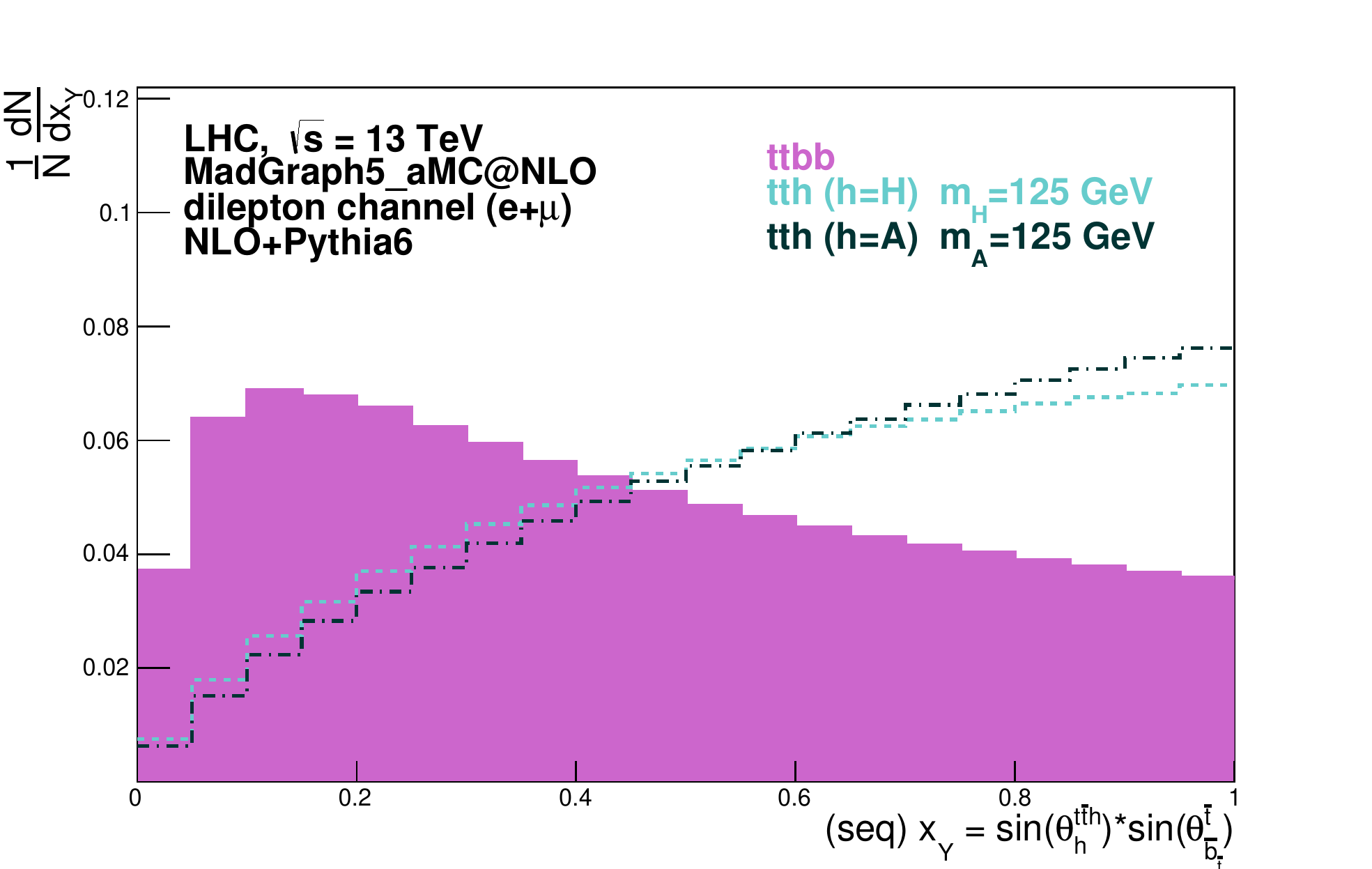,height=3.8cm,clip=}  &
 			 \epsfig{file=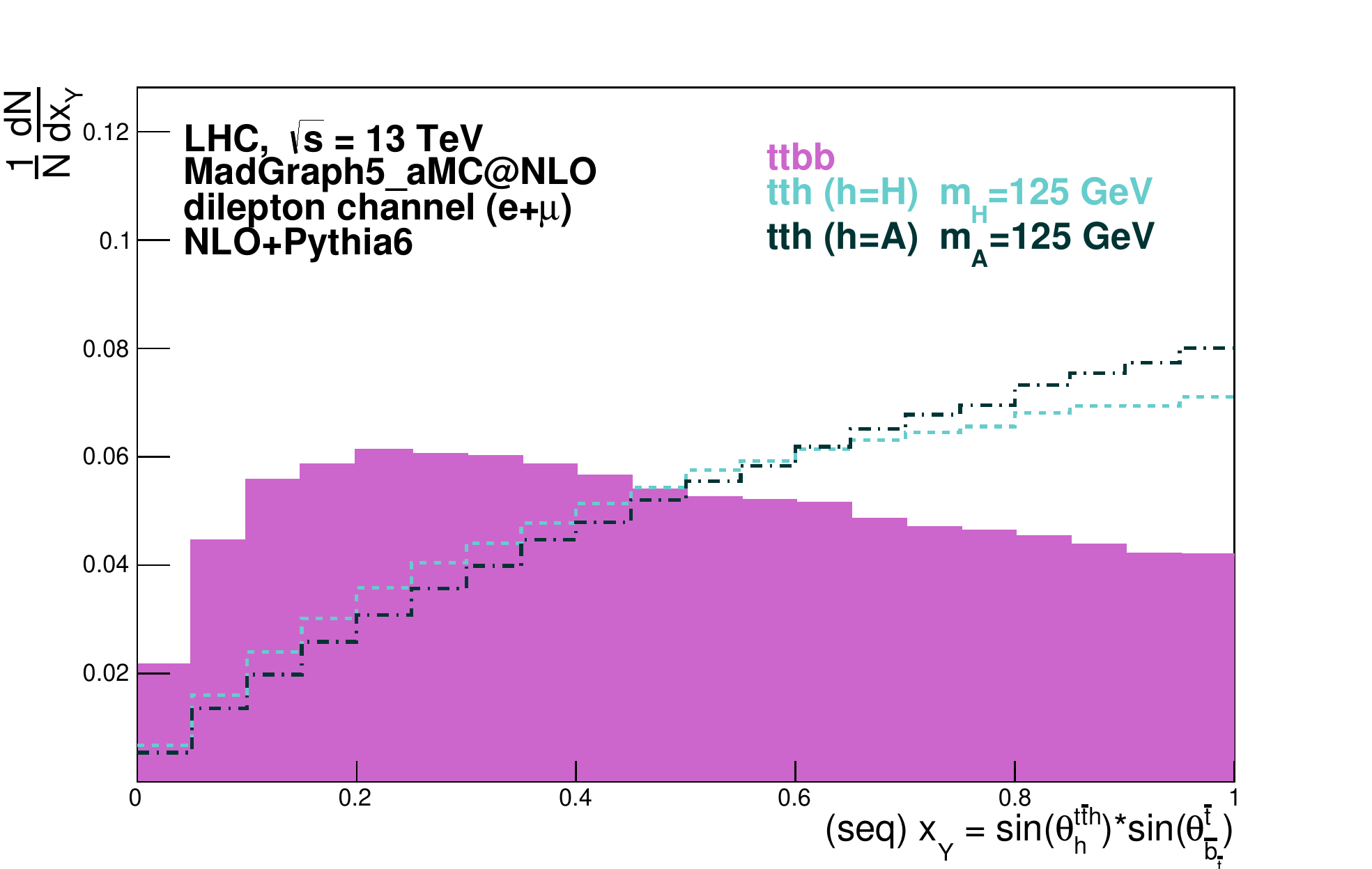,height=3.8cm,clip=} &  
			 \epsfig{file=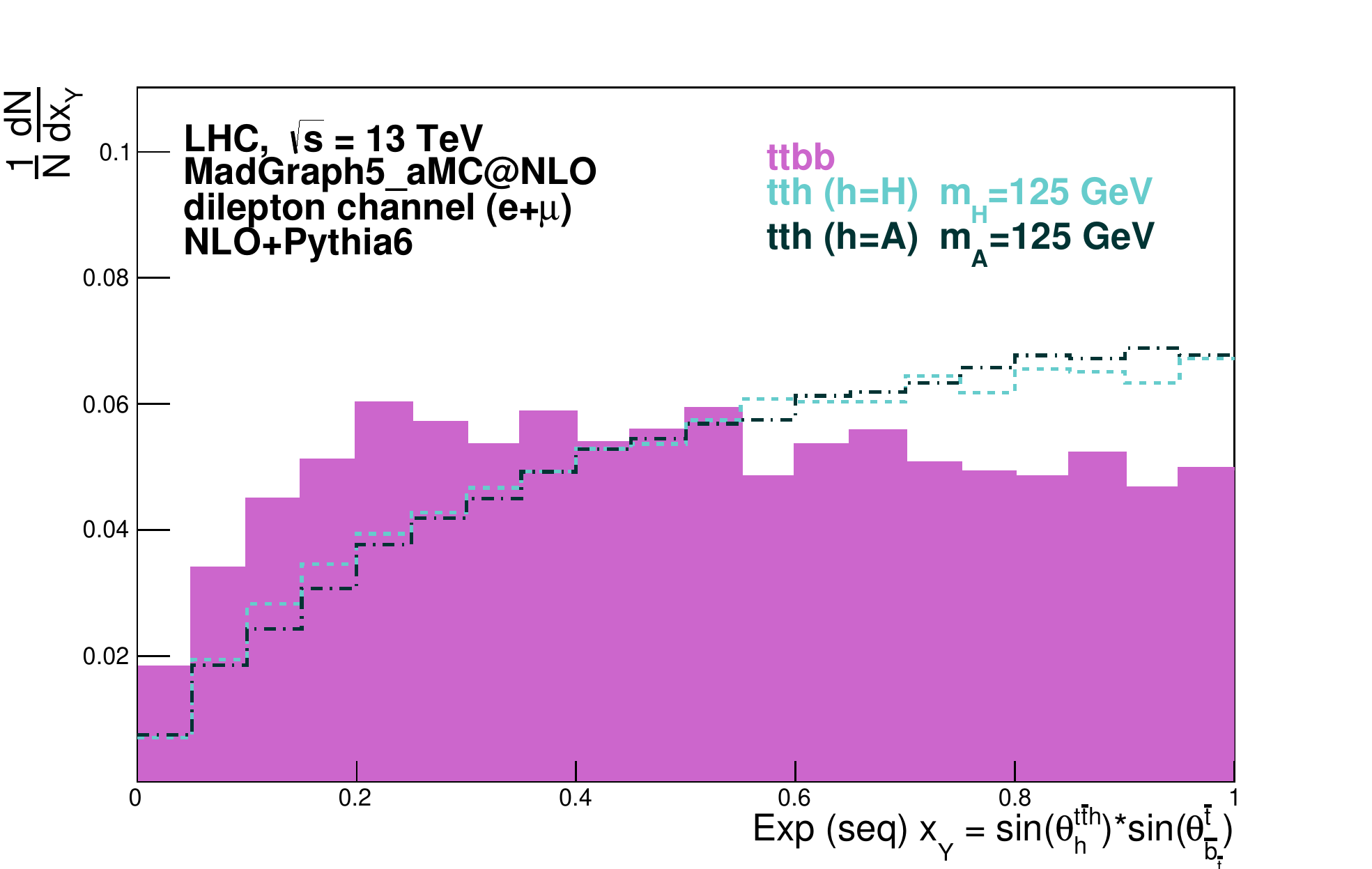,height=3.8cm,clip=}  \\[0mm]
\hspace*{-5mm} \epsfig{file=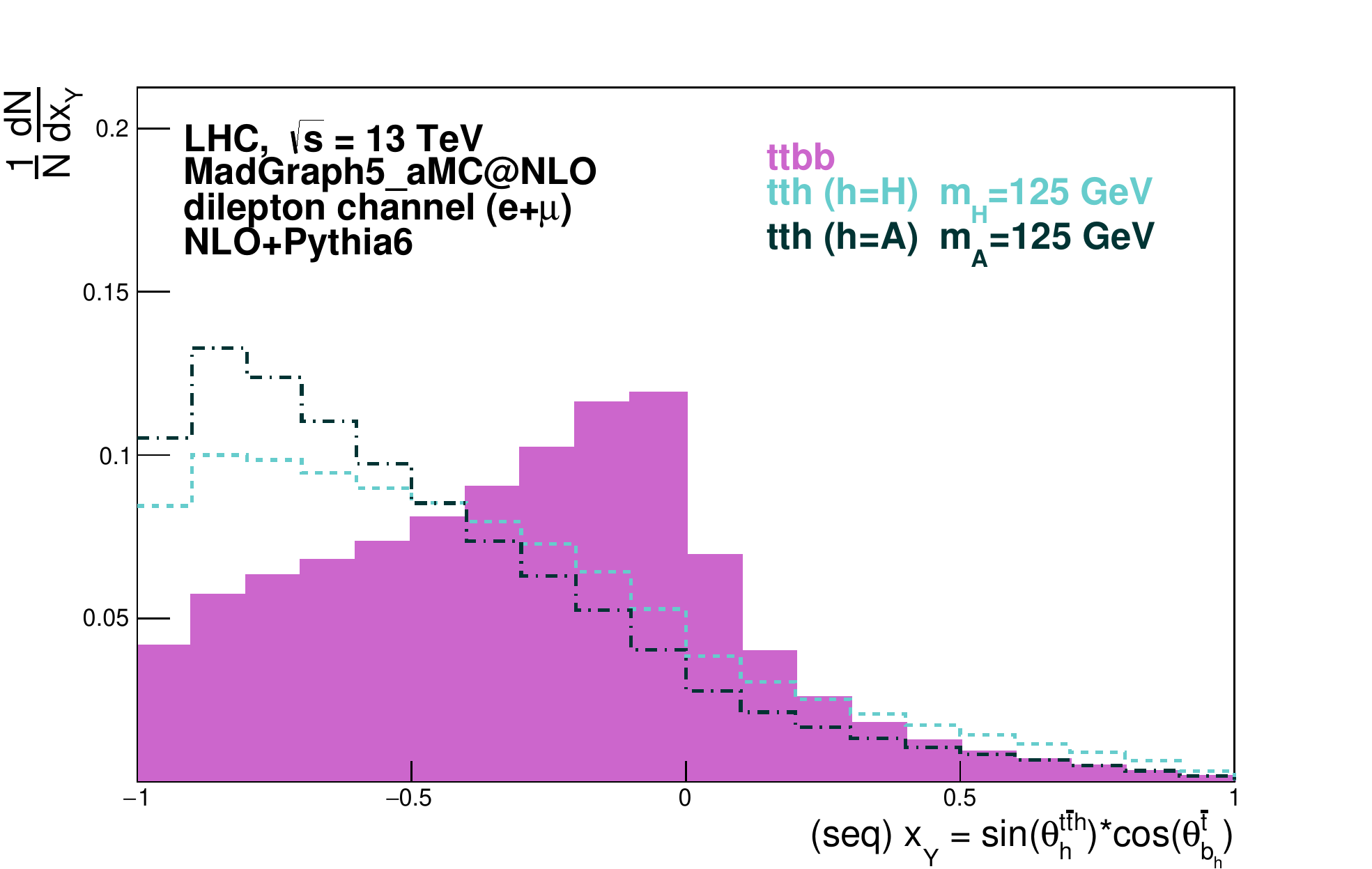,height=3.8cm,clip=} & 
			 \epsfig{file=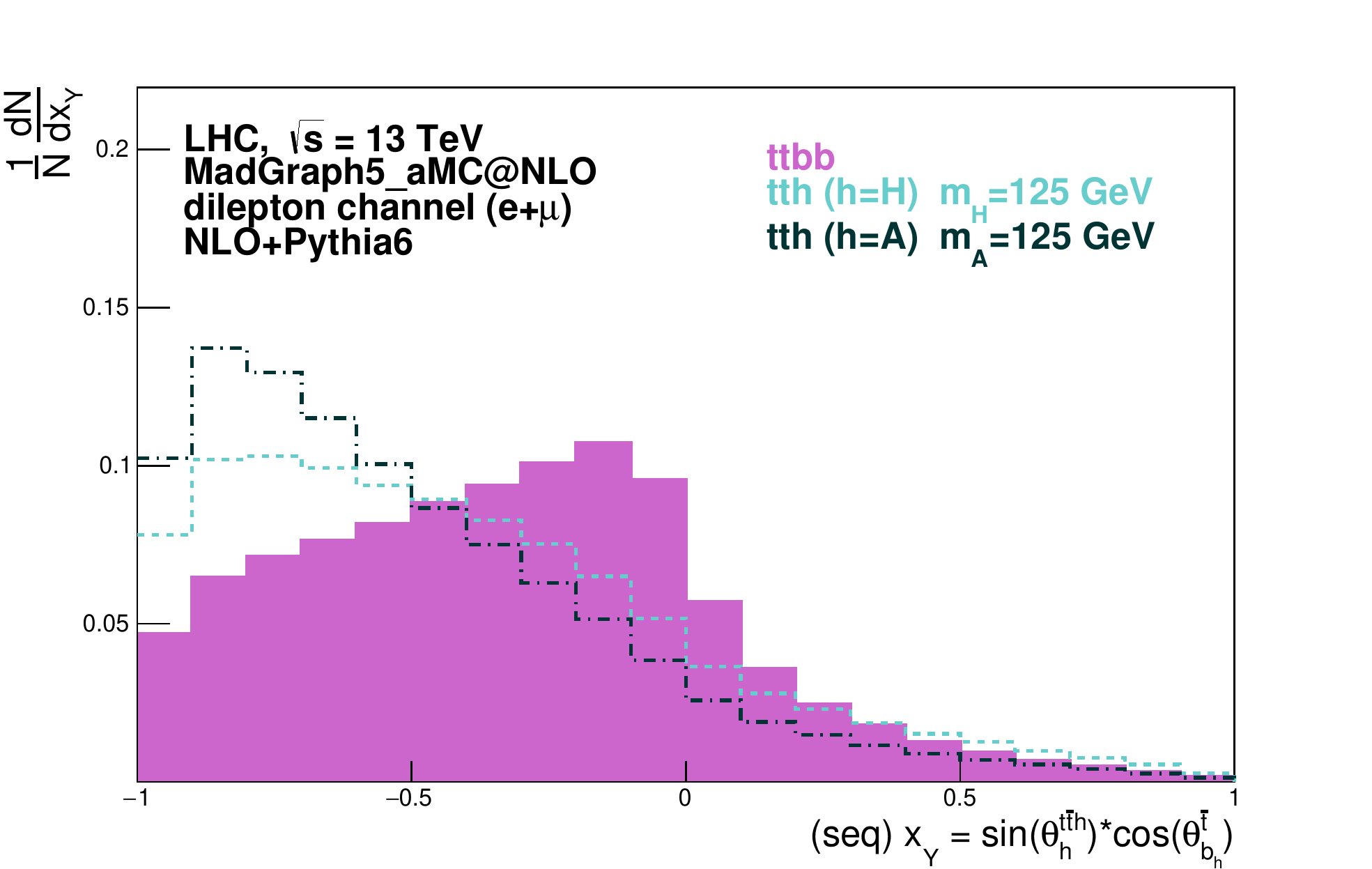,height=3.8cm,clip=} &
			 \epsfig{file=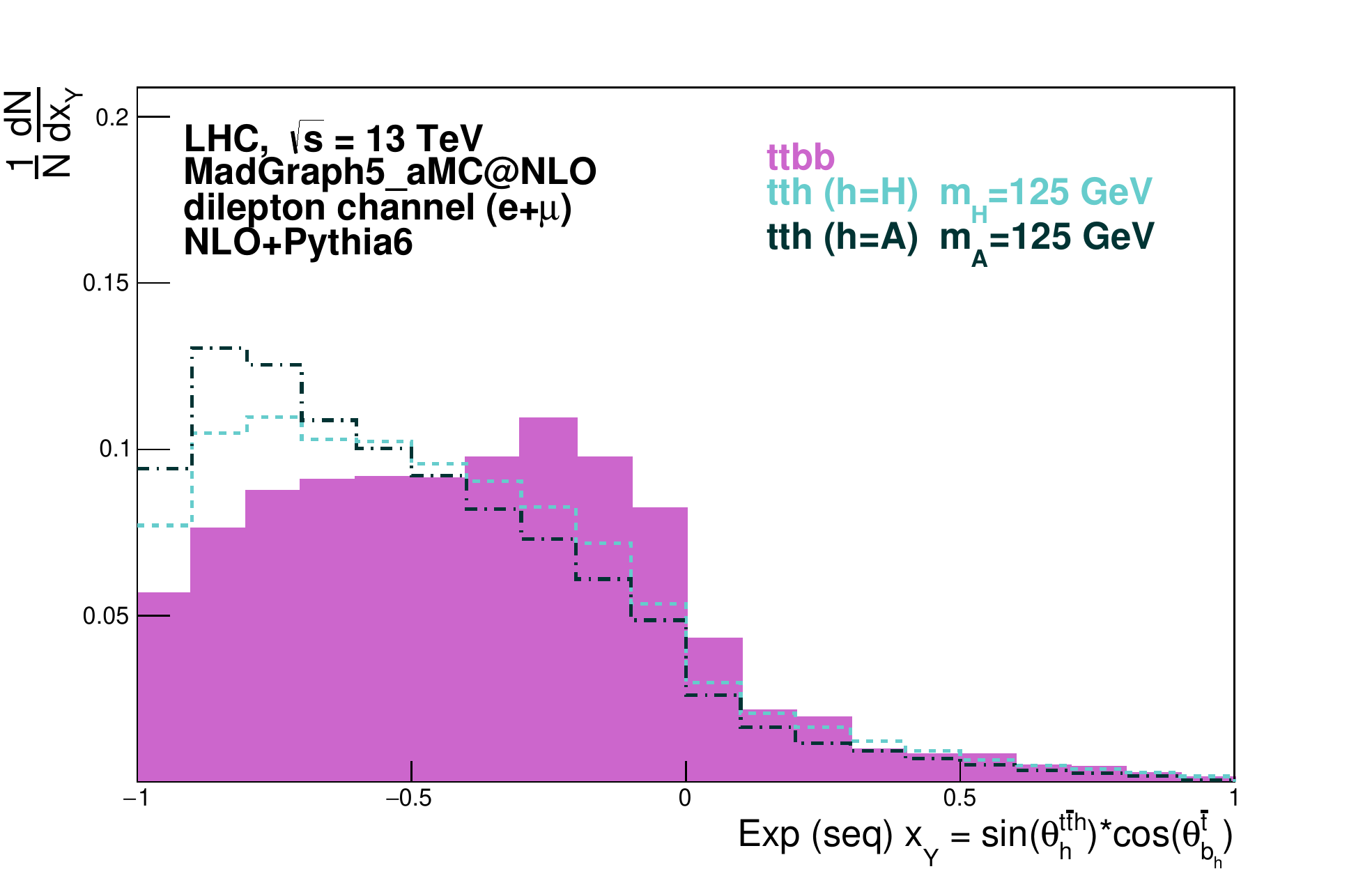,height=3.8cm,clip=}\\[-4mm]
\end{tabular}
\caption{Distributions of $x_Y$=$\sin{(\theta^{t\bar{t}h}_{h})}\sin{(\theta^{\bar{t}}_{\bar{b}_{\bar{t}}})}$ (top) and $x_Y$=$\sin{(\theta^{t\bar{t}h}_{h})}\cos{(\theta^{\bar{t}}_{b_{h}})}$ (bottom). The distributions at NLO+Shower (left), after cuts (middle) and after cuts and full kinematic reconstruction (right), are shown. The dashed line represents the $t\bar{t}h$ SM model signal ($h=H$ and $CP=+1$) and the dashed-dotted line corresponds to the pure pseudo-scalar distribution $t\bar{t}h$ ($h=A$ and $CP=-1$). The shadowed region corresponds to the NLO+Shower $t\bar{t}b\bar{b}$ dominant background. The laboratory four-momentum of $b$ quarks is boosted sequentially to the Higgs centre of mass system (see text for details). }
\label{fig:NewAng01}
\end{center}
\end{figure*}

\begin{figure*}
\begin{center}
\begin{tabular}{ccc}
\epsfig{file=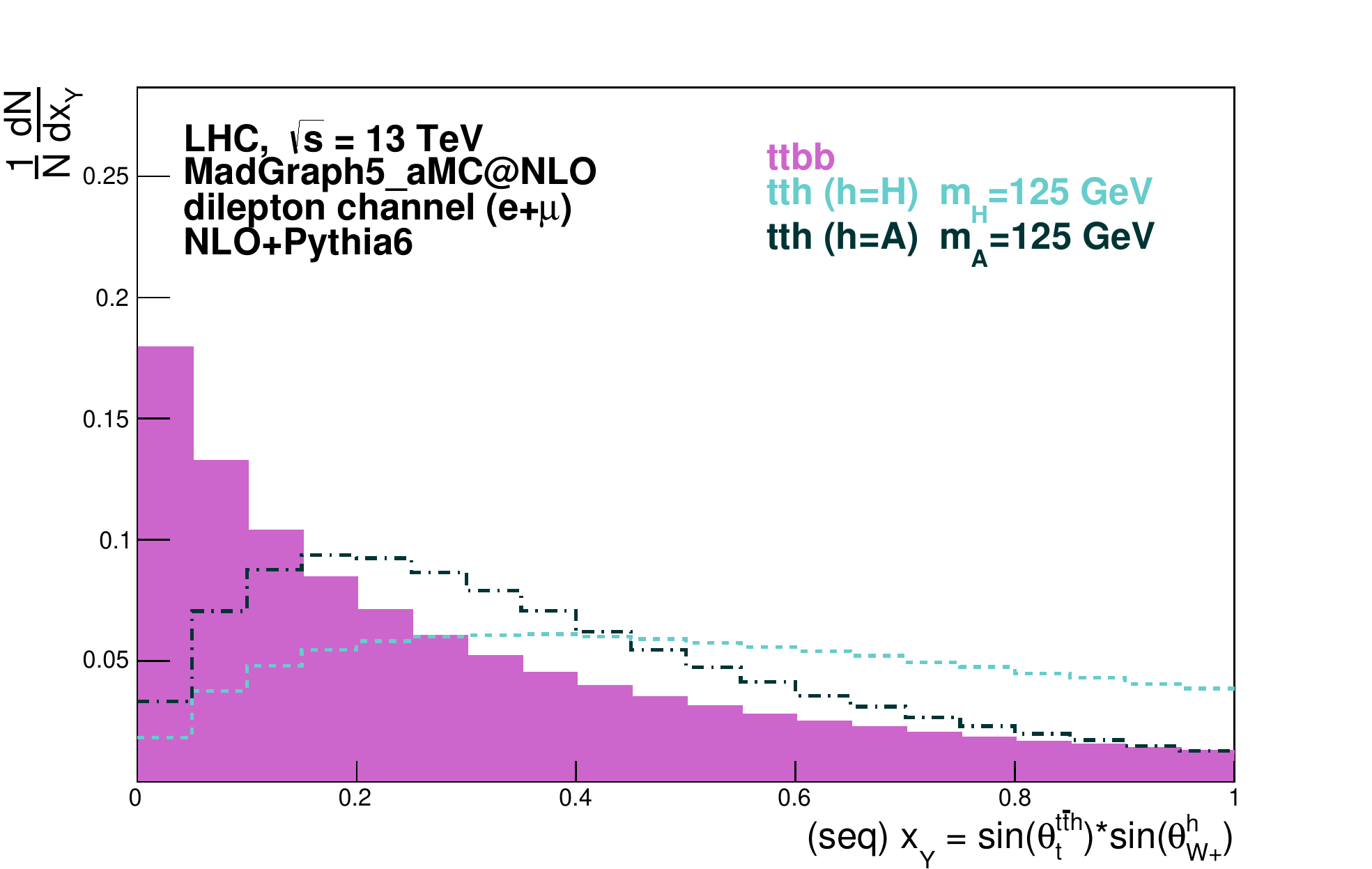,height=5.5cm,clip=} & \quad & 
\epsfig{file=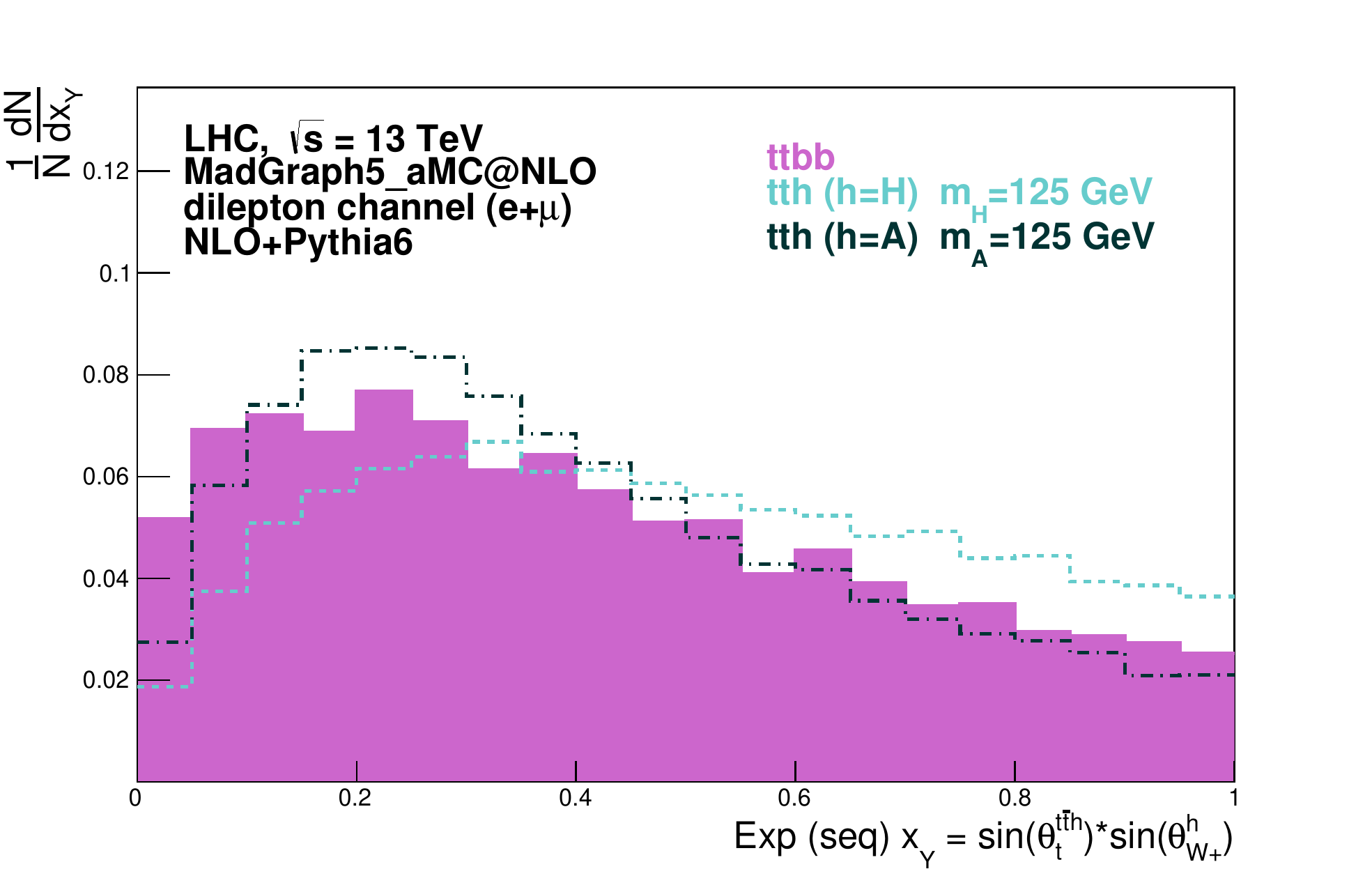,height=5.5cm,clip=}\\[-1mm]
\epsfig{file=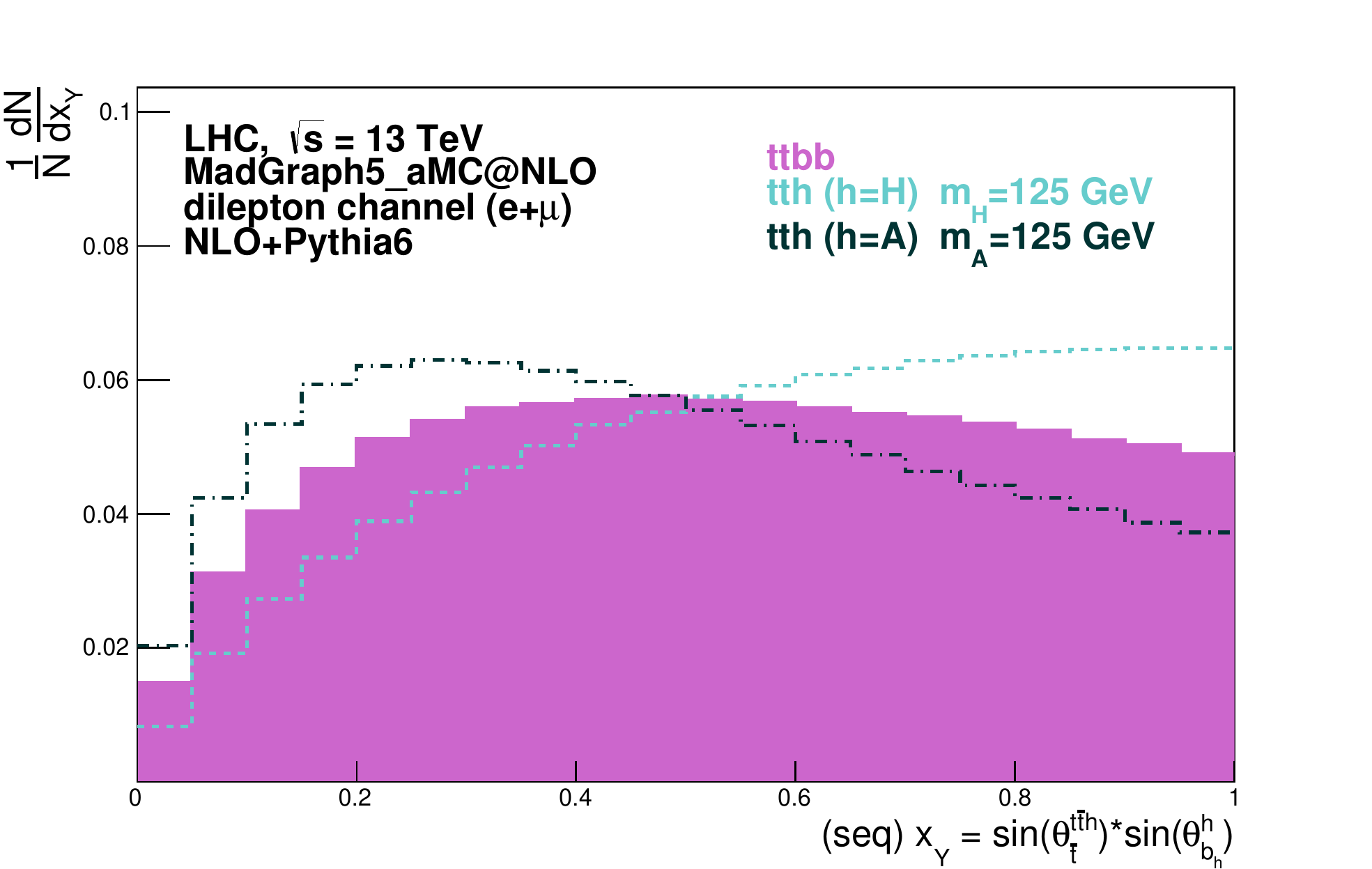,height=5.5cm,clip=} & \quad & 
\epsfig{file=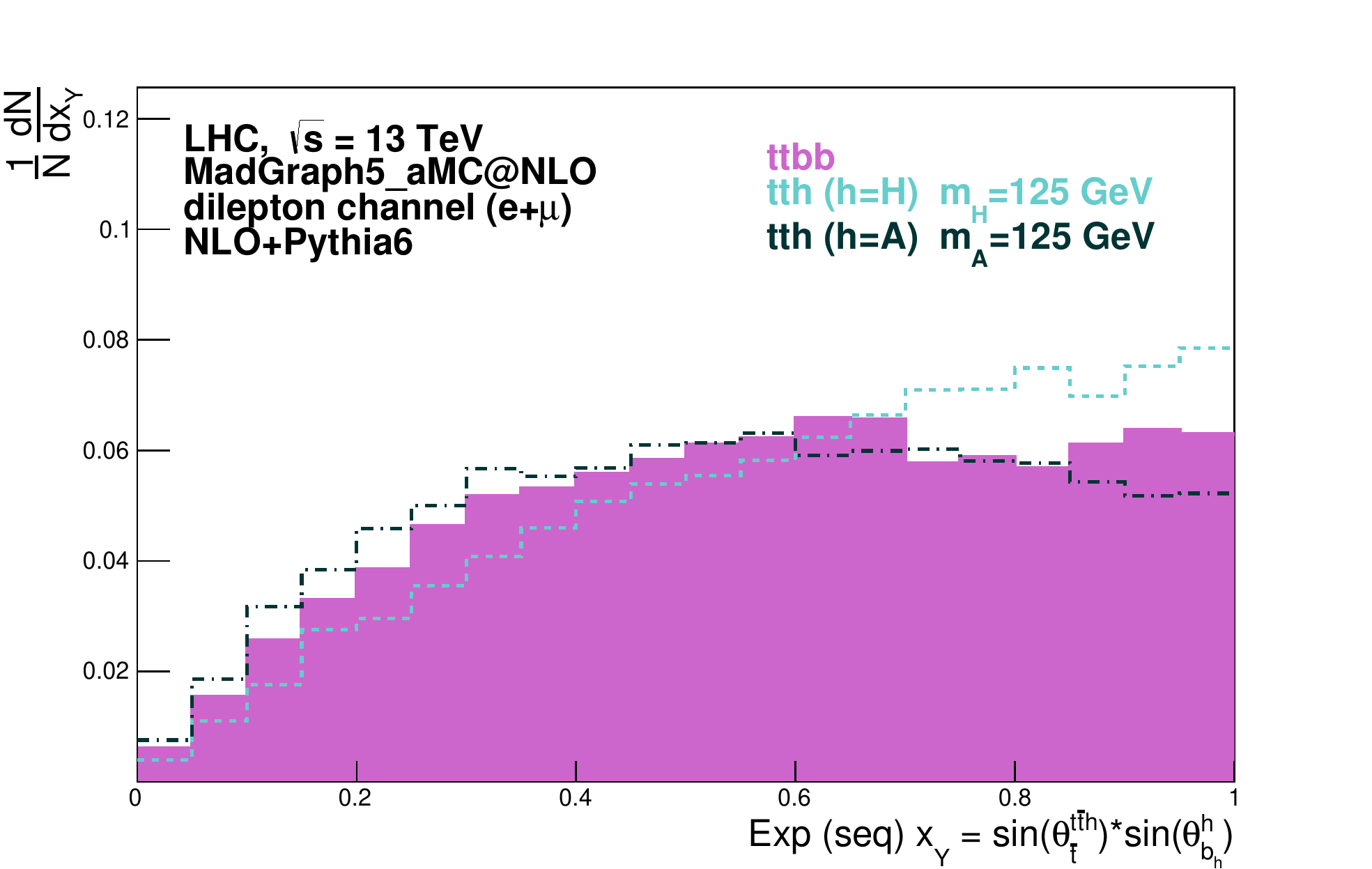,height=5.5cm,clip=} \\[-1mm]
\epsfig{file=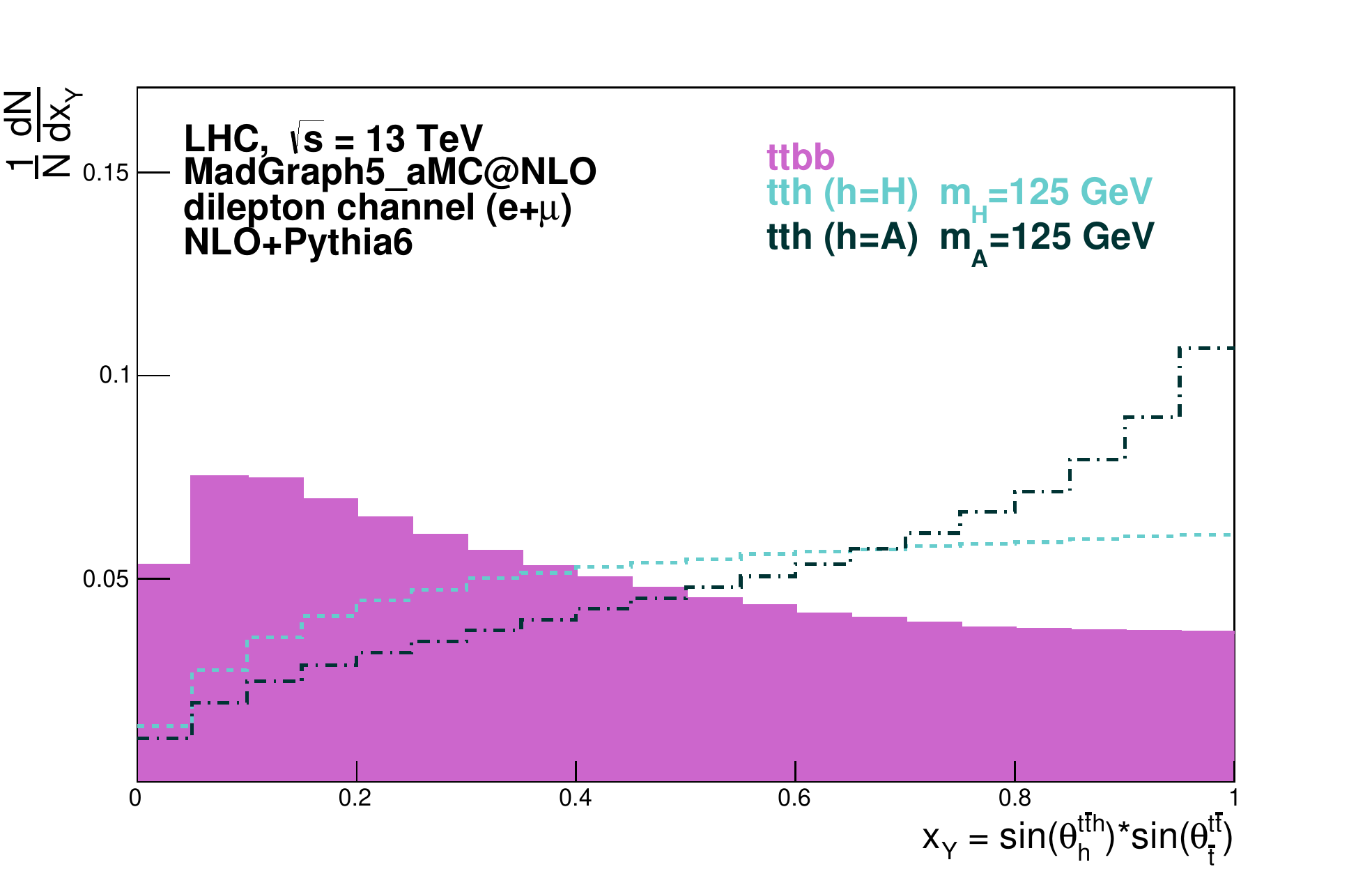,height=5.5cm,clip=} & \quad & 
\epsfig{file=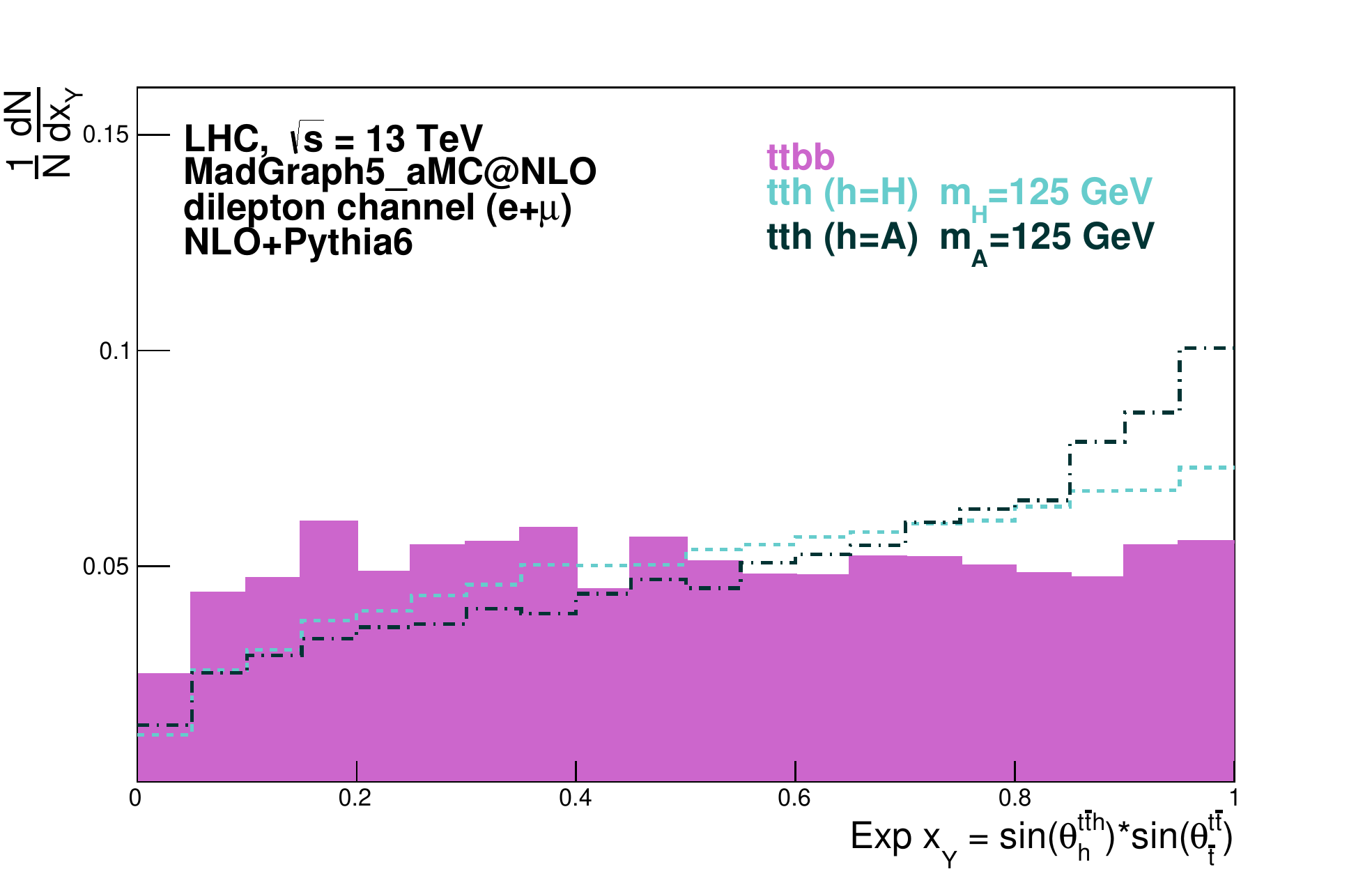,height=5.5cm,clip=} \\[-1mm]
\end{tabular}
\caption{NLO+Shower angular distributions at parton level before selection cuts (left) and after all cuts and full kinematic reconstruction (right) of: 
(top) $x_Y$=$\sin{(\theta^{t\bar{t}H}_{t})}\sin{(\theta^{H}_{W+})}$,
(middle) $x_Y$=$\sin{(\theta^{t\bar{t}H}_{\bar{t}})}\sin{(\theta^{H}_{b_H})}$
and (bottom) $x_Y$=$\sin{(\theta^{t\bar{t}H}_{H})}\sin{(\theta^{t\bar{t}}_{\bar{t}})}$. The dashed line represents the $t\bar{t}h$ SM model signal ($h=H$ and $CP=+1$), the dashed-dotted line corresponds to the pure pseudo-scalar distribution $t\bar{t}h$ ($h=A$ and $CP=-1$) and the shadowed region corresponds to the NLO+Shower $t\bar{t}b\bar{b}$ dominant background (see text for details).}
\label{fig:NewDist01}
\end{center}
\end{figure*}

The angular distributions from which each asymmetry was computed are represented in Figures~\ref{fig:ExpAng1} and Figures~\ref{fig:NewCPAsym}-\ref{fig:NewDist01}. In Table~\ref{tab:AsymGenExp} we show the NLO+Shower values of the asymmetries without any selection applied and after full kinematic reconstruction.

\begin{table}[h]
\renewcommand{\arraystretch}{1.3}
\begin{center}
  \begin{tabular}{c|ccc|cc}
    \toprule
    Asymmetries                										& \multicolumn{2}{c}{NLO+Shower}              	& & \multicolumn{2}{c}{After selection and} \\[-1mm]
              													& \multicolumn{2}{c}{(no cuts applied)}              & & \multicolumn{2}{c}{reconstruction} \\
                                                       								& $t\bar{t}H$/$t\bar{t}A$           				&   $t\bar{t}b\bar{b}$  & & $t\bar{t}H$/$t\bar{t}A$         & $t\bar{t}b\bar{b}$ \\
    \midrule 
      \hspace*{-3mm}$A^{\ell-(h)}_{FB}$    							&    	$+0.37$/$+0.41$	&    $+0.17$   		& &   	$+0.42$/$+0.39$	& $+0.24$  \\
      \hspace*{-3mm}$A^{b_4}_{FB}$    							&    	$+0.35$/$-0.10$	&    $+0.33$  		& &   	$+0.16$/$-0.17$	& $+0.12$  \\               
      \hspace*{-3mm}$A^{\bar{b}_{\bar{t}}({\bar{t}})}_{FB}$({\it seq.}~boost)	&    	$+0.28$/$+0.33$	&    $-0.17$  		& &   	$+0.25$/$+0.28$	& $+0.03$  \\
      \hspace*{-3mm}$A^{b_h({\bar{t}})}_{FB}$({\it seq.}~boost) 			&    	$-0.65$/$-0.77$	&    $-0.62$  		& &   	$-0.78$/$-0.83$	& $-0.76$  \\       
      \hspace*{-3mm}$A^{W+({h})}_{FB}$({\it seq.}~boost)				&    	$-0.03$/$-0.46$	&    $-0.60$  		& &   	$+0.17$/$-0.06$	& $-0.04$  \\
      \hspace*{-3mm}$A^{b_h(h)}_{FB}$({\it seq.}~boost)    				&    	$+0.25$/$-0.08$	&    $+0.07$  		& &   	$+0.37$/$+0.16$	& $+0.23$  \\
      \hspace*{-3mm}$A^{\bar{t}(t\bar{t})}_{FB}$    					&    	$+0.16$/$+0.37$	&    $-0.21$  		& &   	$+0.23$/$+0.31$	& $+0.01$  \\    
    \bottomrule
  \end{tabular}
\caption{{Asymmetry values for $t\bar{t}H$, $t\bar{t}A$ and $t\bar{t}b\bar{b}$ at NLO+Shower (without any cuts) and after applying the selection criteria and kinematic reconstruction, are shown.}}
\label{tab:AsymGenExp}
\end{center}
\end{table}

\section{Observables sensitive to the CP nature of the top Yukawa coupling}

In the previous sections, we identified angular observables for which the distributions of $t\bar{t}b\bar{b}$ events and signal ($t\bar{t}H$ and $t\bar{t}A$) events show important differences. For many such observables, the distributions of the $t\bar{t}H$ and $t\bar{t}A$ samples are very similar (see the plot on the right of Figure \ref{fig:ExpAng1} as an example). These observables are ideal to implement a search for (or set limits to) the total $t\bar{t}h$ production cross-section, since they have the desirable feature of being insensitive to the CP nature of the Higgs-top coupling. However, within the set of new angular observables, many result in incompatible distributions between $t\bar{t}H$ and $t\bar{t}A$ samples at reconstruction level without truth-match. This suggests that they are useful for experimentally measuring (or setting limits to) a pseudo-scalar component of the top Yukawa coupling.

Observables in $t\bar{t}h$ events with this same purpose have been previously proposed, for example, in \citep{Artoisenet:2012st, Gunion:1996xu, tth_spin}. The observables proposed in these works, for the $t\bar{t}H$ and $t\bar{t}A$ signal samples as well as for the $t\bar{t}b\bar{b}$ background, were studied in reconstructed events. For brevity, we show results for two of the most compelling observables. The authors of \citep{Boudjema:2015nda} proposed the observable $\beta_{b\bar{b}}\Delta\theta^{\ell h}(\ell^+,\ell^-)$, where $\theta^{\ell h}(\ell^+,\ell^-)$ is the angle between the $\ell^+$ and $\ell^-$ directions, projected onto the plane perpendicular to the $h$ direction in the lab frame, and $\beta$ is defined as the sign of $(\vec{p_b}-\vec{p_{\overline{b}}})\cdot(\vec{p_{\ell^-}}\times\vec{p_{\ell^+}})$ ($b$ and $\bar{b}$ result from the $t$ and $\bar{t}$ decays, respectively). The other observable is $b_4$, already introduced in the previous section, and first proposed in \citep{Gunion:1996xu}. An important remark is that $b_4$, like many other observables in the referred publications, requires the reconstruction of the $t$ and $\bar{t}$ four-momenta, only achievable through a kinematic fit such as the one used in this work. In Figure \ref{fig:NewCPAsym}, distributions are presented of 
$\beta_{b\bar{b}}\Delta\theta^{\ell h}(\ell^+,\ell^-)$ (top) and $b_4$ (bottom), for $t\bar{t}H$, $t\bar{t}A$ and $t\bar{t}b\bar{b}$ samples. They are shown at NLO+Shower without cuts (left), with cuts (middle), and at reconstruction level without truth-match, after additionally requiring at least 3 $b$-tagged jets and $\vert m_{\ell\ell}-m_Z\vert >10$~GeV (right). While it is evident that detector simulation and reconstruction degrade the discriminating power of these observables, the most dramatic effect on the distribution shapes comes from applying the acceptance cuts. After these cuts, the distributions at NLO+Shower already exhibit roughly the same behaviour as the distributions after reconstruction. Optimisation of the selection criteria is thus quite important, but stays largely outside the scope of this paper.

In figure \ref{fig:NewAng01}, distributions of $\sin{(\theta^{t\bar{t}h}_{h})}\sin{(\theta^{\bar{t}}_{\bar{b}_{\bar{t}}})}$ (top) and $\sin{(\theta^{t\bar{t}h}_{h})}\cos{(\theta^{\bar{t}}_{b_{h}})}$ (bottom) are shown. These are among the investigated angular observables for which the $t\bar{t}b\bar{b}$ background sample was least compatible with both $t\bar{t}H$ and $t\bar{t}A$ samples. The distributions are represented at NLO+Shower without cuts (left), after selection cuts (middle), and after full kinematic reconstruction and the additional requirements of $\vert m_{\ell\ell}-m_Z\vert >10$~GeV and at least 3 $b$-tagged jets (right). The dashed line represents the $t\bar{t}H$ distribution and the dashed-dotted line corresponds to $t\bar{t}A$. The shadowed region corresponds to the $t\bar{t}b\bar{b}$ dominant background.

Figure \ref{fig:NewDist01} shows distributions of three angular observables among the ones for which the $t\bar{t}H$ and $t\bar{t}A$ samples were least compatible at the reconstruction level without truth-match. They are represented at NLO+shower without cuts (left) and at reconstruction level without truth-match, after the previously mentioned cuts on $b$-tag multiplicity and $m_{\ell\ell}$ (right). Distributions of $t\bar{t}b\bar{b}$ events are also included for completeness. The discriminating performance of these observables is comparable to that of the ones proposed in the literature. Computing the angular observables also requires full reconstruction of $t$ and $\bar{t}$. Again, applying the acceptance cuts, detector simulation and kinematic reconstruction visibly degrades the discrimination between $t\bar{t}H$ and $t\bar{t}A$ samples.

\section{Analysis and Results}

In order to estimate the experimental sensitivity of an analysis employing the observables under study, further selection criteria was applied, as mentioned previously. Depletion of the $Z$+jets background is accomplished by selecting events with a dilepton invariant mass $m_{\ell\ell}$ such that $|m_{\ell^+\ell^-}-m_Z|>10$~GeV. This selection was applied in all dilepton flavour categories ($ee$, $\mu\mu$ and $e\mu$). Most backgrounds, notably $t\bar{t}$+jets, are then mitigated by selecting events with at least 3 $b$-tagged jets.

Table~\ref{tab:cutflow} shows the expected effective cross-sections in $fb$, at several levels of the event selection, for dileptonic signal and SM backgrounds. The $t\bar{t}A$ pseudo-scalar signal was scaled to the $t\bar{t}H$ scalar cross-section for comparison purposes. 

\begin{table}[h]
\renewcommand{\arraystretch}{1.3}
\begin{center}
  \begin{tabular}{l|c|c|c|c|c}
    \toprule
                    						&   ~$N_{jets} \ge$~4~     	   &   Kinematic   	&   $m_Z$   	&   $N_b$  	&     $N_b$    \\[-1mm]
                    						&   ~$N_{lep} \ge$~2~    	   &    Fit      		&      cut      	&   $\ge$ 3        &     $\ge$ 4        \\[-1mm]
    \midrule  
     $t\bar{t}$+$c\bar{c},t\bar{t}$+lf    		&    	     2160    	           &         1300          	&           1110 	&   	4.78     	&   0.06  \\         
     $t\bar{t}$+$b\bar{b}$    			&    	      87.1    	           &          51.9          	&            44.5 	&   	2.91     	&   0.27  \\     
     $t\bar{t}$+$V(V$=$Z,W)$    			&    	        7.9    	           &            4.5          	&             3.9 	&   	0.09     	&   0.01  \\ 
     Single $t$    						&    	         54      	           &             26          	&              23 	&      0.12     	&   0.00  \\
     $V$+jets $(V$=$W,Z)$    			&    	     2700      	           &         1200         	&            200	&      0.00     	&   0.00  \\              
     $V$+$b\bar{b}(V$=$W,Z)$    		&    	       570      	           &           280     	&              20	&      0.00    	&   0.00  \\              
     Diboson    						&             130      	           &             53         	&              14	&      0.00     	&   0.00  \\         
    \midrule  
     Total back.    					&    	     5700    	           &         2900          	&           1410 	&   	7.90     	&  0.34  \\             
    \midrule  
     $t\bar{t}H$    					&            4.04      	           &          2.49         	&           2.15	&      0.26   	&  0.033  \\      
     $t\bar{t}A$    					&            4.43      	       	   &          2.69         	&           2.36	&      0.31   	&  0.041  \\     
    \bottomrule
  \end{tabular}
\caption{{Expected cross-sections (in $fb$) as a function of selection cuts, at 13~TeV, for dileptonic signal and background events at the LHC.}}
\label{tab:cutflow}
\end{center}
\end{table}

\begin{figure*}
\begin{center}
\begin{tabular}{ccc}
\epsfig{file=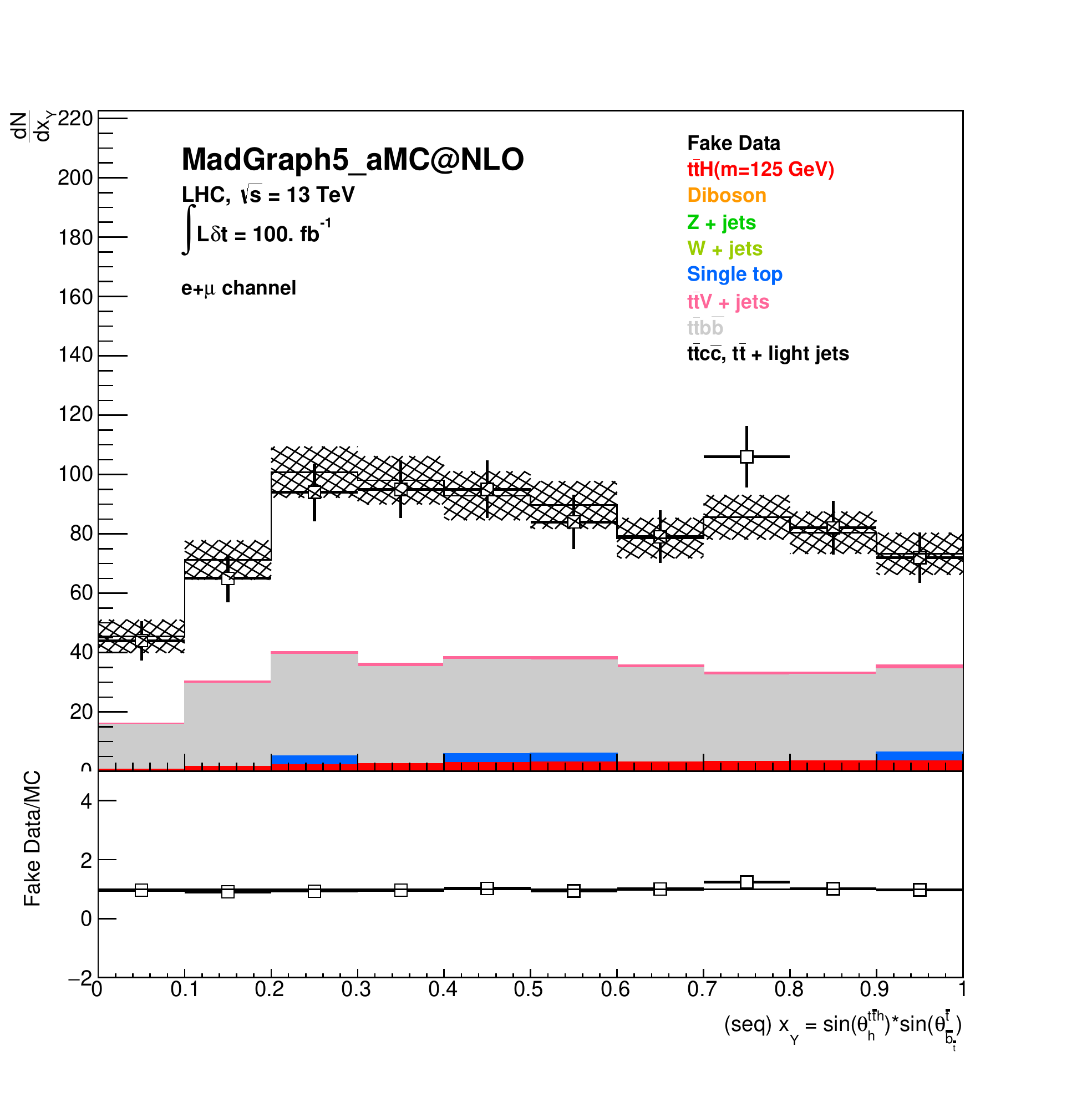,height=7.7cm,clip=} & \quad & 
\epsfig{file=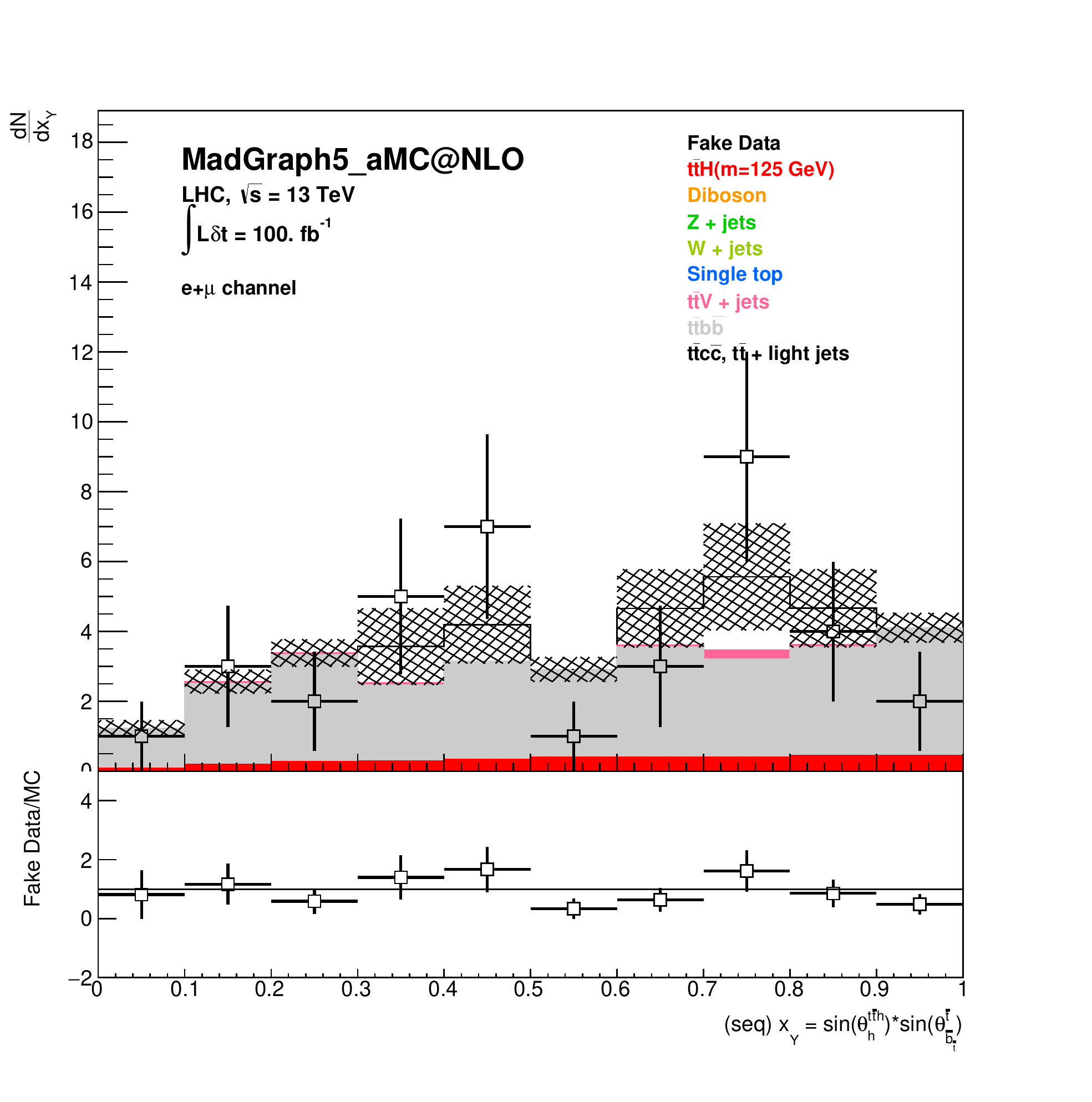,height=7.7cm,clip=}\\[-4mm]
\epsfig{file=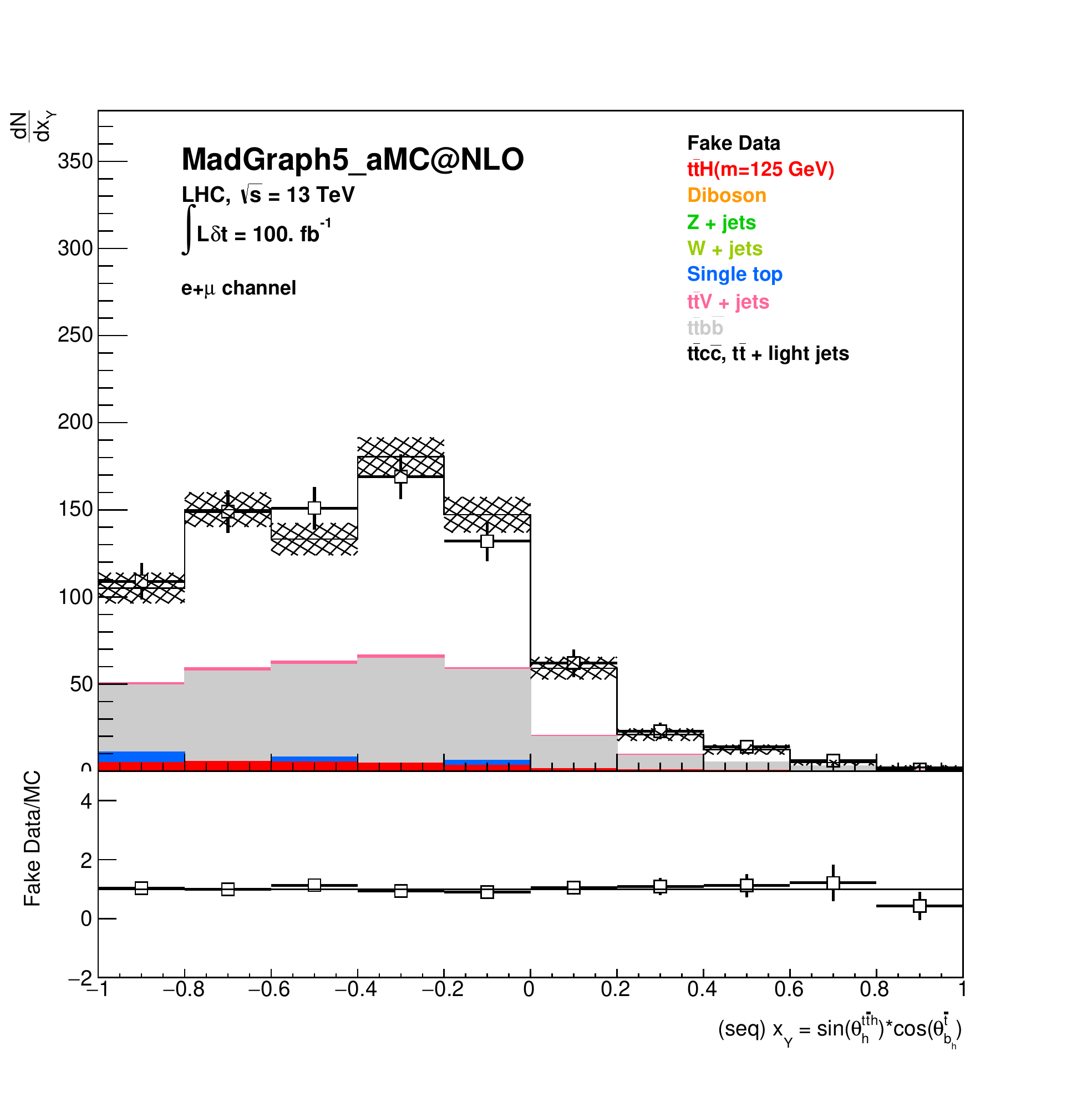,height=7.7cm,clip=} & \quad & 
\epsfig{file=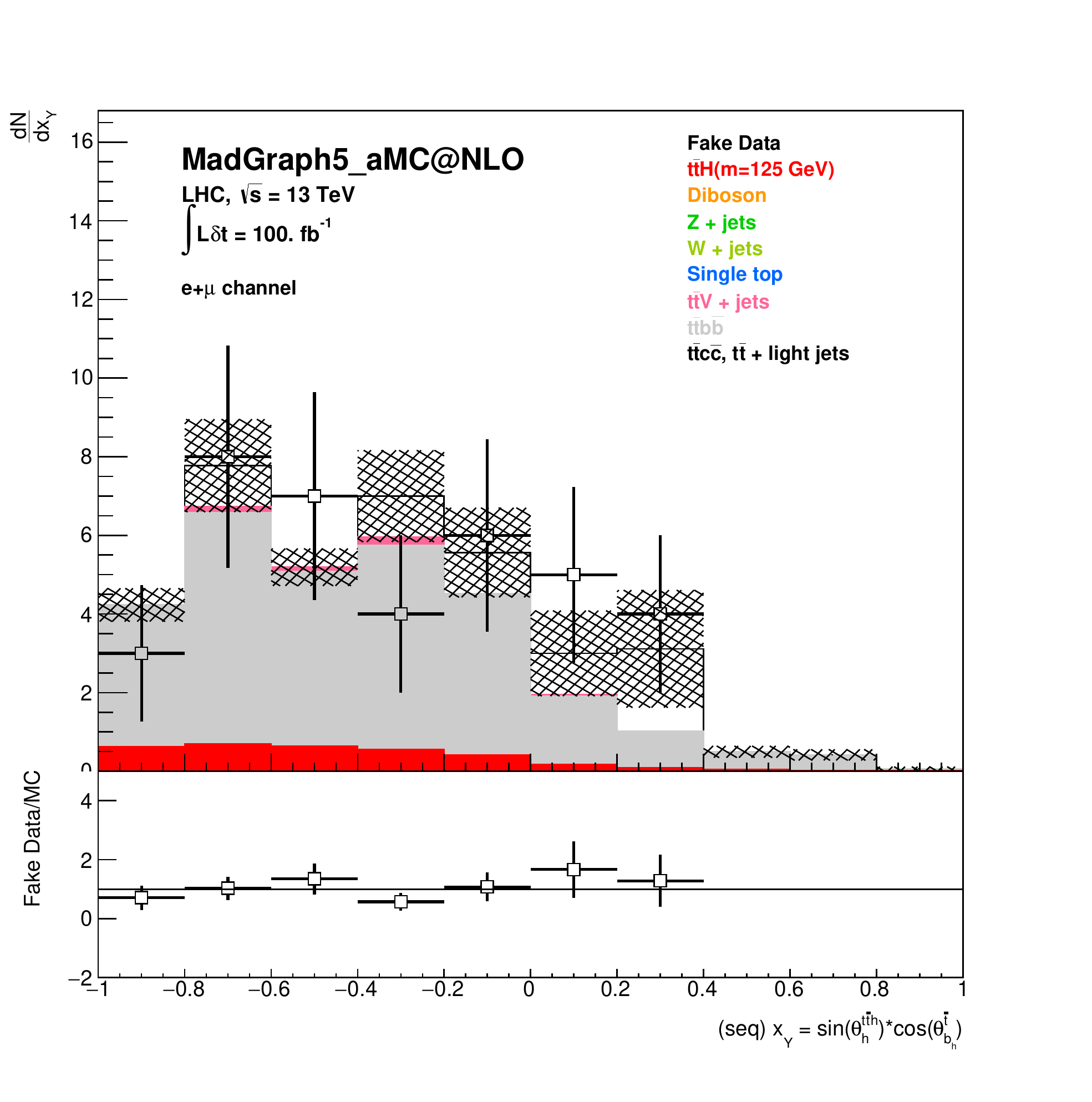,height=7.7cm,clip=} \\[-4mm]
\epsfig{file=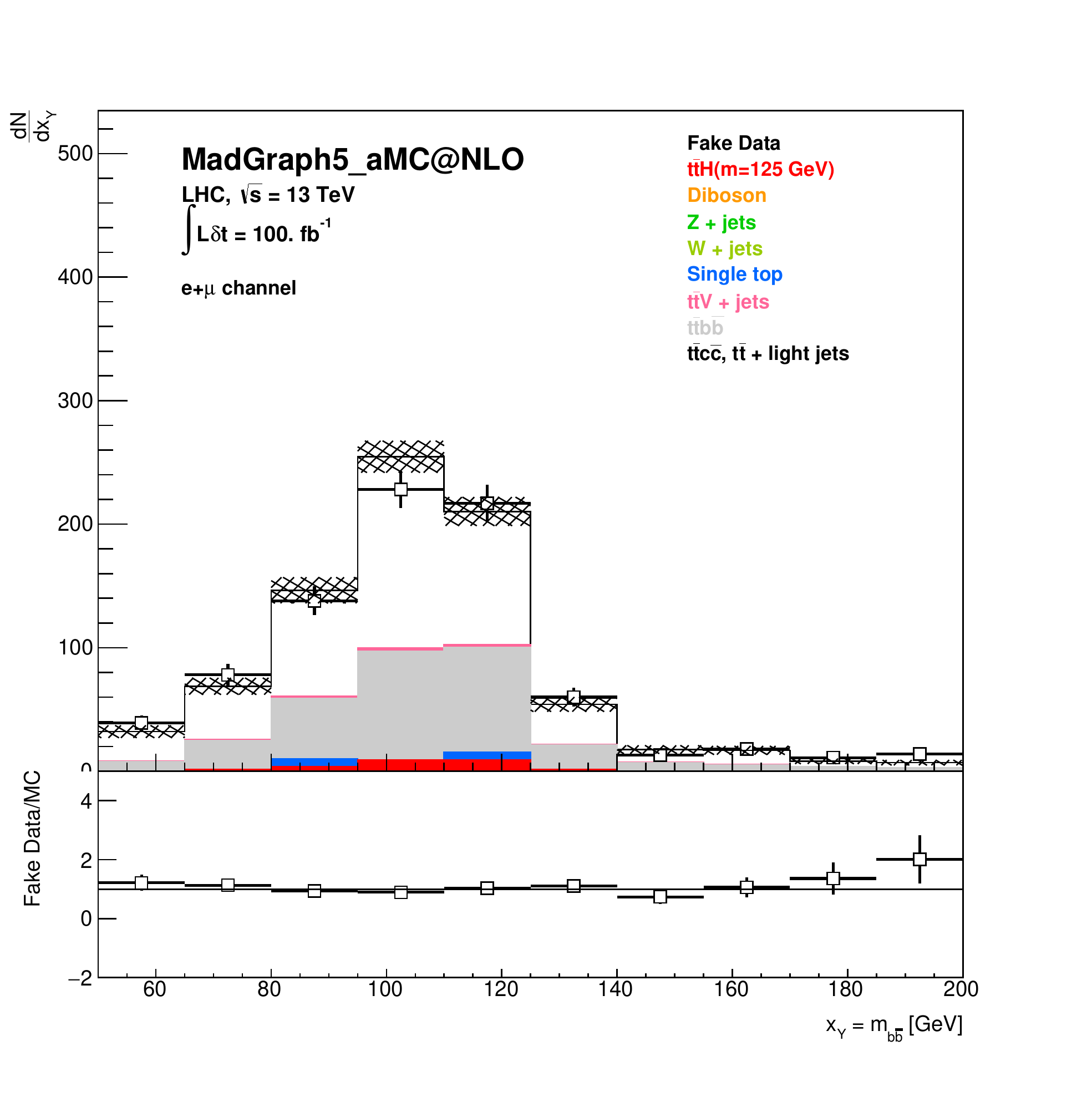,height=7.7cm,clip=} & \quad & 
\epsfig{file=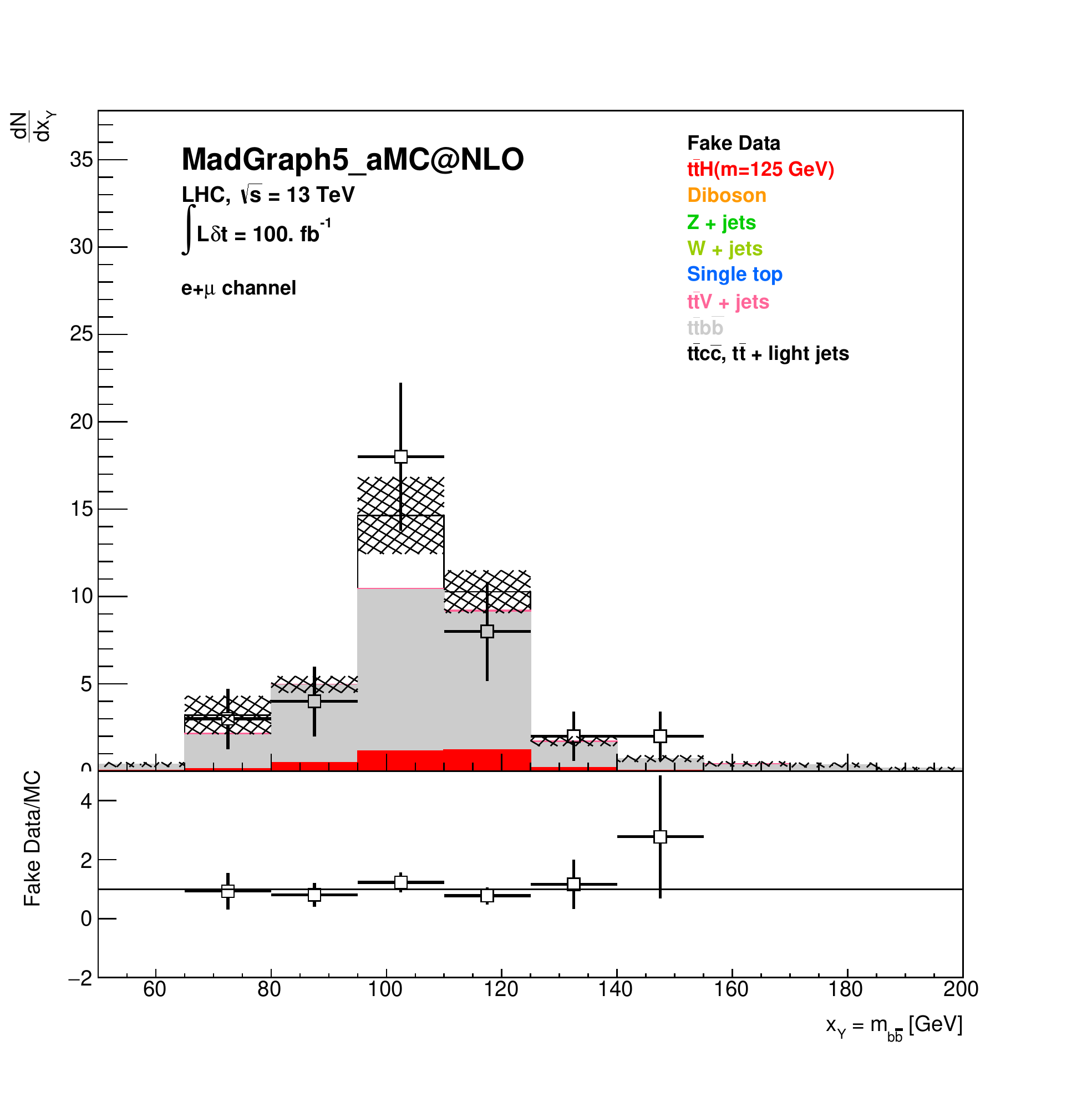,height=7.7cm,clip=} \\[-4mm]
\end{tabular}
\caption{Distributions of $x_Y$=$\sin{(\theta^{t\bar{t}H}_{H})}\sin{(\theta^{\bar{t}}_{\bar{b_{\bar{t}}}})}$ (top), 
$x_Y$=$\sin{(\theta^{t\bar{t}H}_{H})}\cos{(\theta^{\bar{t}}_{b_{H}})}$ (middle) and 
$x_Y$=$m_{b\bar{b}}$ (bottom) after final selection at 13 TeV for 100~fb$^{-1}$. The distributions on the left (right) corresponds to events with at least 3 (4) jets from the hadronization of $b$-quarks.}
\label{fig:NewAnal01}
\end{center}
\end{figure*}

In Figure~\ref{fig:NewAnal01}, the expected number of events from the different SM processes are shown, including the Higgs signal, for a luminosity of 100~fb$^{-1}$ at the LHC, for events with at least 3 $b$-jets (left) and at least 4 $b$-jets (right). As expected, the composition of backgrounds changes quite significantly after event selection. 

The fake data points correspond to one particular pseudo-experiment randomly created from the expected Standard Model $t\bar{t}H$ signal and background distributions. Its purpose is only to guide the reader through the total number of expected events and related statistical uncertainties, after event selection and full reconstruction. 

Several kinematic properties of the events, including the new angular distributions introduced in this paper, were tested with several multivariate methods. A boosted decision tree with gradient boost (BDTG) has the best performance among the methods investigated. Its output was used to test the analysis sensitivity to probe the scalar versus pseudo-scalar component of the top-Higgs couplings, as a function of $\cos{\alpha}$. From the long set of variables tried, the 15 best ranked by the multivariate method, after reconstruction, were: 
the $b_4$ and Higgs mass ($m_{b\bar{b}}$); 
the angular distributions with direct boost i.e., $\cos(\theta^{\bar{t}h}_{h})\cos(\theta^{h}_{\ell^-})$, $\sin(\theta^{t\bar{t}h}_{h})\sin(\theta^{t\bar{t}}_{\bar{t}})$ and the variables with sequential boost $\sin(\theta^{t\bar{t}h}_{\bar{t}})\sin(\theta^{h}_{b_{h}}) (seq.)$, $\sin(\theta^{t\bar{t}h}_{h})\cos(\theta^{\bar{t}}_{b_{h}}) (seq.)$, $\sin(\theta^{t\bar{t}h}_{h})\sin(\theta^{\bar{t}}_{\bar{b}_{\bar{t}}}) (seq.)$, $\sin(\theta^{t\bar{t}h}_{t})\sin(\theta^{h}_{W^+}) (seq.)$; 
the $\Delta\eta$ between the jets with maximum $\Delta\eta$ ($\Delta\eta_{jj}^{\max\Delta\eta}$) and the invariant mass of the two $b$-tagged jets with lowest $\Delta R$ ($m_{bb}^{\min\Delta R}$); 
the $\Delta R$ between the Higgs candidate and the closest ($\Delta R_{hl}^{\min\Delta R}$) and farthest ($\Delta R_{hl}^{\max\Delta R}$) leptons; 
the $\Delta R$ between the $b$-tagged jets with highest $p_T$ ($\Delta R_{bb}^{\max p_{T}}$) and the invariant mass of the two jets with closest value to the Higgs mass ($m_{jj}^{\textrm{\small closest to 125~GeV}}$); 
the jets aplanarity.

\begin{figure*}
\begin{center}
\begin{tabular}{ccc}
\epsfig{file=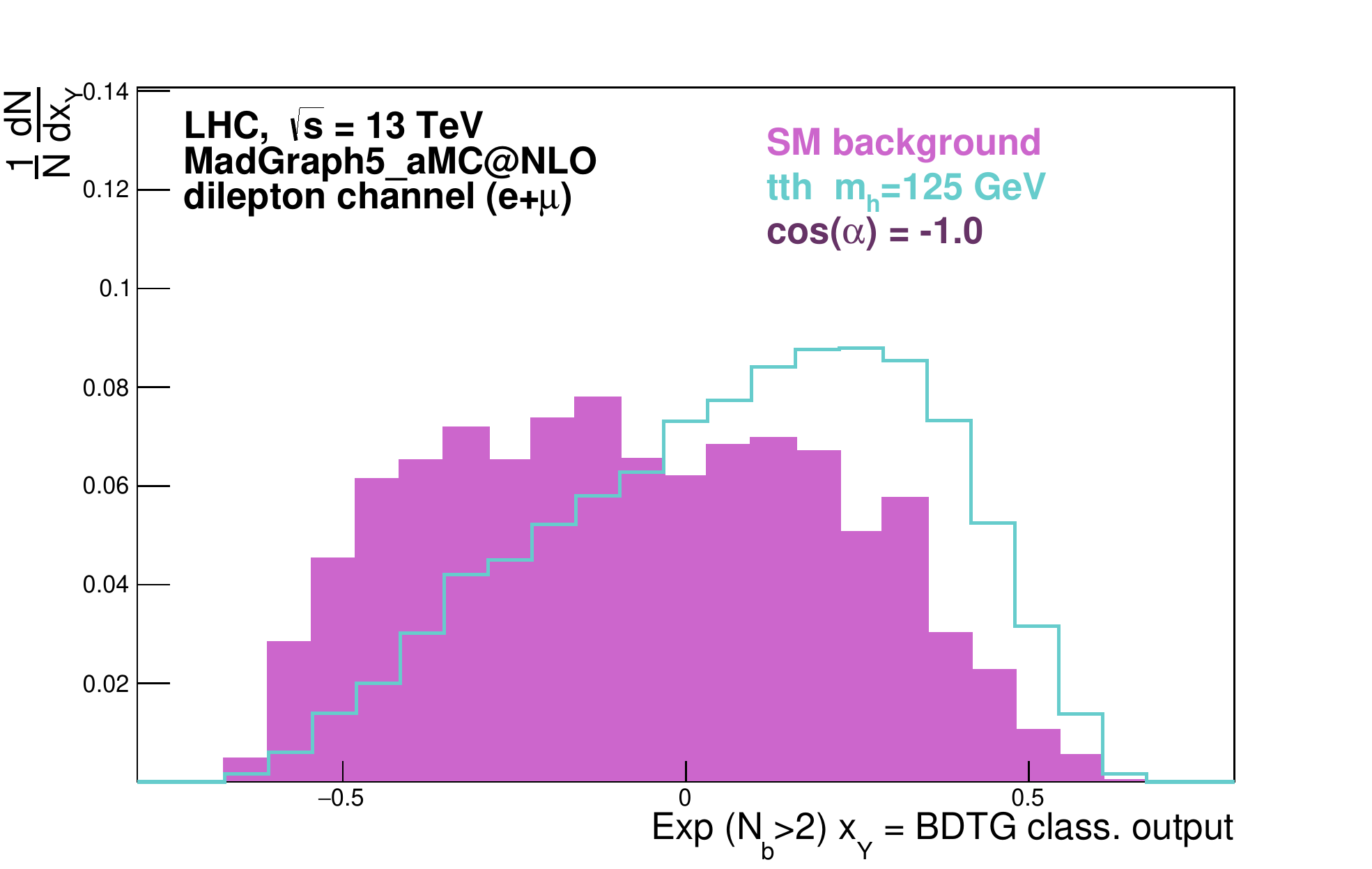,height=5.0cm,clip=} & \quad & 
\epsfig{file=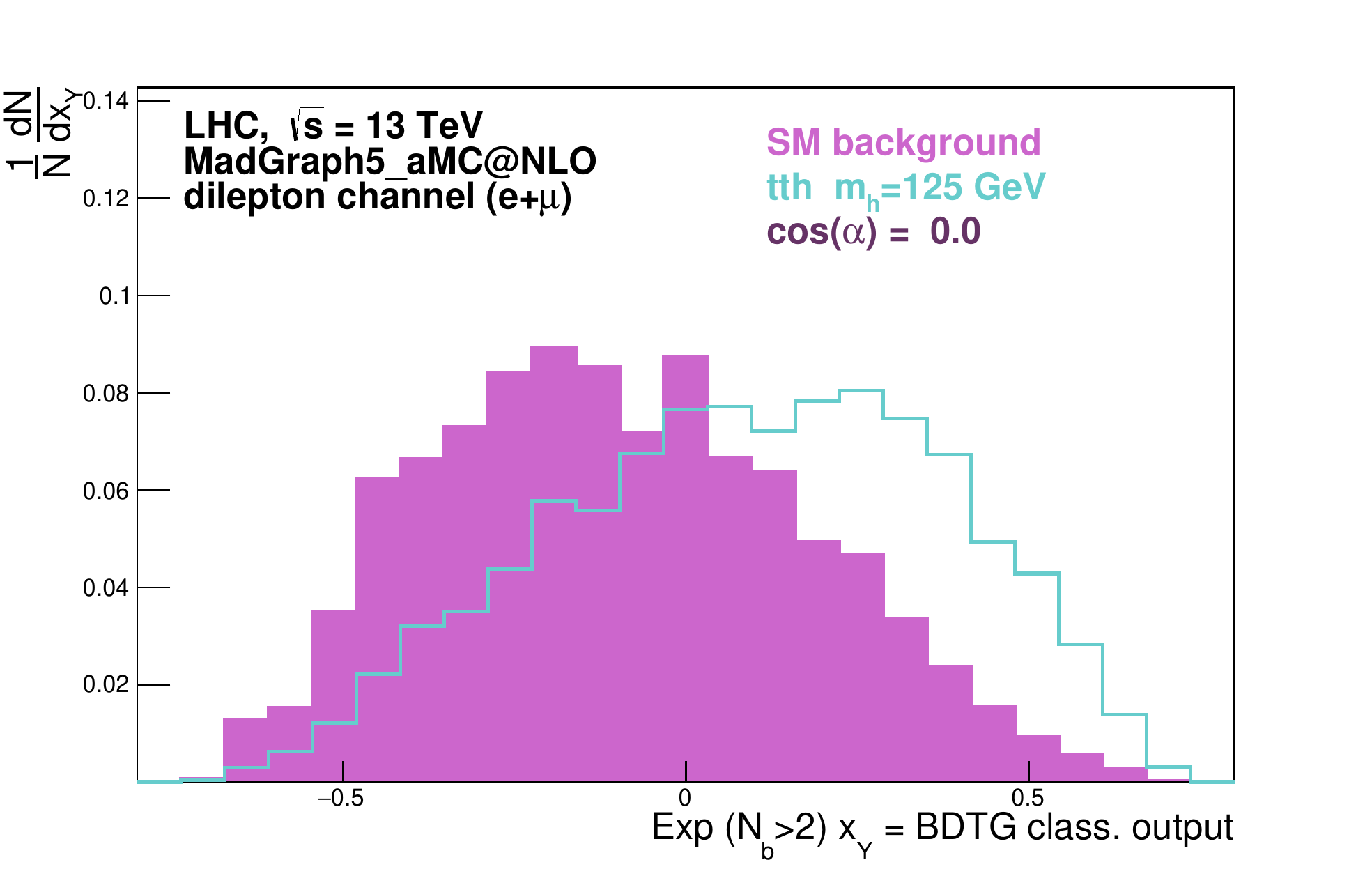,height=5.0cm,clip=}\\[-.5mm]
\epsfig{file=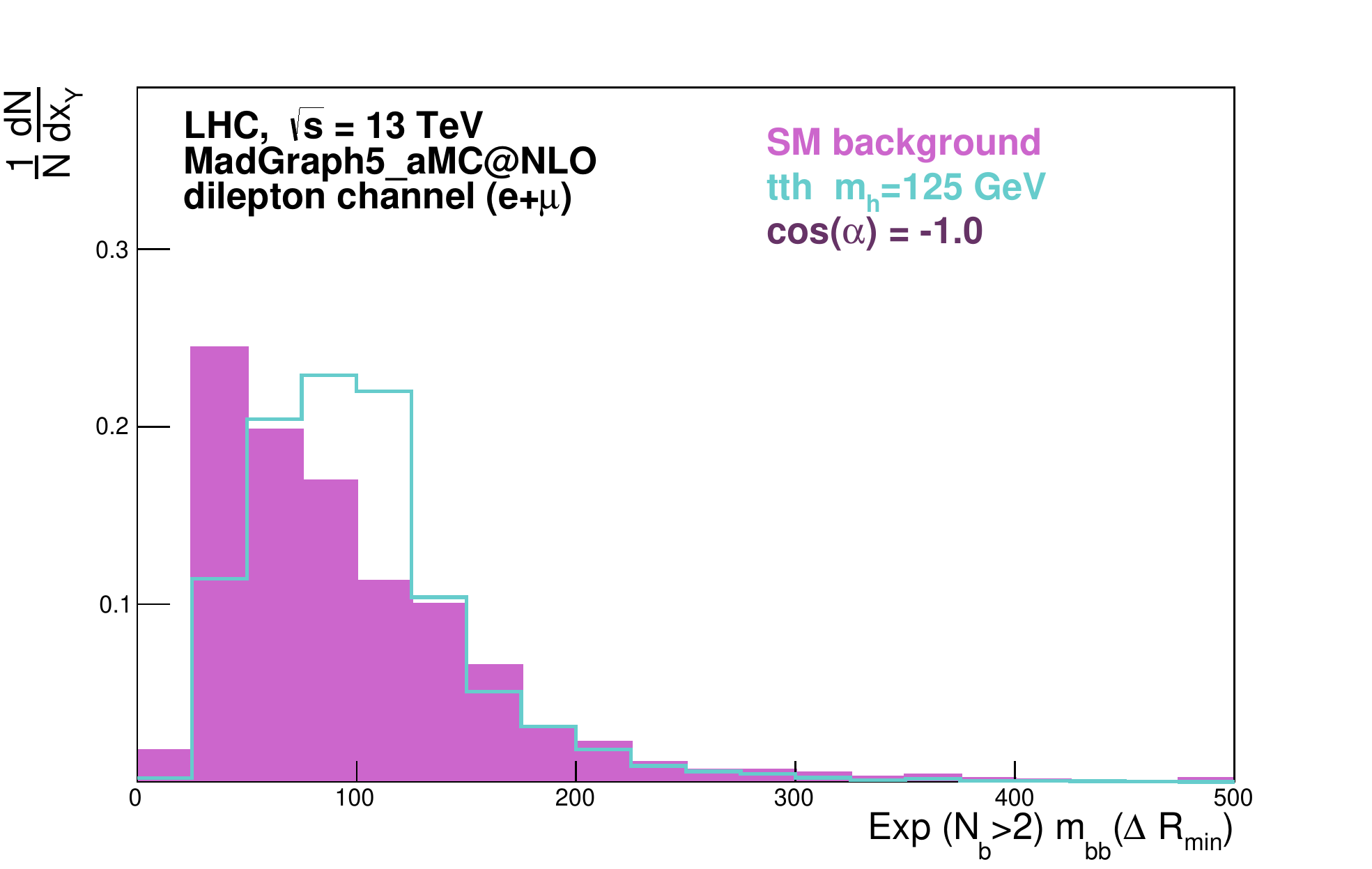,height=5.0cm,clip=} & \quad & 
\epsfig{file=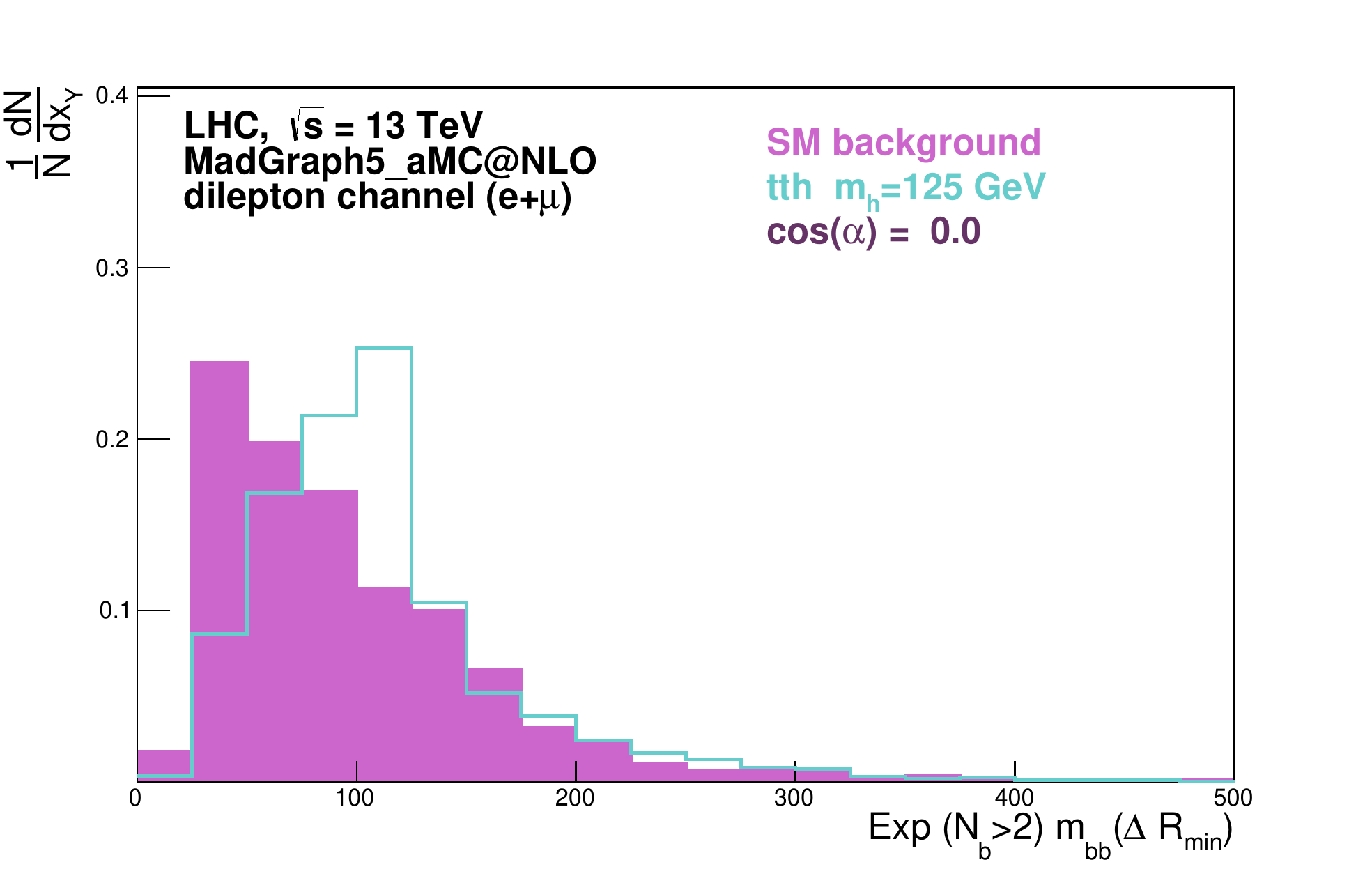,height=5.0cm,clip=} \\[-.5mm]
\epsfig{file=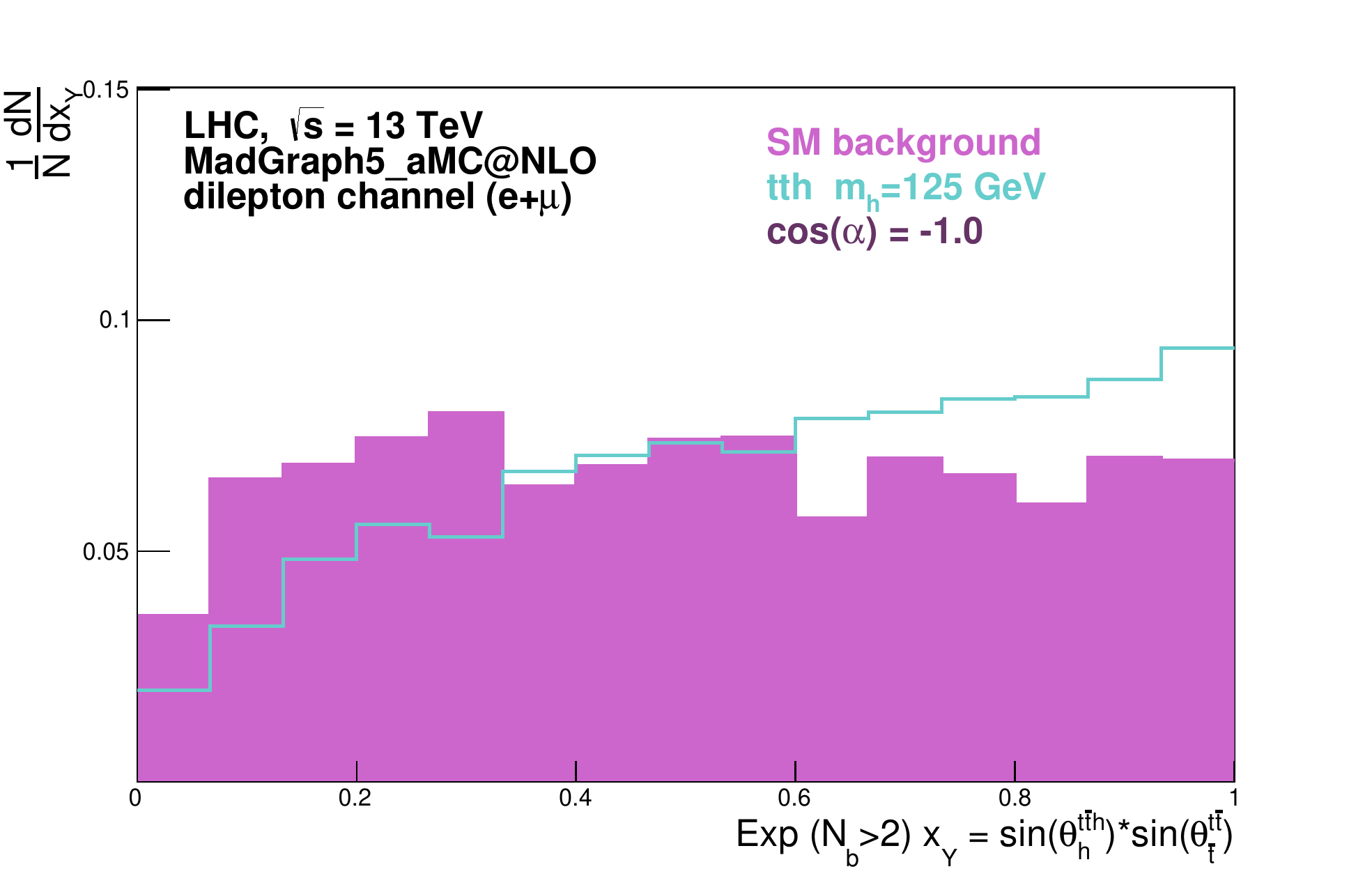,height=5.0cm,clip=} & \quad & 
\epsfig{file=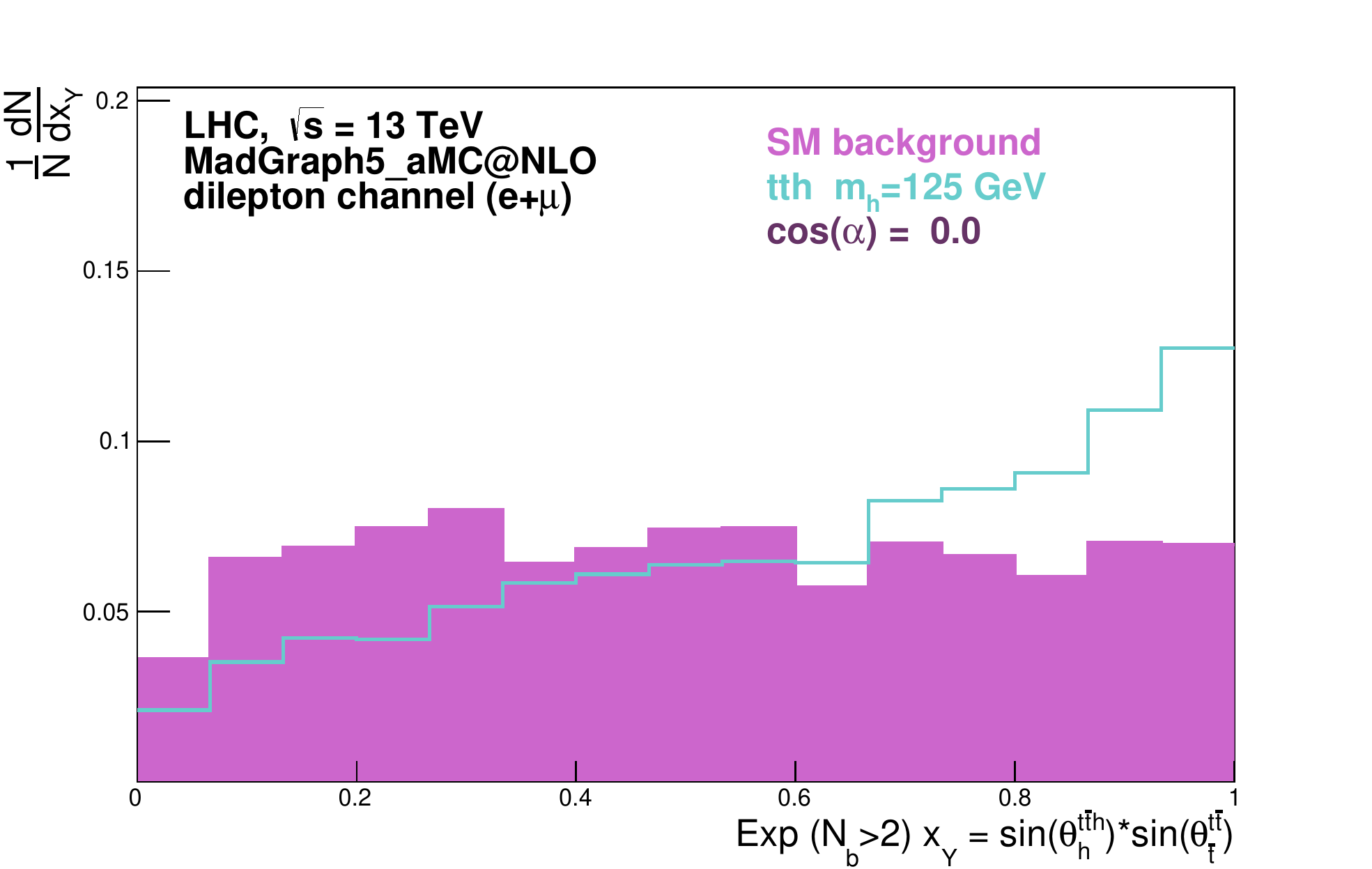,height=5.0cm,clip=} \\[-.5mm]
\epsfig{file=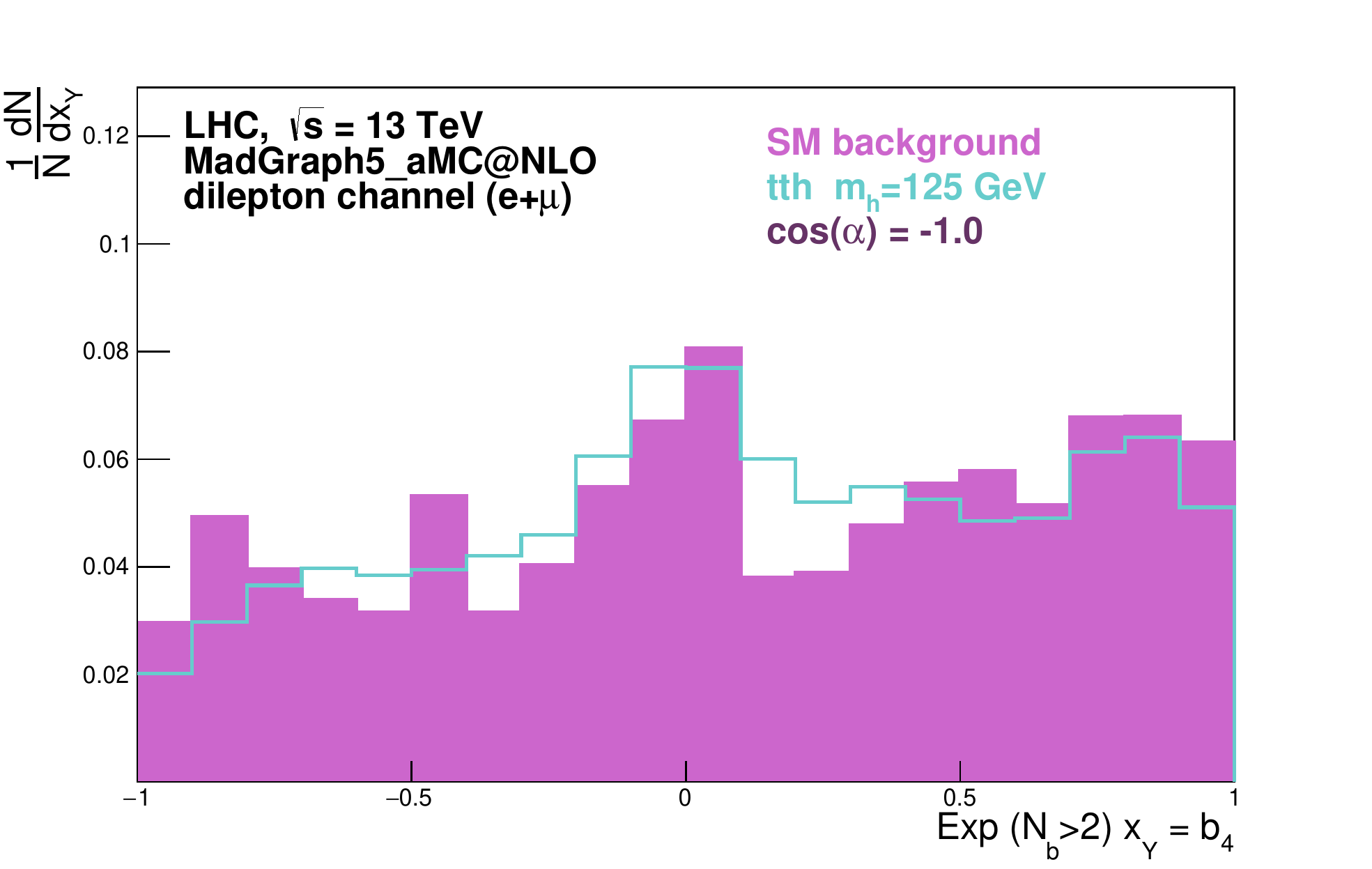,height=5.0cm,clip=} & \quad & 
\epsfig{file=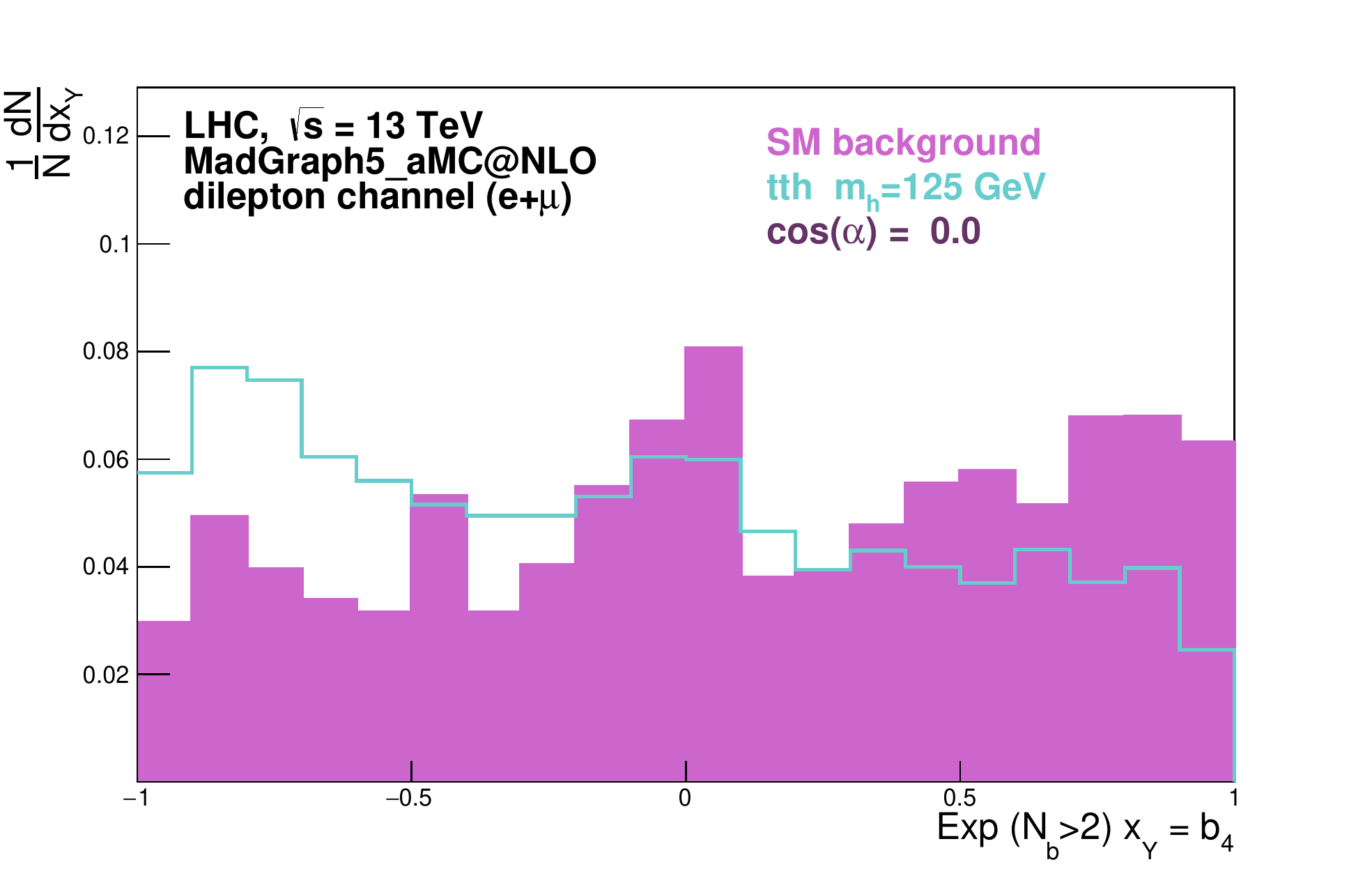,height=5.0cm,clip=} \\[-.5mm]
\end{tabular}
\caption{Normalized distributions of the BDTG output discriminant variable (first row),
the invariant mass of the two $b$-tagged jets with minimum $\Delta R$ ($m_{bb}^{\min\Delta R}$) (second row), the
$sin(\theta^{t\bar{t}h}_{h})sin(\theta^{t\bar{t}}_{\bar{t}})$ (third row) and the $b_4$ variable (fourth row),
after final selection at 13 TeV. The distributions on the left (right) corresponds to pure scalar (pseudo-scalar) Higgs bosons.}
\label{fig:NewDiscVar01}
\end{center}
\end{figure*}

In Figure~\ref{fig:NewDiscVar01}, normalized distributions of the BDTG output classifier (first row) and three of the input variables (remaining rows) used in the multivariate method are shown for the pure scalar (left plots) and pseudo-scalar (right plots) Higgs bosons. It should be noted that the BDTG used for the limit extraction at a given $\cos(\alpha)$ has been trained on a signal sample generated with that same value $\cos(\alpha)$. This justifies the different SM background shapes between the left and right plots of the first row in Figure~\ref{fig:NewDiscVar01}. The invariant mass of the two $b$-tagged jets with minimum $\Delta R$ ($m_{bb}^{\min\Delta R}$) (second row), the $sin(\theta^{t\bar{t}h}_{h})sin(\theta^{t\bar{t}}_{\bar{t}})$ (third row) and the $b_4$ variable (fourth row) are also shown for completeness. Shape differences between signal and background are clearly visible, and they are different for the scalar and pseudo-scalar cases. In these figures, the line corresponds to the signal distribution and the shaded region corresponds to the full SM background at the LHC.

\begin{figure*}
\begin{center}
\begin{tabular}{ccc}
\epsfig{file=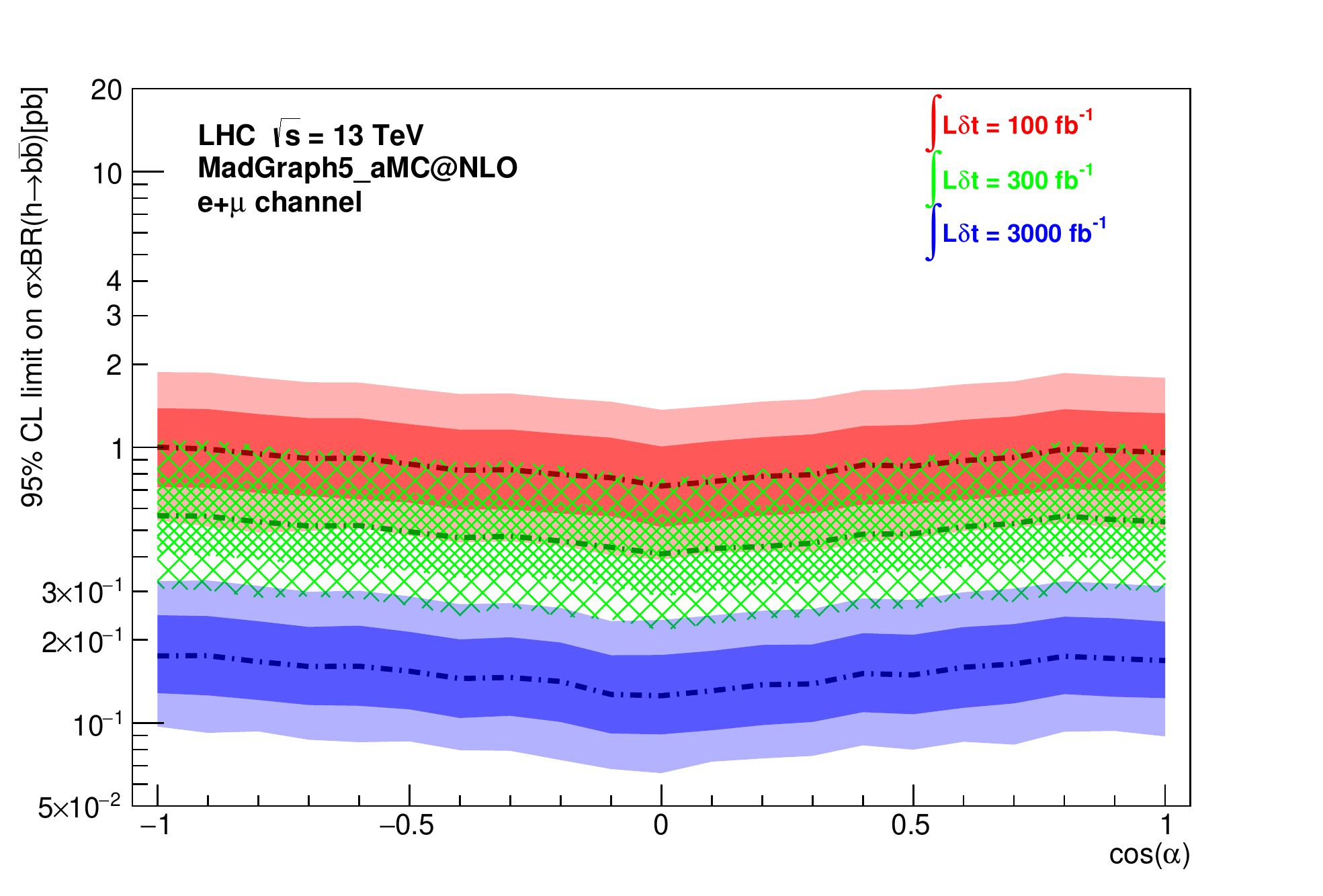,height=6.0cm,clip=} & \quad & 
\epsfig{file=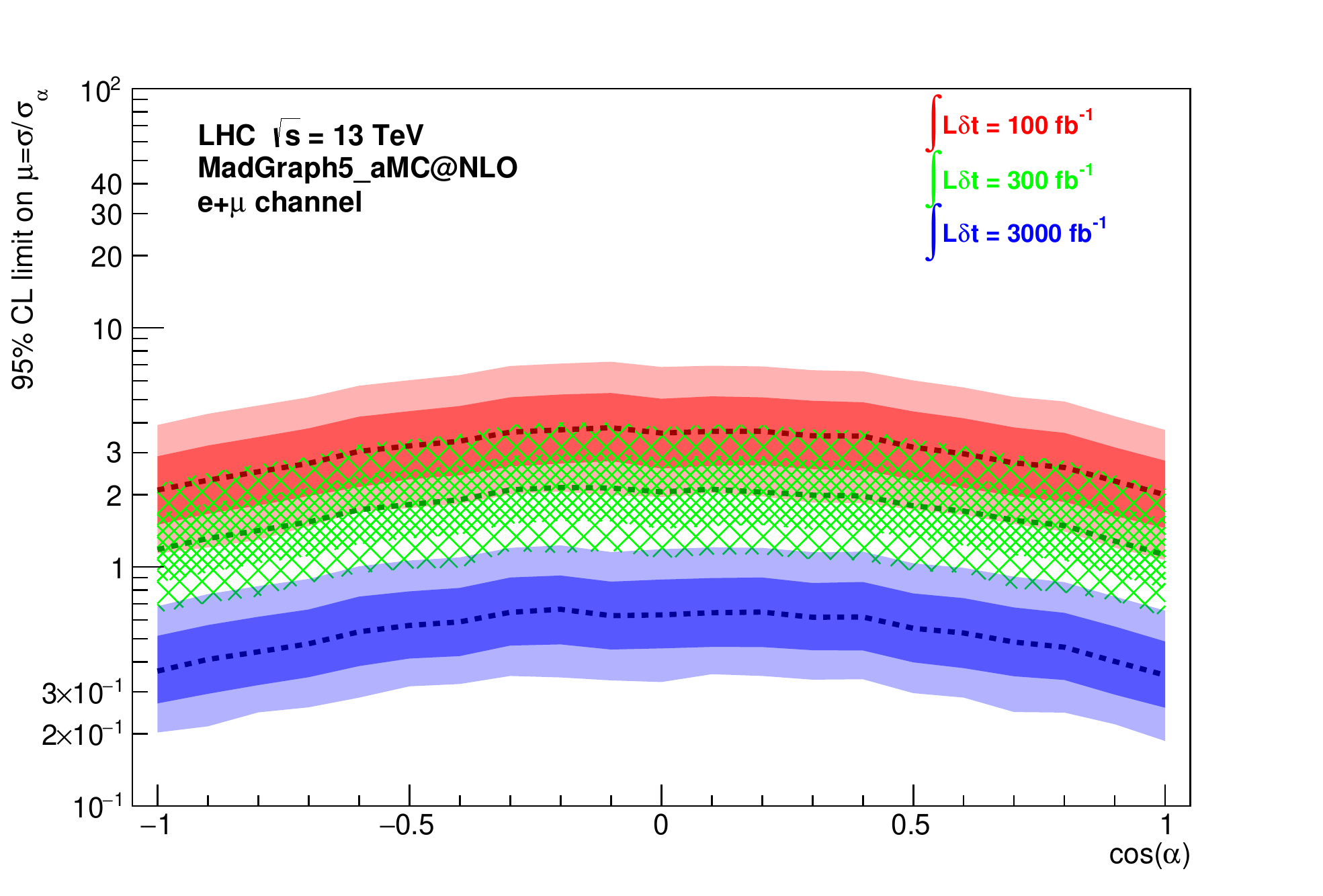,height=6.0cm,clip=}\\[-.5mm]
\epsfig{file=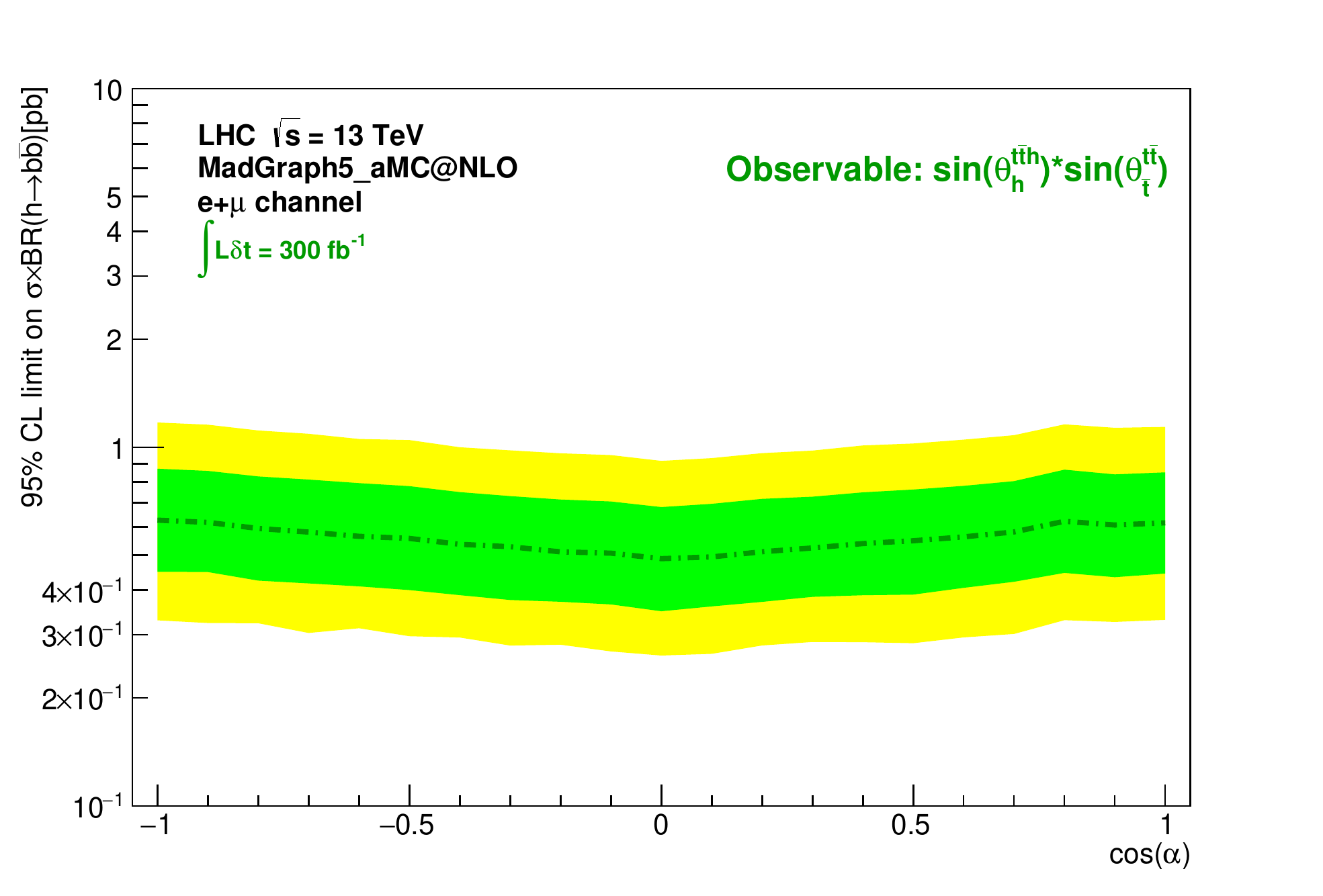,height=6.0cm,clip=} & \quad & 
\epsfig{file=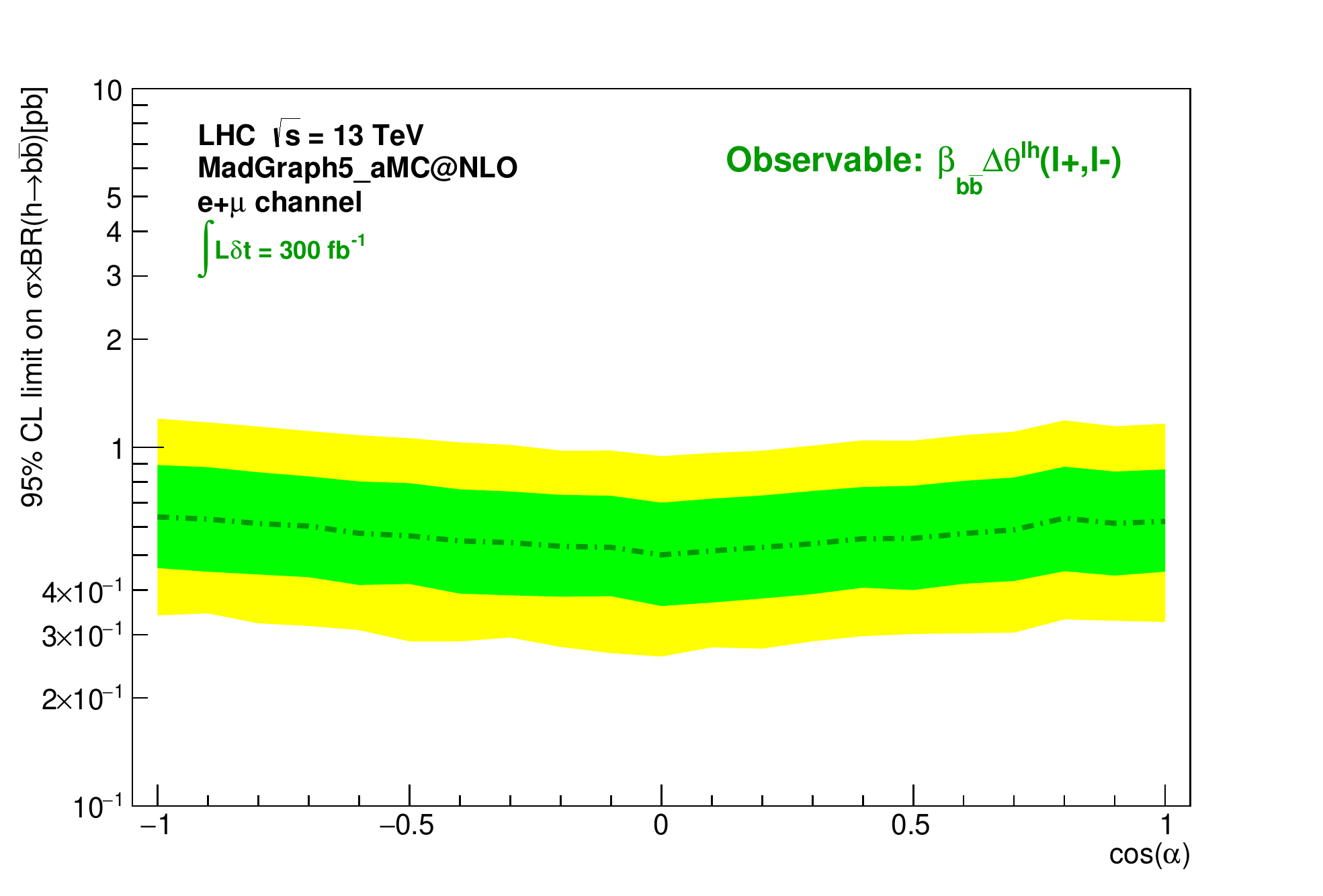,height=6.0cm,clip=} \\[-.5mm]
\epsfig{file=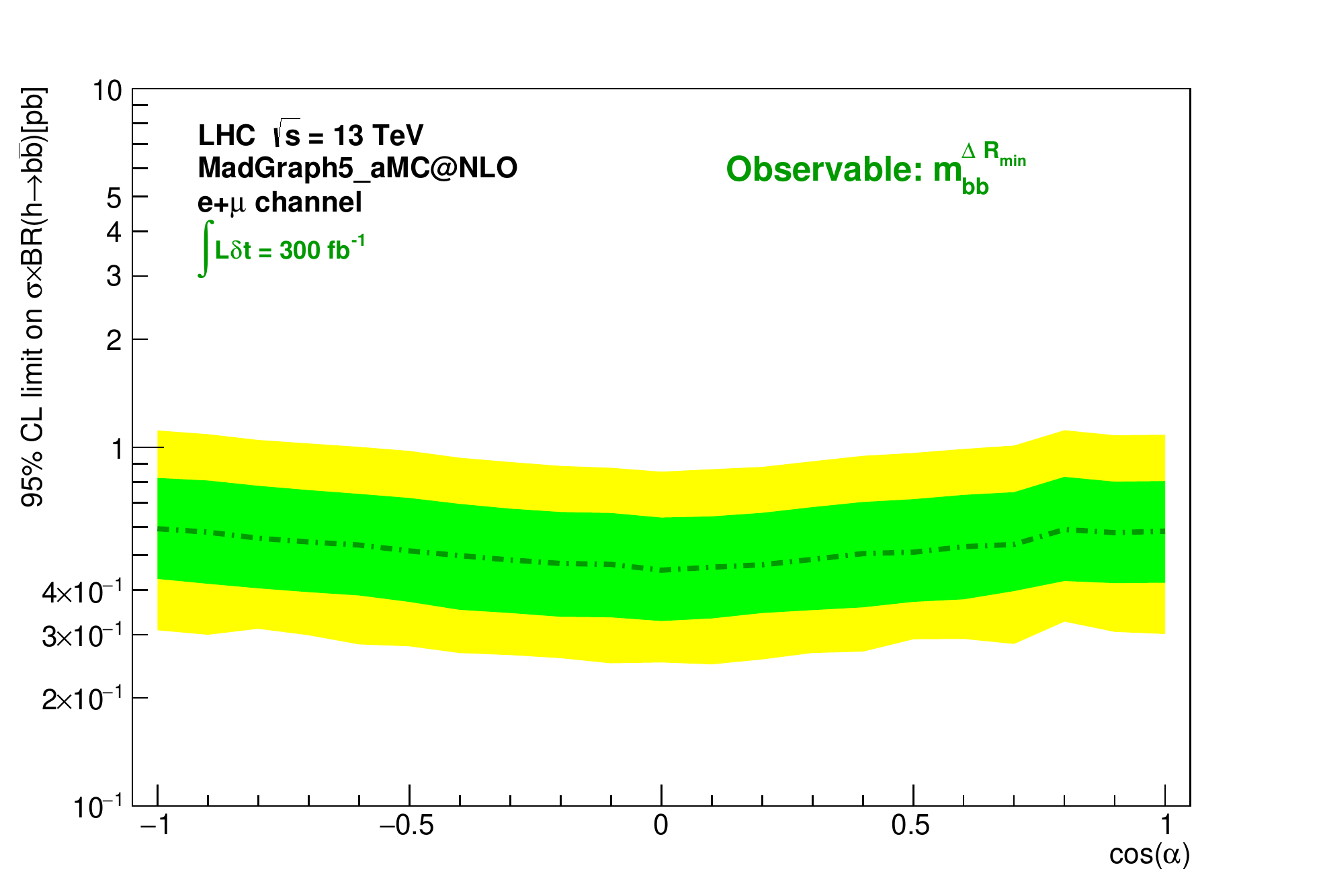,height=6.0cm,clip=} & \quad & 
\epsfig{file=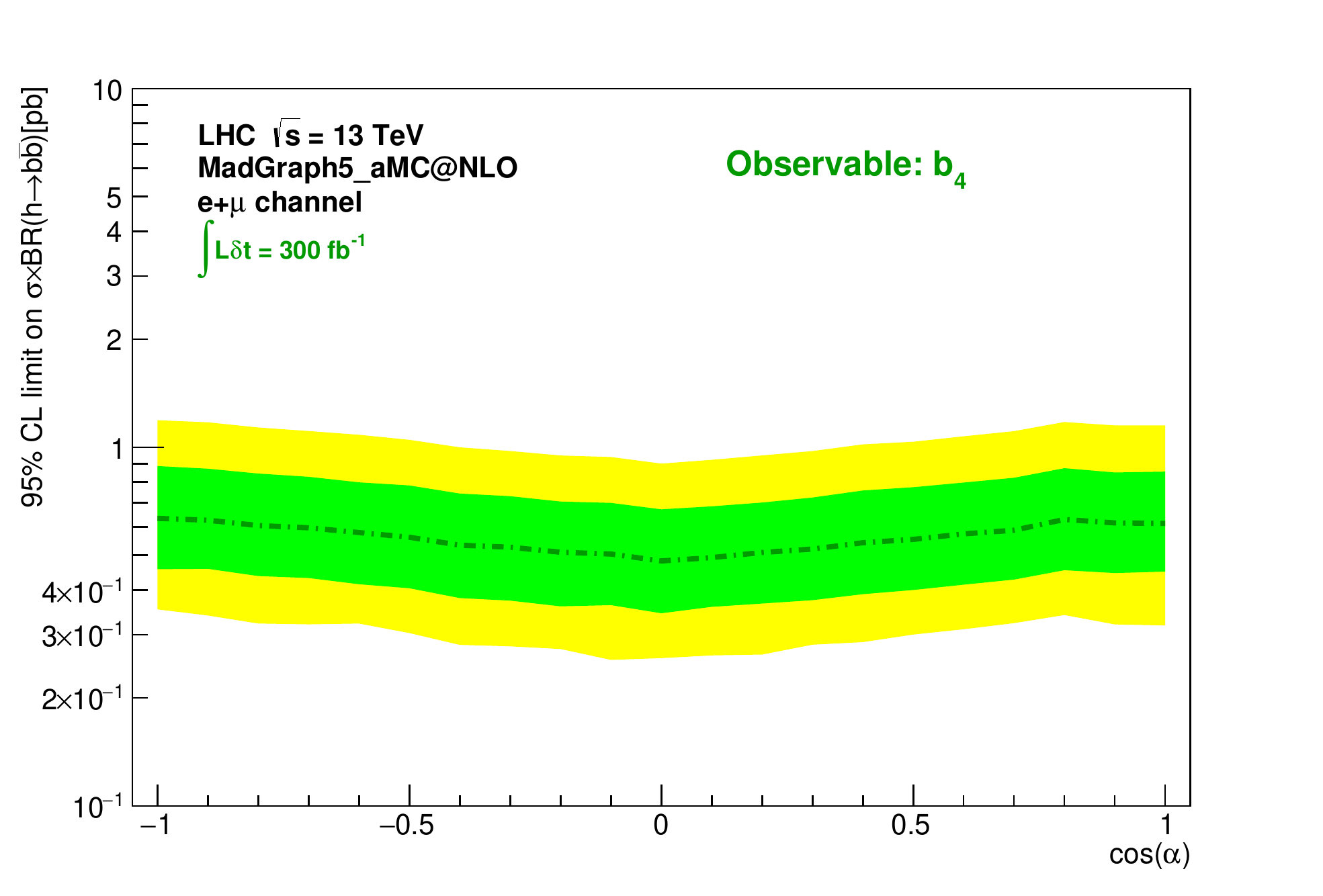,height=6.0cm,clip=} \\[-.5mm]
\end{tabular}
\caption{Expected limits at 95\% CL in the background-only scenario, as a function of $\cos(\alpha)$. Limits on $\sigma\times BR(h\rightarrow b\bar{b})$ (top left) and $\mu$ (top right) obtained with the BDTG output discriminant for integrated luminosites of 100, 300 and 3000 fb$^{-1}$. The lines correspond to the median, while the narrower (wider) bands correspond to the 1$\sigma$(2$\sigma$) intervals. Limits on $\sigma\times BR(h\rightarrow b\bar{b})$, at 300 fb$^{-1}$, using individual observables: $sin(\theta^{t\bar{t}h}_{h})sin(\theta^{t\bar{t}}_{\bar{t}})$ (center left) and $\beta_{b\bar{b}}\Delta\theta^{\ell h}(\ell^+,\ell^-)$ (center right); $m_{bb}^{\min\Delta R}$ (bottom left) and $b_4$ (bottom right).}
\label{fig:Limits01}
\end{center}
\end{figure*}

\begin{figure*}
\begin{center}
\begin{tabular}{ccc}
\epsfig{file=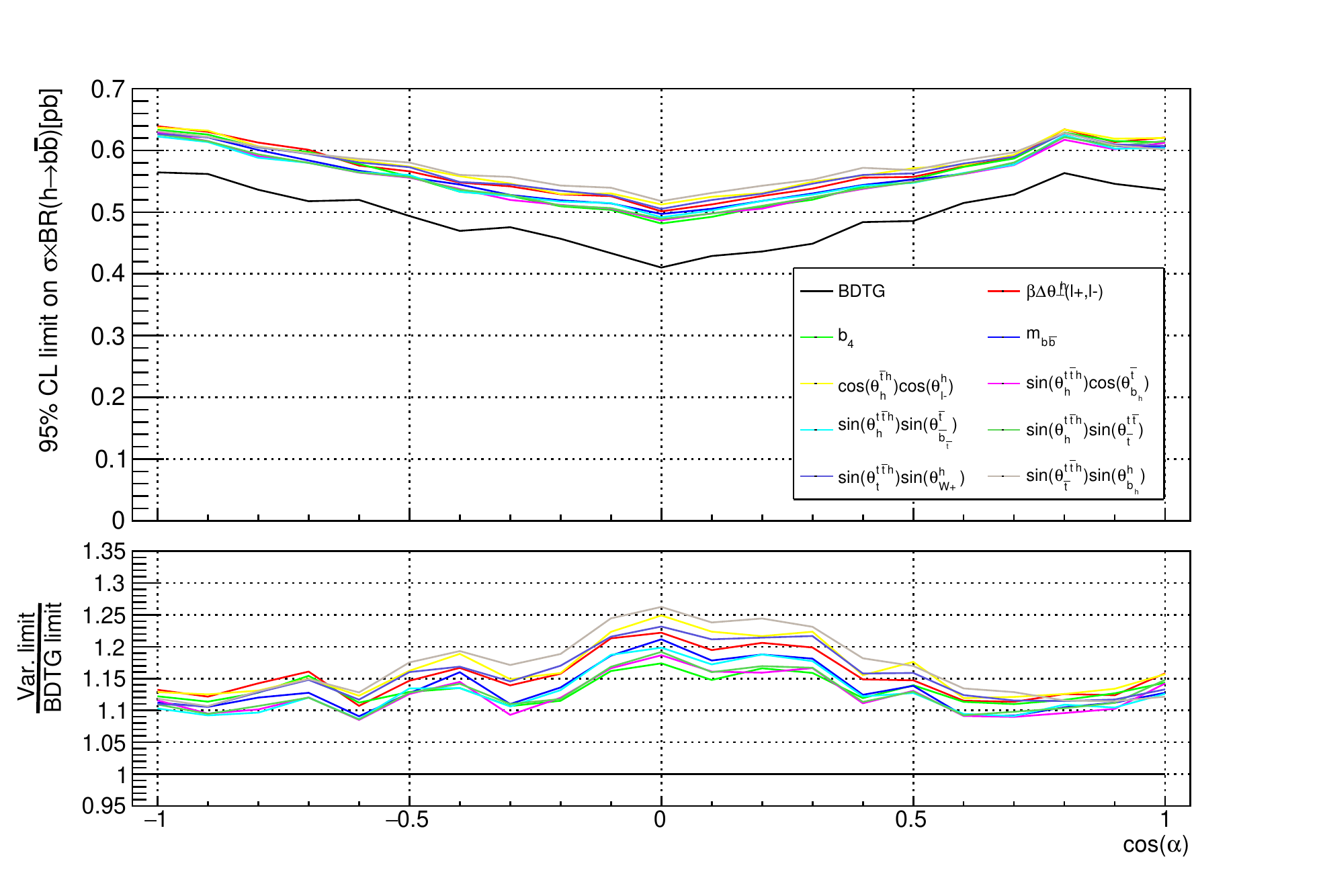,height=6.0cm,clip=} & \quad & 
\epsfig{file=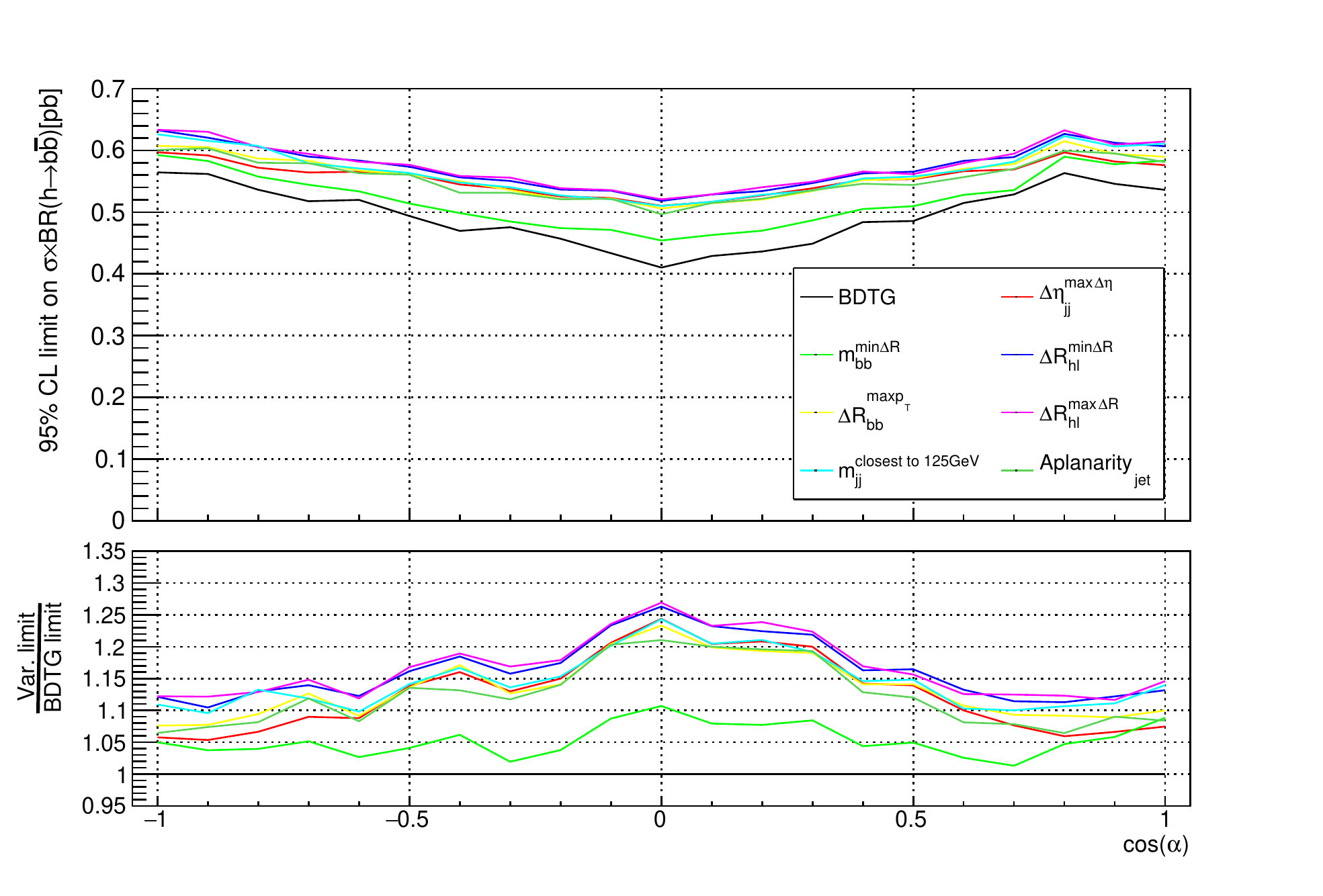,height=6.0cm,clip=} \\[-2mm]
\end{tabular}
\caption{Comparison between limits on $\sigma\times BR(h\rightarrow b\bar{b})$, at 300 fb$^{-1}$, obtained from each one of the individual distributions used in the BDTG: $\beta_{b\bar{b}}\Delta\theta^{\ell h}(\ell^+,\ell^-)$, $b_4$, $m_{b\bar{b}}$ and angular distributions (left), and remaining distributions used as input for the BDTG (right). The ratios with respect to the limit obtained from the BDTG distribution are also represented.}
\label{fig:Limits02}
\end{center}

\end{figure*}

Expected limits at 95\% confidence level (CL) for $\sigma\times BR(h\rightarrow b\bar{b})$ and for signal strength $\mu$, in the background-only scenario, were extracted, using the BDTG output distribution. Several signal samples were used, with values of $\cos(\alpha)$ ranging from -1 to 1 (in steps of 0.1). Figure~\ref{fig:Limits01}, first row, shows these limits, for integrated luminosities of 100, 300 and 3000 fb$^{-1}$. Although data taking for large values of luminosity is expected to occur with $\sqrt{s}$=14~TeV, we show the results at 3000 fb$^{-1}$ for comparison. Sensitivity to SM $t\bar{t}H$ production at the $\mu$=1 should be attained shortly after the 300 fb$^{-1}$ milestone, using this channel alone. Combining the dileptonic channel with other decay channels should allow to decrease significantly the luminosity necessary to probe the structure of the top quark Yukawa couplings to the Higgs boson. Figure~\ref{fig:Limits01}, second and third row, show limits on $\sigma\times BR(h\rightarrow b\bar{b})$ at 300 fb$^{-1}$, obtained from fits to the following individual distributions: $sin(\theta^{t\bar{t}h}_{h})sin(\theta^{t\bar{t}}_{\bar{t}})$ (center left), $\beta_{b\bar{b}}\Delta\theta^{\ell h}(\ell^+,\ell^-)$ (center right); $m_{bb}^{\min\Delta R}$ (bottom left), and $b_4$ (bottom right). The results show that the different distributions used as input to the BDTG, although with the same general dependence on $\cos(\alpha)$, can have different sensitivities. A common feature to all variables is a better 95\% CL limit on $\sigma\times BR(h\rightarrow b\bar{b})$ as we approach the pure pseudo-scalar region. In Figure~\ref{fig:Limits02}, a comparison is shown between limits on $\sigma\times BR(h\rightarrow b\bar{b})$, at 300 fb$^{-1}$, obtained from each one of the individual distributions used in the BDTG multivariate discriminant. Additionally, the limits corresponding to the BDTG itself are shown, as well as those from the $\beta_{b\bar{b}}\Delta\theta^{\ell h}(\ell^+,\ell^-)$ distribution, which is not included in the BDTG, and is only shown for completeness. Figure~\ref{fig:Limits02} (left) includes the limits from the angular observables, $\beta_{b\bar{b}}\Delta\theta^{\ell h}(\ell^+,\ell^-)$, $b_4$ and $m_{b\bar{b}}$. Figure~\ref{fig:Limits02} (right) shows the limits from all the other individual observables used as input for the BDTG method. The ratios with respect to the limit obtained from the BDTG distribution are also represented. While most individual angular variables result in limits 15 to 20\% worse than the BDTG method for the pseudo-scalar case, the other variables tend to be in the 20 to 25\% region, with the exception of $m_{bb}^{\min\Delta R}$, which clearly shows a better discriminating power (as expected from the plots in Figure~\ref{fig:NewDiscVar01}).

\section{Conclusions}

In this paper, studies of the $t{\bar t}h$ production, for scalar and pseudo-scalar Higgs bosons, at a centre-of-mass energy of 13~TeV at the LHC, are considered for different luminosities. Dileptonic final states from $t{\bar t}h$ decays ($t\rightarrow bW^+ \rightarrow b \ell^+\nu_\ell$, $\bar{t}\rightarrow \bar{b}W^- \rightarrow \bar{b} \ell^-\bar{\nu}_\ell$ and $h\rightarrow b\bar{b}$) are fully reconstructed by means of a kinematic fit that reconstructs the four momenta of the undetected neutrinos. New angular distributions and asymmetries are proposed to allow better discrimination between signals of different nature (scalar or pseudo-scalar) and backgrounds at the LHC. Using fully reconstructed $t\bar{t}h$ events, it is possible to obtain relevant spin information of signal and background processes, through the measurements of new angular distributions and asymmetries. Even after event selection and full kinematical reconstruction, the spin information is largely preserved, opening a window for spin measurements and a better understanding of the nature of the top-Higgs Yukawa coupling and $t\bar{t}h$ production at the LHC. Expected limits at 95\% CL were extracted on the $\sigma\times BR(h\rightarrow b\bar{b})$ and signal strength $\mu$ using a boosted decision tree. A comparison between the sensitivities of the individual variables as a function of $\cos(\alpha)$ was also performed, showing that a multivariate method combining all the variables can improve the individual limits up to 25\%. It should be stressed that some of the angular distributions investigated in this work were used in addition to the kinematical distributions commonly discussed in the literature, yielding at least the same sensitivity to the nature of the top quark Yukawa coupling to Higgs boson, if not better. The fact the expected limits do not exhibit a too strong dependence on the particular choice of the CP-phase ($\alpha$), makes the analysis of the SM Higgs case (CP-even) a good starting point for any other case, where mixtures with CP-odd contributions are probed. Also, it was found that the invariant mass distribution of the two $b$-tagged jets with the lowest $\Delta R$ between them shows a particularly interesting behaviour. All results presented so far were obtained using the dileptonic final states of $t\bar{t}h$ events alone. These are expected to be improved when other decay channels are combined, using fully reconstructed final states. 

\section*{Acknowledgements}

This work was partially supported by Funda\c{c}\~ao para a Ci\^encia e Tecnologia, FCT (projects CERN/FIS-NUC/0005/2015 and CERN/FP/123619/2011, grant SFRH/BPD/100379/2014 and contract IF/01589/2012/CP0180/CT0002). The work of R.S. is supported in part by HARMONIA National Science Center - Poland project UMO-2015/18/M/ST2/00518. The work of R.F.~is supported by the Alexander von Humboldt Foundation in the framework of the Sofja Kovalevskaja Award Project ``Event Simulation for the Large Hadron Collider at High Precision''. Special thanks goes to our long term collaborator Filipe Veloso for the invaluable help and availability on the evaluation of the confidence limits discussed in this paper.


%
\newpage
%
%
%
\newpage
%
%
\newpage
%
%
\newpage
%
%
\newpage
%
%

\newpage
%

\newpage

%
%
\newpage
%
%
\newpage

\end{document}